\documentclass[useAMS,usenatbib]{mnras}    

\usepackage[caption=false]{subfig}
\usepackage[final]{graphicx}
\usepackage{epstopdf}
\usepackage{mathptmx}
\usepackage{float}
\usepackage{fixltx2e}
\usepackage{natbib}
\usepackage{amsmath}
\usepackage{amssymb}
\usepackage[usenames,dvipsnames]{color}
\usepackage{appendix}
\usepackage[mathscr]{euscript}
\usepackage{afterpage}

\graphicspath{{figures/}}

\newcommand{\mvir}{M_{\text{200c}}^\text{z = 0}}
\newcommand{\mhostzz}{M_{\text{200c}}^\text{z = 0}}
\newcommand{\rvir}{r_{200c}}

\newcommand{\msun}{\mathrm{M}_\odot}

\newcommand{\mpeak}{M_\text{tot}^\text{peak}}
\newcommand{\mgalpeak}{M_\text{tot}^\text{peak}}
\newcommand{\mgalpeakbr}{M_\text{tot}^\text{peak}}
\newcommand{\mgalzz}{M_\text{tot}^\text{z = 0}}
\newcommand{\mgalz}{M_\text{tot}^{z = 0}}
\newcommand{\mhost}{\mvir}

\newcommand{\zacc}{z_\text{acc}}

\newcommand{\mgalstar}{M_{\star}^\text{peak}}
\newcommand{\mgalstarpeak}{M_{\star}^\text{peak}}
\newcommand{\mgalstarzz}{M_{\star}^\text{z = 0}}
\newcommand{\mgalstarpeakbr}{M_{\star}^\text{peak}}
\newcommand{\mtot}{M_\text{tot}}

\newcommand{\eagle}{EAGLE}

\newcommand{\subfind}{\textsc{Subfind}}
\newcommand{\spiderweb}{\textsc{Spiderweb}}

\title[Survival of satellite galaxies]{Disruption of satellite galaxies in simulated groups and clusters: the roles of accretion time, baryons, and pre-processing}

\author[Y.~Bah\'{e} et al.]{\parbox[t]{\textwidth}{
Yannick M.~Bah\'{e}$^{1}$\thanks{\href{mailto:bahe@strw.leidenuniv.nl}{bahe@strw.leidenuniv.nl}}, 
Joop Schaye$^{1}$, David J.~Barnes$^{2}$, Claudio Dalla Vecchia$^{3, 4}$, \mbox{Scott T.~Kay$^{5}$}, Richard G.~Bower$^{6}$, Henk Hoekstra$^{1}$, Sean L.~McGee$^{7}$, and Tom Theuns$^{6}$ \vspace*{12pt}}
\\
$^1$ Leiden Observatory, Leiden University, PO Box 9513, 2300 RA Leiden, The Netherlands\\
$^2$ Department of Physics, Kavli Institute for Astrophysics and Space Research, Massachusetts Institute of Technology, Cambridge, MA 02139, USA\\
$^3$ Instituto de Astrof\'{i}sica de Canarias, C/V\'{i}a L\'{a}ctea s/n, E-38205 La Laguna, Tenerife, Spain\\
$^4$ Departamento de Astrof\'{i}sica, Universidad de La Laguna, Av. del Astrof\'{i}sico Francisco S\'{a}nchez s/n, E-38206 La Laguna, Tenerife, Spain\\
$^5$ Jodrell Bank Centre for Astrophysics, School of Physics and Astronomy, The University of Manchester, Manchester M13 9PL, UK\\
$^6$ Institute for Computational Cosmology, Department of Physics, University of Durham, South Road, Durham DH1 3LE, UK\\
$^7$ School of Physics and Astronomy, University of Birmingham, Edgbaston, Birmingham, B15 2TT, UK\\
}

\date{Accepted 2019 January 31. Received 2019 January 10; in original form 2018 October 30}

\pubyear{2019}

\begin{document}
\label{firstpage}
\pagerange{\pageref{firstpage}--\pageref{lastpage}}
\maketitle

\begin{abstract}
We investigate the disruption of group and cluster satellite galaxies with total mass (dark matter plus baryons) above $10^{10}\,\msun$ in the Hydrangea simulations, a suite of 24 high-resolution cosmological hydrodynamical zoom-in simulations based on the EAGLE model. The simulations predict that $\sim$50 per cent of satellites survive to redshift $z = 0$, with higher survival fractions in massive clusters than in groups and only small differences between baryonic and pure $N$-body simulations. For clusters, up to 90 per cent of galaxy disruption occurs in lower-mass sub-groups (i.e., during pre-processing); 96 per cent of satellites in massive clusters that were accreted at $z < 2$ and have not been pre-processed survive. Of those satellites that are disrupted, only a few per cent merge with other satellites, even in low-mass groups. The survival fraction changes rapidly from less than 10 per cent of those accreted at high $z$ to more than 90 per cent at low $z$. This shift, which reflects faster disruption of satellites accreted at higher $z$, happens at lower $z$ for more massive galaxies and those accreted onto less massive haloes. The disruption of satellite galaxies is found to correlate only weakly with their pre-accretion baryon content, star formation rate, and size, so that surviving galaxies are nearly unbiased in these properties. These results suggest that satellite disruption in massive haloes is uncommon, and that it is predominantly the result of gravitational rather than baryonic processes.
\end{abstract}



\section{Introduction}
\label{sec:introduction}

A key prediction of the concordance $\Lambda$ Cold Dark Matter ($\Lambda$CDM) cosmology is that dark matter structures form hierarchically: small objects collapsed first and then built up successively more massive structures through mergers (e.g.,~\citealt{Press_Schechter_1974, Searle_Zinn_1978, White_Rees_1978}). Galaxy groups and clusters represent the highest level of this hierarchy at the present day, built up from the largest number of individual accreted galaxies\footnote{We here use the term `galaxy' to refer to distinct self-bound objects, irrespective of their mass or composition. A galaxy therefore includes the dark matter halo as well as stellar component and gas reservoir, where they exist.}. Once accreted, galaxies are subject to mass loss due to tidal forces and ram pressure stripping, while dynamical friction can drive them towards the centre of their host halo and therefore enhance the mass loss yet further (see, e.g., \citealt{Binney_Tremaine_2008}). In this way, the galaxy may be reduced to a mass below a given detection threshold, or even disrupted completely (e.g., \citealt{Hayashi_et_al_2003}).

Understanding the extent to which satellite galaxies survive this mass loss is desirable for a number of reasons. It allows measuring the halo mass from the abundance of galaxies (see, e.g.,~\citealt{Rozo_et_al_2009, Budzynski_et_al_2012, Rykoff_et_al_2014, Andreon_2015, Saro_et_al_2015}) or kinematics (e.g.,~\citealt{Zhang_et_al_2011, Bocquet_et_al_2015, Sereno_Ettori_2015}; see also \citealt{Armitage_et_al_2018}). Detailed characterisation of substructure is one of the most promising avenues to constrain the nature of dark matter (e.g.,~\citealt{Randall_et_al_2008, Lovell_et_al_2012, Vegetti_et_al_2012, Harvey_et_al_2015, Robertson_et_al_2018}). Finally, satellite galaxies differ from isolated galaxies of the same stellar mass in key aspects, such as their colour (e.g.,~\citealt{Peng_et_al_2010}), star formation rate (e.g.,~\citealt{Kauffmann_et_al_2004, Wetzel_et_al_2012}), and morphology (e.g.,~\citealt{Dressler_1980}). The detailed origins of these differences are still unsolved puzzles, which also requires understanding to what extent satellites survive at all: if, for example, survival correlates with galaxy properties prior to infall, this may (partly) explain the aforementioned differences. 

Because of its complexity, this problem needs to be addressed with numerical simulations (see, e.g.,~\citealt{vanDenBosch_Ogiya_2018}). Since the late 1990s, these have achieved sufficiently high resolution to avoid ubiquitous numerical dissolution of satellite galaxies (or `subhaloes'; e.g.,~\citealt{Ghigna_et_al_1998, Moore_et_al_1999, Springel_et_al_2001b, Springel_et_al_2008, Gao_et_al_2012}), which prompted a multitude of studies that analysed their evolution and survival in detail (e.g.,~\citealt{Tormen_et_al_1998, DeLucia_et_al_2004, Gao_et_al_2004, Weinberg_et_al_2008, Dolag_et_al_2009, Xie_Gao_2015, Chua_et_al_2017, vanDenBosch_2017}). The qualitatively consistent conclusion from these studies is that subhaloes survive for a limited amount of time, with the lowest survival rate (i.e.,~fastest disruption) at both the highest and lowest ends of the subhalo mass range. The  majority of surviving subhaloes in massive clusters were therefore accreted relatively recently, at $z < 1$ \citep{DeLucia_et_al_2004, Gao_et_al_2004}. Of those that were accreted earlier, only a small fraction was typically predicted to survive to $z = 0$: \citet{Gao_et_al_2004} and \citet{Jiang_vanDenBosch_2017}, for instance, both found that only 10 per cent of simulated subhaloes accreted at $z = 2$ could still be identified at $ z = 0$. 

An inherent limitation in all numerical studies is that limited resolution precludes the identification of subhaloes below a limiting mass, even if they are physically not completely disrupted. If survival is defined as the subhalo retaining at least a given number of particles (e.g.,~\citealt{Gao_et_al_2004, Xie_Gao_2015, vanDenBosch_2017}) or a minimum mass set by the resolution of the simulation (e.g.,~\citealt{Chua_et_al_2017}), simulations with higher resolution predict higher survival fractions: for example, \citet{Xie_Gao_2015} found that in the Phoenix dark matter only galaxy cluster simulations \citep{Gao_et_al_2012}, which resolve each cluster with $\sim$$10^8$ particles, more than half of all subhaloes with mass above $10^{10}\, \msun$ accreted at $z = 2$ survive to the present day.  

A more subtle consequence of numerical resolution has been pointed out in a recent series of papers by \citet{vanDenBosch_2017}, \citet{vanDenBosch_et_al_2018}, and \citet{vanDenBosch_Ogiya_2018}: they found that the complete disruption of subhaloes should, physically, be extremely rare and that numerical artefacts can occur even well above the nominal resolution limit of a simulation. Through a suite of idealised $N$-body experiments, \citet{vanDenBosch_Ogiya_2018} demonstrated that inadequate force softening -- i.e.,~spatial resolution -- and particle numbers -- i.e.,~mass resolution -- both act to accelerate the tidal disruption of subhaloes, even when they are `well resolved' with $\gtrsim\,$100 particles. Due to the extremely demanding resolution requirements found to be necessary to prevent such numerical disruption, \citet{vanDenBosch_Ogiya_2018} argued that this constitutes a serious road-block on the path to understanding the evolution of satellite galaxies.

Another limitation in many of the aforementioned simulations is the neglect of baryons. Ram pressure can efficiently remove gas from infalling galaxies \citep{Gunn_Gott_1972}, making them more susceptible to disruption (e.g.,~\citealt{Saro_et_al_2008}), while gas cooling and star formation may have a stabilising effect through the formation of dense cores, which are more difficult to disrupt. Non-radiative hydrodynamical simulations have given discrepant answers about the impact of gas removal on subhalo survival, with some finding it to be more relevant \citep{Saro_et_al_2008, Dolag_et_al_2009} than others \citep{Tormen_et_al_2004}.  

The modelling of additional baryonic effects, such as gas cooling, star formation, and its associated energy feedback remains uncertain (see, e.g.,~\citealt{Scannapieco_et_al_2012} and the discussion in \citealt{Schaye_et_al_2015}) and cosmological hydrodynamical simulations accounting for them have long struggled to produce even realistic \emph{isolated} galaxies. They have therefore, perhaps unsurprisingly, led to a variety of contradictory conclusions about the net effect of baryons on satellite survival: \citet{Weinberg_et_al_2008} found that their inclusion increases survival, particularly in low-mass galaxies, while \citet{Dolag_et_al_2009} concluded that the effect of gas cooling and star formation is largely cancelled by the disruptive effect of gas stripping. The Illustris simulation \citep{Vogelsberger_et_al_2014} predicts a net disruptive effect of baryons \citep{Chua_et_al_2017}.

With an improved implementation of energy feedback that largely overcomes numerical cooling losses \citep{DallaVecchia_Schaye_2012}, and by calibrating the uncertain subgrid prescriptions against observational relations in the local Universe, the EAGLE project \citep{Schaye_et_al_2015} has produced a population of galaxies that match not only these calibration diagnostics, but also their evolution to high redshift \citep{Furlong_et_al_2015, Furlong_et_al_2017} and a wide range of other observables including galaxy colours \citep{Trayford_et_al_2015, Trayford_et_al_2017}, star formation rates \citep{Schaye_et_al_2015}, and neutral gas content \citep{Lagos_et_al_2015, Bahe_et_al_2016, Marasco_et_al_2016, Crain_et_al_2017}. This model therefore provides realistic initial conditions to study the evolution of satellite galaxies. 

The Hydrangea simulation suite applies this successful model to the scale of galaxy clusters by combining it with the zoomed initial conditions technique (e.g., \citealt{Katz_White_1993}). Despite some tensions in the mass of their simulated central cluster galaxies \citep{Bahe_et_al_2017b} and hot gas fractions \citep{Barnes_et_al_2017b}, the $z = 0.1$ satellite stellar mass function agrees remarkably well with observations, down to stellar masses far below that of the Milky Way \citep{Bahe_et_al_2017b}. This suggests that the fraction of satellites that survive to the present day is modelled correctly. The Hydrangea suite therefore allows us to study the evolution of satellites in a realistic way, over a wide range of host and galaxy masses.

With this tool, we revisit the question of satellite survival in massive haloes. We aim to address in particular the following three questions: (i) What fraction of accreted satellites survive to $z = 0$, and how does this depend on accretion time, galaxy mass, and host mass?  How important, therefore, is satellite disruption\footnote{Throughout this paper, we use `disruption' as antonym to `survival'. It therefore refers to the dispersal of galaxies into their host halo as well as to mergers with another galaxy.} in a simulation suite that is characteristic of the current state of the art in cosmological hydrodynamical simulations that include massive clusters (see also, e.g.,~\citealt{Pillepich_et_al_2018} and \citealt{Tremmel_et_al_2019})? (ii) What is the predicted effect of baryons on galaxy survival? (iii) What is the role of environmental effects on galaxies prior to accretion onto their (final) halo? This `pre-processing' step (e.g.,~\citealt{Fujita_2004, Berrier_et_al_2009, McGee_et_al_2009, Balogh_McGee_2010}) has been identified as a key stage in the evolution of cluster galaxies (e.g.,~\citealt{Zabludoff_Mulchaey_1998, Berrier_et_al_2009, McGee_et_al_2009, Bahe_et_al_2013, Wetzel_et_al_2013, Han_et_al_2018}), but to our knowledge no study has so far examined its role in satellite disruption. 

The remainder of this paper is structured as follows. Section \ref{sec:sims} summarises the key aspects of the Hydrangea simulations and the relevant post-processing steps, including an overview of our new method to trace simulated galaxies through time. The predicted survival fractions are presented in Section \ref{sec:survival}, followed by an analysis of the roles of pre-processing, satellite--satellite mergers, and galaxy accretion time in Section \ref{sec:influence}. We investigate the influence of galaxy properties prior to accretion on their survival in Section \ref{sec:bias}, and summarize our conclusions in Section \ref{sec:summary}. In appendices, we provide a detailed description of our new tracing method (Appendix \ref{app:spiderweb}), a verification of the robustness of our results against numerical limitations (Appendix \ref{app:robustness}), and a comparison to the numerical experiments of \citet[Appendix \ref{sec:vdbo}]{vanDenBosch_Ogiya_2018}. A companion study (Paper II; Bah\'{e} et al., in prep.) examines the mechanisms of galaxy disruption and its role in building central group/cluster galaxies and their extended haloes.

Throughout, we assume the same flat $\Lambda$CDM cosmology as the EAGLE project, with parameters as determined by \citet{Planck_2014}: Hubble parameter $h \equiv \text{H}_0 / (100\,\text{km}\,\text{s}^{-1}\, \text{Mpc}^{-1}) = 0.6777$, dark energy density parameter $\Omega_\Lambda = 0.693$ (dark energy equation of state parameter $w = -1$), matter density parameter $\Omega_\text{M} = 0.307$, and baryon density parameter $\Omega_\text{b} = 0.04825$. All galaxy stellar, dark matter, and total masses are computed as the sum of all gravitationally bound particles of the respective type as identified by the \subfind{} code (see Section \ref{sec:subfind}).


\section{Simulations and post-processing}
\label{sec:sims}

\subsection{The Hydrangea simulations}
\label{sec:hydrangea}
The Hydrangea simulations are part of the C-EAGLE project, a suite of cosmological hydrodynamical zoom-in smoothed particle hydrodynamics (SPH) simulations of 30 massive galaxy clusters \citep{Bahe_et_al_2017b, Barnes_et_al_2017b}. They were run with the `AGNdT9' variant of the EAGLE model (see Table 3 of \citealt{Schaye_et_al_2015}), with initial particle masses $m_\text{DM} = 9.7 \times 10^6\, \msun$ and $m_\text{gas} = 1.8 \times 10^6\, \msun$ for dark matter and gas, respectively. The (spatially constant, Plummer-equivalent) gravitational softening length of the simulations is $\epsilon = 0.7$ proper kpc at $z < 2.8$. Here, we provide a succinct summary of their key features and refer to \citet{Bahe_et_al_2017b} and \citet{Barnes_et_al_2017b} for more details.

The 30 clusters of the C-EAGLE project were chosen from a low-resolution $N$-body simulation \citep{Barnes_et_al_2017a}, in the mass range\footnote{$\mvir$ denotes the total mass within a sphere of radius $\rvir$, centred on the potential minimum of the cluster, within which the average density equals 200 times the critical density.}  $14.0 > \log_{10} (M_{200c}^{z = 0} / \msun) > 15.4$ at $z = 0$ and without a more massive halo closer than max(20 $r_{200c}$, 30 Mpc) at $z = 0$. 24 clusters -- the Hydrangea suite -- were simulated with a high-resolution region extending to at least $10\,\rvir$ from the centre of the target cluster (defined as the location of its potential minimum). Within these large zoom-in regions, they contain a multitude of additional lower-mass groups and clusters on the outskirts of the main target cluster. 

The EAGLE code \citep{Schaye_et_al_2015} that was used for the zoom-in resimulations is a substantially modified version of the \textsc{Gadget-3} code (last described in \citealt{Springel_2005}). The changes include updates to the hydrodynamics scheme collectively referred to as `\textsc{Anarchy}' \citep{Schaller_et_al_2015c, Schaye_et_al_2015} and a large number of subgrid physics models to simulate unresolved astrophysical processes, which are described in detail by \citet{Schaye_et_al_2015}. They include models for radiative cooling, photoheating, and reionization \citep{Wiersma_et_al_2009a}; star formation based on the Kennicutt-Schmidt relation cast as a pressure law \citep{Schaye_DallaVecchia_2008} but with a metallicity-dependent star formation threshold \citep{Schaye_2004}; a pressure floor corresponding to $P \propto \rho^{4/3}$ imposed on gas with $n_\mathrm{H} \geq 10^{-1} \mathrm{cm}^{-3}$ to prevent the formation of an inadequately modelled cold gas phase; mass and metal enrichment of gas due to stellar outflows based on \citet{Wiersma_et_al_2009b}; energy feedback from star formation in thermal stochastic form based on \citet{DallaVecchia_Schaye_2012}; and seeding, growth of, and energy feedback from supermassive black holes based on \citet{Springel_et_al_2005b}, \citet{Rosas-Guevara_et_al_2015}, and \citet{Schaye_et_al_2015}. 

Particularly relevant to this study is that those sub-grid parameters that are not well-constrained by observations -- primarily the efficiency scaling of star formation feedback -- were calibrated so that the simulated \emph{field} galaxy population matches low-redshift observations in terms of the stellar mass function and stellar sizes (as described by \citealt{Crain_et_al_2015}). These are crucial prerequisites for meaningful predictions about the survival of cluster galaxies, because an overly massive or overly compact stellar component may make the simulated galaxies artificially resilient against disruption (and vice versa).

In addition to the main simulation with hydrodynamics and baryon physics, each volume was also simulated in $N$-body only mode, i.e.,~starting from the same initial conditions but assuming that all matter is dark. These `DM-only' simulations allow us to directly quantify the net impact of baryons (see also \citealt{Armitage_et_al_2018}).

\subsection{Structure identification}
\label{sec:subfind}

The primary output from each simulation consists of 30 snapshots, which are mostly spaced equidistant in time between $z = 14.0$ and $z = 0$ with $\Delta t = 500 \text{Myr}$. In each of these outputs, structures were identified with the \subfind{} code \citep{Springel_et_al_2001b, Dolag_et_al_2009} in a two-step process. 

First, spatially disjoint groups of particles were found with a friends-of-friends (FoF) algorithm with a linking length of $b = 0.2$ times the mean inter-particle separation. As shown by \citet{More_et_al_2011}, this linking length corresponds approximately (within a factor of $\approx\,$2) to a limiting isodensity contour of $\delta \equiv \rho/\rho_\text{mean} = 82$. The FoF algorithm is applied only to DM particles; baryon particles are attached to the FoF group (if any) of their nearest DM neighbour particle \citep{Dolag_et_al_2009}. Groups with less than $N_\text{FoF} = 32$ DM particles are deemed unresolved and discarded. 

Within each FoF group, \subfind{} then identifies gravitationally self-bound `subhaloes'. This procedure is described in detail by \citet{Springel_et_al_2001b} and \citet{Dolag_et_al_2009}. Candidate subhaloes are identified as locally over-dense regions, limited by the isodensity contour at the density saddle point that separates the candidate subhalo from the local background. Within each candidate, gravitationally unbound particles are iteratively removed and candidates retaining more than 20 particles (excluding gas) are identified as genuine subhaloes. Finally, all particles in the FoF group that are not part of any subhalo are collected into the `background' subhalo, provided that they are gravitationally bound to it.

In the following, we will refer to all subhaloes as `galaxies', including the background subhalo (which is typically the most massive one in any FoF group). The latter will be referred to as `central' and all others as `satellites'. This nomenclature is independent of the stellar content of a subhalo (which may be zero); unless specifically stated otherwise, we define galaxies as including all particle types, including their gaseous and dark matter haloes. 

Previous work has shown that the subhalo identification step of \subfind{} tends to incorrectly assign particles near the edge of satellites to the central subhalo (e.g.,~\citealt{Muldrew_et_al_2011}). In idealised tests, \citet{Muldrew_et_al_2011} have shown that this can artificially suppress the mass of even massive subhaloes ($M = 10^{12}\, \msun$) by as much as 90 per cent near the centre of a galaxy cluster; in extreme cases, it may be lost altogether. Our tracing procedure accounts for this spurious, temporary ``disruption'' where possible (see below), and we have verified that only a minute fraction of galaxies missing from the $z = 0$ \subfind{} catalogue still exist as self-bound structures (see Appendix \ref{app:SubfindRobustness}). One must, however, bear in mind that the \emph{masses} of satellite subhaloes calculated by \subfind{} may be (substantially) underestimated.

\subsection{Tracing galaxies through time}
\label{sec:spiderweb}

The subhalo catalogues returned by \subfind{} describe the simulated structures at one point in time. In order to follow individual simulated galaxies -- physical objects that appear at some point in time and potentially disappear later -- these outputs must be linked together as an additional post-processing step. We accomplish this with the `\textsc{Spiderweb}' algorithm, a substantially modified version of the procedure outlined in \citet{Bahe_et_al_2017b}. A full description of the code elements and their physical motivation is provided in Appendix \ref{app:spiderweb}; the following is a brief summary of its main aspects. 

\spiderweb{} follows a galaxy through time by identifying the sequence of subhaloes in subsequent snapshots that share the highest fraction of particles. Although this is conceptually straightforward, subtleties arise due to interactions between galaxies, particularly in the dense environments of groups and clusters. We therefore consider multiple candidate descendants for each subhalo in a given snapshot ($i$), namely all those in the subsequent snapshot ($j$) that are `linked' to the original subhalo by sharing at least one particle. In the case of multiple links from one subhalo in $i$, the highest priority is given to the one that contains the largest number of its 5 per cent most bound collisionless particles (its `core'). The other links are reserved as backup in case this highest priority link leads to a subhalo in $j$ that already overlaps more closely with another subhalo in $i$: this may, for example, happen if the galaxy is undergoing severe stripping so that most of its (core) particles are transferred to another galaxy between two snapshots. We note that this approach differs from other `merger tree' algorithms (e.g.,~\citealt{Rodriguez-Gomez_et_al_2015, Qu_et_al_2017}), which only consider one possible descendant for each subhalo.

To account for instances of a galaxy temporarily not being identified at all by \subfind{}, \spiderweb{} attempts to re-connect lost galaxies after up to 5 snapshots (corresponding to a maximum gap of 2.5 Gyr at our standard snapshot spacing). Our code also gives special consideration to the treatment of mergers, by explicitly accounting for prior mass transfers between galaxies when selecting the main progenitor of a subhalo in $j$ that is linked to multiple subhaloes in $i$. 

If no descendant can be found for a subhalo in $i$, its galaxy is treated as disrupted and `merged' onto the galaxy that contains the largest number of its core particles. By following these target galaxies (possibly over multiple mergers), \spiderweb{} identifies a unique `carrier' galaxy at $z = 0$ as the endpoint of every galaxy that has ever existed in the simulation. For a comprehensive description and justification of these methods, the interested reader is referred to Appendix \ref{app:spiderweb}.

\subsection{Sample selection}
\label{sec:sample}

Galaxies are characterised by the peak (total) subhalo mass they have ever attained, which we denote as $\mgalpeakbr$. In contrast to the equivalent mass at $z = 0$ ($\mgalzz$), this can be homogeneously computed for both surviving and disrupted galaxies, and compared to the \emph{stellar} peak mass $\mgalstarpeakbr$, it allows a direct comparison between hydrodynamical and DM-only simulations. There is a fairly tight relation between $\mpeak$ and $\mgalstar$ (see also \citealt{Moster_et_al_2013} and \citealt{Behroozi_et_al_2018}), with a $1\sigma$ scatter of typically only $\approx\,$0.5 dex: $\mpeak = 10^{10}$ ($10^{11.5}$, $10^{12.5}$) $\msun$ corresponds approximately to $\mgalstar = 10^{7.7}$ ($10^{10.1}$, $10^{11}$) $\msun$. 

Here, we analyse galaxies with $\mpeak > 10^{10}\,\msun$ ($\mgalstar \gtrsim 5\times10^{7}\,\msun$), i.e., those that have at some point been resolved by $>$ 1000 particles. Many \emph{baryonic} $z = 0$ properties of our simulated galaxies are already unconverged or in tension with observations at $\mpeak < 10^{11.5}\,\msun$, including stellar masses (at $\mpeak < 5\times 10^{10}\,\msun$), sizes, quenched fractions (both at $\mpeak < 10^{11}\,\msun$), metallicities \citep{Schaye_et_al_2015}, and neutral gas content \citep{Crain_et_al_2017}. We include these low-mass galaxies here to test the predicted survival fractions in this poorly converged regime, but emphasize that they should be interpreted with caution, at least to the extent that they deviate between hydrodynamical and DM-only simulations.  

We exclude a small number of galaxies ($\ll$ 1 per cent at $\mpeak > 10^{10}\, \msun$) that are formed predominantly from particles that were previously associated with another galaxy. These `spectres' typically correspond to substructures within a more massive galaxy (e.g.,~a dense part of a spiral arm) that are temporarily identified as a separate subhalo (see Appendix \ref{app:spiderweb} for further details). 

Because the Hydrangea simulations use the zoom-in technique, some subhaloes in each snapshot lie close to the edge of the high-resolution region and may be subject to numerical artefacts. We therefore exclude all galaxies from our analysis whose potential minimum lies closer than 5 comoving Mpc from a low-resolution boundary particle in any snapshot. We also exclude a very small population of low-mass galaxies ($<$ 0.1 per cent at $\mpeak < 10^{11} \msun$) that have no identifiable carrier at $z = 0$ because all their particles became unbound when they were disrupted. 

\subsection{Satellite accretion times}
\label{sec:satAccTimes}

As a final step, we need to identify galaxies that have been accreted by a group or cluster at some point in their lives. Not all of these are satellites at $z = 0$: some may have been disrupted completely, and others may have temporarily or permanently escaped as `backsplash' galaxies (see, e.g.,~\citealt{Gill_et_al_2005}). For each galaxy, we therefore first identify the snapshots in which it is a satellite\footnote{As noted in Section \ref{sec:subfind}, we define satellite status and accretion times in terms of a galaxy's membership to an FoF group: it is a satellite if it is not the central subhalo of the FoF group to which it belongs. Not all of these satellites are necessarily within $\rvir$ from the central, particularly in highly aspherical groups.}; if there are none, the galaxy is discarded. In each of these snapshots, we then find the corresponding central galaxy. The FoF group containing this central (or its carrier, in case the central itself has merged) at $z = 0$ is a candidate host of the galaxy under consideration. If there are multiple candidates (from different snapshots), we select the one which is a candidate in the largest number of snapshots and, in the event of a tie, the one from the earliest snapshot. By definition, all hosts correspond to FoF groups at $z = 0$ and can therefore be classified by their present-day $\mvir$. 

\begin{figure}
  \centering
    \includegraphics[width=\columnwidth]{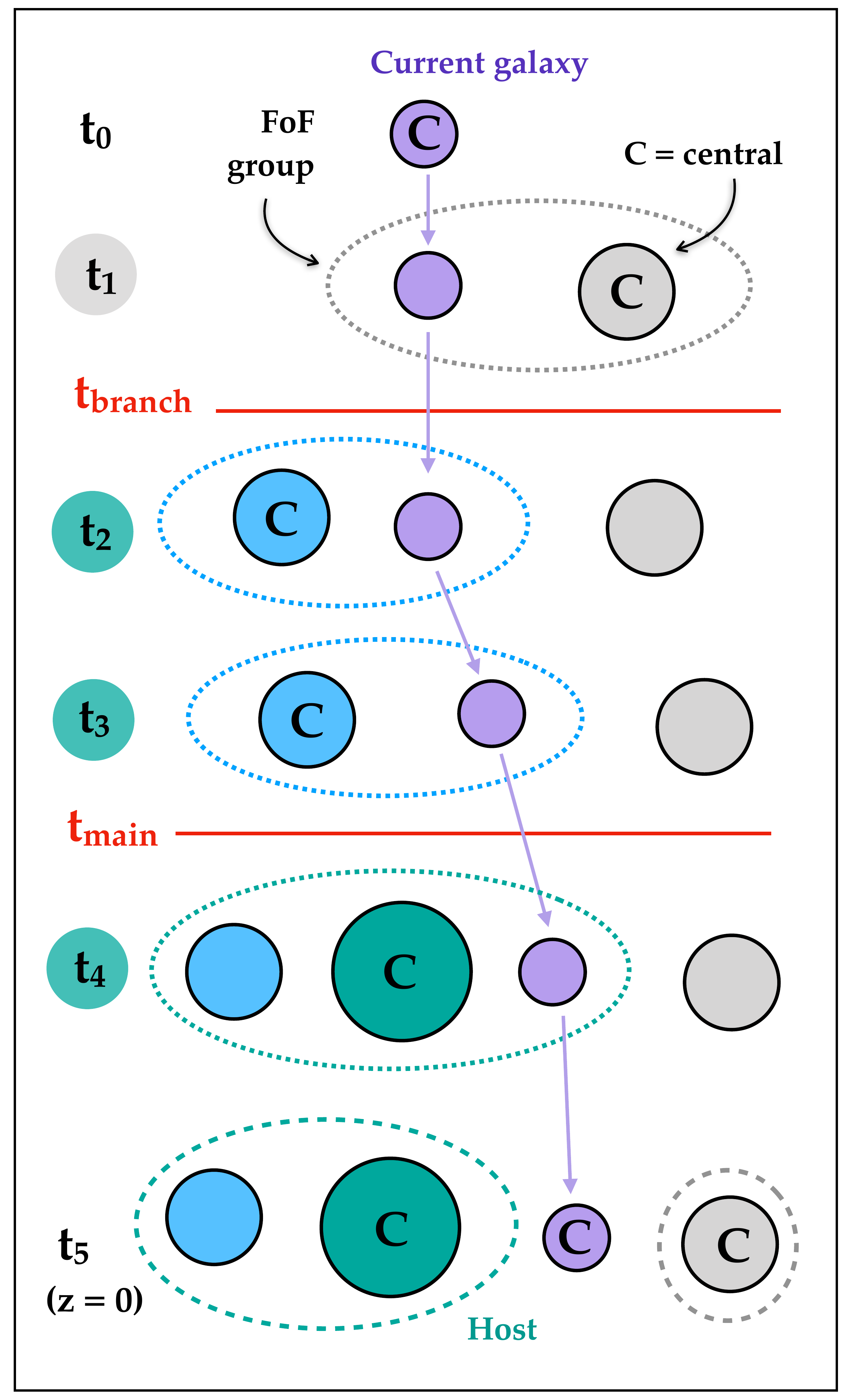}
       \caption{An illustration of our host assignment scheme. Shown are six consecutive snapshots at times $t_0$ to $t_5$ (the latter corresponding to $z = 0$). Coloured circles represent four individual galaxies, of which the purple one is currently under consideration. Although it is a central at redshift $z = 0$, it was a satellite in four previous snapshots ($t_1$--$t_4$), with the respective groups outlined by dotted ellipses. Their centrals lie in two FoF groups at $z = 0$, indicated by the grey and green dashed ellipses in the bottom row. These are the two candidate hosts of the galaxy, and the shading behind each snapshot label (on the far left) indicates to which one it was associated at this point. Because it is associated most often to the green candidate, this is selected as the galaxy's host. The two horizontal red lines indicate the two accretion times used in this paper, corresponding to first infall into the host itself ($t_\text{main}$) and one of its progenitor branches ($t_\text{branch}$).}  
         \label{fig:accTimesDemo}
\end{figure}

An illustration of our host assignment scheme is provided in {\color{Blue}Fig.~\ref{fig:accTimesDemo}}. This follows one galaxy (represented by purple circles) through six consecutive snapshots at times $t_0$--$t_5$ (different rows from from top to bottom), with the last row at $t_5$ corresponding to $z = 0$. Circles in other colours represent other galaxies. The purple galaxy is a satellite in four snapshots ($t_1$--$t_4$), during which it is a member of the FoF groups indicated with dotted ellipses in the colour of their centrals (which are denoted with a `C'). Because one of these (blue) is itself a satellite (of the green one) at $z = 0$, there are only two candidate host groups, indicated with the green and grey dashed ellipses in the bottom row. The galaxy under consideration (purple) was associated to the green candidate in three snapshots ($t_2$--$t_4$) and to the grey candidate in only one ($t_1$). The former is therefore selected as its host, even though the purple galaxy is, in this example, not actually part of it\footnote{Fig.~\ref{fig:accTimesDemo} deliberately depicts the non-standard situation of a galaxy that has escaped from its host at $z = 0$, to highlight that our host assignment scheme does not depend (exclusively) on $z =0$ group membership. The choice of host and accretion times would be exactly the same in the (more typical) situation of the purple galaxy being part of the green group at $z = 0$, or having merged with one of its members.} at $z = 0$.

We exclude galaxies that are the central galaxy of their own host at $z = 0$, which can occur as a result of satellite--central swaps. This only affects 0.1 per cent of our galaxies, but because these all have\footnote{There are small differences between $\mpeak$ and $\mhost$ even for galaxies that are their own host, because the latter excludes particles beyond $\rvir$, but also includes unbound particles and those in satellites within this radius.} $\mpeak \approx \mhost$, the fraction is almost 50 per cent within the most extreme combination of high $\mpeak$ ($ > 10^{12.5}\,\msun$) and low $\mhost$ ($= 10^{12.5}$--$10^{13.5}\,\msun$). Our final sample contains \mbox{165 566} galaxies with $\mpeak > 10^{10}\,\msun$ that are associated with a host of $\mhost > 10^{12.5}\,\msun$, including \mbox{3 433} with $\mpeak > 10^{12}\,\msun$.

With a host halo selected for each galaxy, we next find their accretion times. We consider two alternative definitions, but note that a plethora of others have been used in the literature (see, e.g., \citealt{Gao_et_al_2004, Xie_Gao_2015, Chua_et_al_2017}). The `branch accretion time' ($t_\text{branch}$) is the middle of the snapshot interval before the galaxy first became a satellite in any progenitor branch of its host halo (in other words, in a halo whose central -- or its carrier -- at $z = 0$ is in the same group as the galaxy's host). The `main accretion time' ($t_\text{main}$) is taken as the analogous point when the galaxy became a satellite in its actual host halo. Galaxies that never reach their host halo, for example because they disrupted in a side-branch (see Section \ref{sec:influence_preProc}), are assigned $t_\text{main} = \infty$. When a galaxy became a satellite and then merged before the next snapshot was written (so that it is never recorded as a satellite), we assign an accretion time half-way between the last snapshot in which the galaxy was detected, and the first in which it was not. 

For the situation depicted in Fig.~\ref{fig:accTimesDemo}, these two definitions of accretion time are indicated by red horizontal lines. We highlight that $t_\text{branch}$ is, in this example, not equivalent to the first time at which the purple galaxy became a satellite, because its (brief) association with the grey group in $t_1$ is not yet part of its accretion into its final host (green). 

\begin{figure}
  \centering
    \includegraphics[width=\columnwidth]{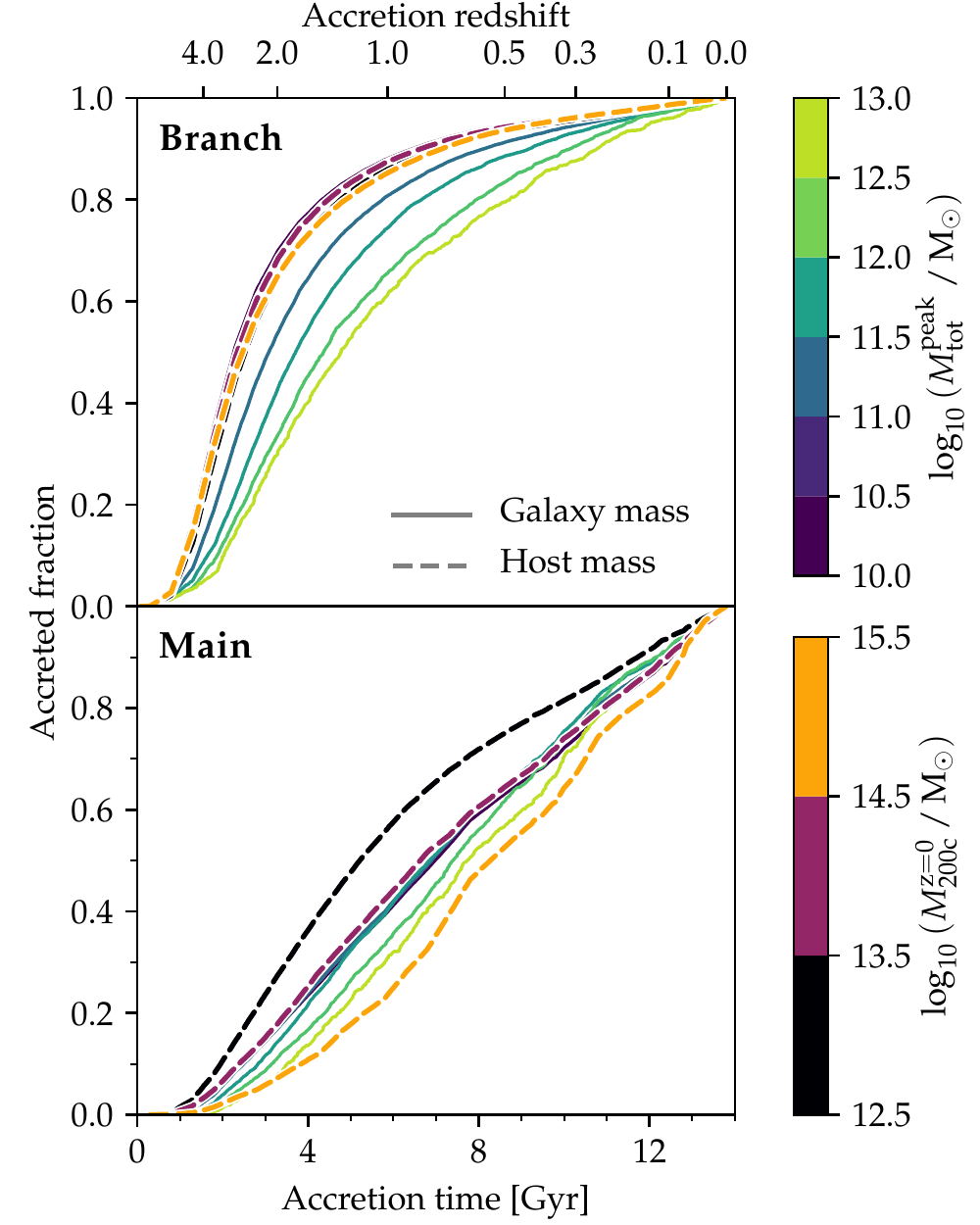}
       \caption{Cumulative distribution of accretion times for different peak total galaxy masses (solid lines) and different host halo masses (dashed lines) in the hydrodynamical simulations. \textbf{Top:} accretion onto any branch of the host group ($t_\text{branch}$), \textbf{bottom:} accretion onto its main progenitor branch ($t_\text{main}$), for galaxies with $t_\text{main} < \infty$. Values of $t_\text{branch}$ are predominantly early (around $z \approx 2$), while $t_\text{main}$ is more evenly spread out. The former depend mostly on galaxy mass, the latter on host mass.} 
    \label{fig:accTimes}
\end{figure}

In Fig.~\ref{fig:accTimes}, we show the cumulative distribution of both branch (top) and main (bottom) accretion times for galaxies with different peak total galaxy masses ($\mpeak$; solid lines in shades of green and blue) and host halo masses ($\mhost$; dashed lines in shades of yellow and red). For the former we divide galaxies into six equal bins in log-space, from $\mpeak = 10^{10}$ to $10^{13}\, \msun$. For the hosts, we distinguish between `massive clusters' ($\mhost > 10^{14.5} \msun$; orange), `low-mass clusters' ($\mhost = 10^{13.5}$--$10^{14.5} \msun$; lilac), and `groups' ($\mhost = 10^{12.5}$--$10^{13.5} \msun$; black). 

Due to the setup of our simulations, the latter two bins are dominated by objects at the periphery of a more massive cluster and therefore not necessarily representative of all haloes in these mass bins. However, we found that the survival fractions shown below only vary by $\loa$ 5 per cent between galaxies with a host at $<$ 5 and 5--10 $\rvir$ from the central cluster of their simulation volume, respectively\footnote{The survival fraction is, in general, slightly higher for galaxies whose host lies closer to the central cluster.}. We are therefore confident that the large-scale environment does not induce a significant bias in our conclusions for lower-mass haloes. For display purposes, all times are offset by a random value of up to $\pm$250 Myr to suppress artificial discreteness due to the finite number of snapshots.

Galaxies are accreted over a wide redshift range, $4 \goa z \geq 0$. The distribution of $t_\text{branch}$ (top; median at $z \approx 1.5$--3) is more concentrated towards high $z$ than that of $t_\text{main}$ (bottom; median at $z \approx 0.5$--1). In addition, $t_\text{branch}$ depends strongly on $\mpeak$ -- more massive galaxies are accreted later (compare the dark blue and yellow-green solid lines) -- but hardly on $\mhost$ (the orange and black dotted lines lie almost on top of each other)\footnote{There is a slight dependence on $\mhost$ when only considering more massive galaxies ($\mpeak > 10^{11.5} \msun$), in the sense that $t_\text{branch}$ is $\approx\,$1 Gyr later for low-mass groups than clusters (for clarity not shown in Fig.~\ref{fig:accTimes}).}. The main accretion time shows the opposite behaviour, with a clear difference between different hosts -- galaxies in clusters (orange dashed) are accreted later than those in groups (black dashed) -- but a much weaker dependence on galaxy mass. There is hardly any difference between hydrodynamical and DM-only simulations (omitted for clarity).

\begin{figure}
  \centering
    \includegraphics[width=1.05\columnwidth]{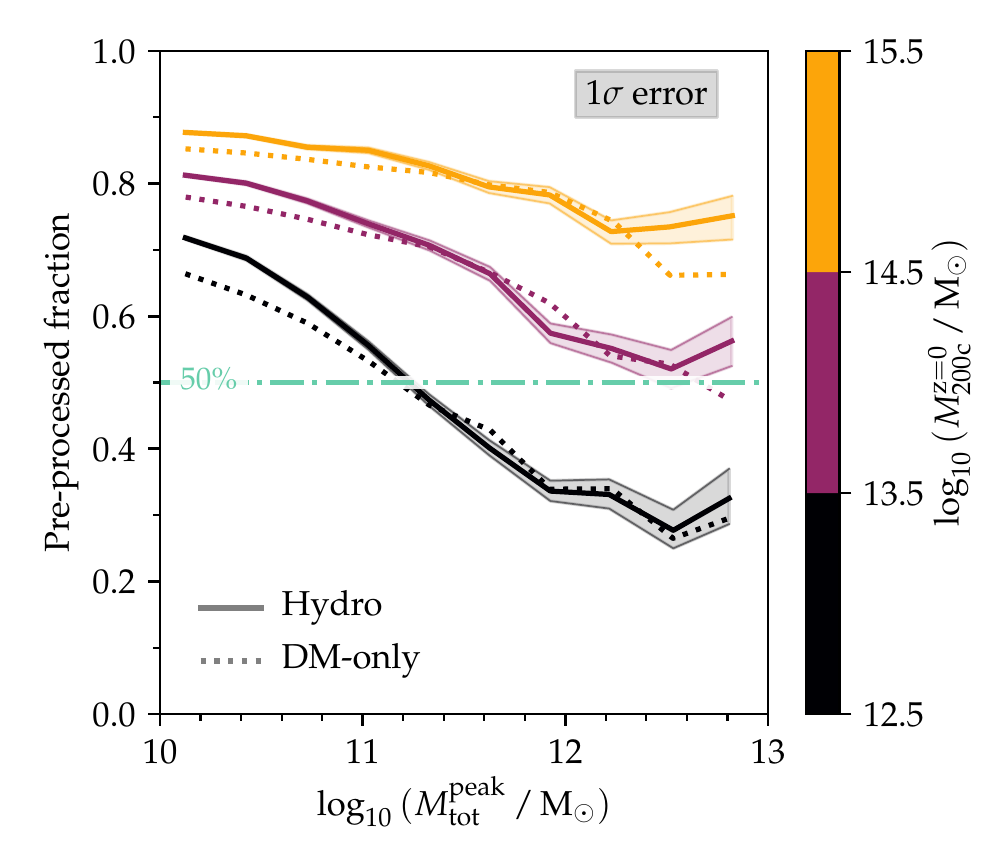}
       \caption{Fraction of satellite galaxies that are pre-processed (first accreted by a subgroup, rather than their final host) as a function of their peak total mass. Hydrodynamical simulations are represented by solid lines (shaded bands indicating their binomial $1\sigma$ uncertainties), the DM-only runs by dotted lines. Different colours represent galaxies in hosts of different mass, as indicated by the colour bar along the right edge. Pre-processing is ubiquitous, especially for low-mass galaxies and those associated with massive clusters (orange).} 
    \label{fig:accMode}
\end{figure}

The gap between the median $t_\text{branch}$ and $t_\text{main}$ implies a significant role of pre-processing, i.e.,~that many galaxies first fall into a sub-group which is later accreted by their final host (e.g., \citealt{Berrier_et_al_2009, McGee_et_al_2009, Balogh_McGee_2010}). This is shown directly in {\color{blue}Fig.~\ref{fig:accMode}}, where we plot the fraction of galaxies with $t_\text{branch} < t_\text{main}$ as a function of $\mpeak$ for the three host mass bins in the hydrodynamical simulations (solid lines; shaded bands indicate binomial 1$\sigma$ uncertainties following \citealt{Cameron_2011} both here and in subsequent figures) and the corresponding DM-only runs (dotted lines).  

The pre-processed fraction is very high: 87 (73) per cent of galaxies in massive clusters with $\mpeak > 10^{10}$ ($10^{12}$) $\msun$, and still $\approx\,$35 per cent of Milky Way analogues ($\mpeak \sim 10^{12}\, \msun$) in groups (black) were first a satellite in a sub-group that was later accreted by their main host. This is notably higher than what previous authors have found for \emph{surviving} galaxies (only $\approx\,$50 per cent even in massive clusters; e.g.,~\citealt{Bahe_et_al_2013, Wetzel_et_al_2013, Han_et_al_2018}). As we show below, this discrepancy arises because many galaxies do not survive the pre-processing stage. 

At $\mpeak > 2\times 10^{11}\,\msun$, the pre-processed fraction in the hydrodynamical simulations agrees closely with the DM-only runs. Only at lower masses is there a small, but consistent, tendency towards slightly higher pre-processing fractions in the hydrodynamical simulations (by $<$ 6 per cent). This could be caused by subtle differences in the halo finder between the two simulation types, or reflect a small impact of baryons on the actual accretion paths of low-mass galaxies.     


\section{Survival fractions of satellites}
\label{sec:survival}

\begin{figure}
  \centering
    \includegraphics[width=1.05\columnwidth]{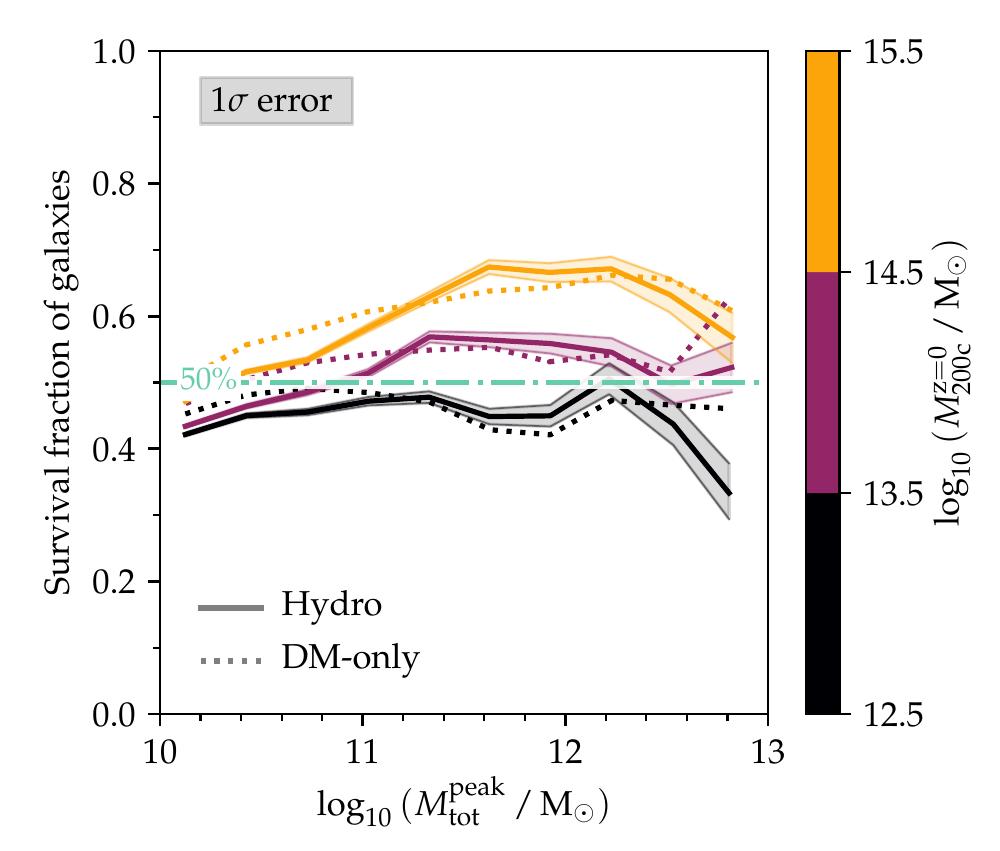}
       \caption{The fraction of all accreted galaxies (including pre-processed ones) that survive with $M_\text{tot}^\text{z = 0} > 5\times 10^8\,\msun$. Different host masses at $z = 0$ are indicated by different colours. Both the hydrodynamical simulations (solid lines, shaded bands indicate $1\sigma$ binomial uncertainties) and the DM-only counterparts (dotted lines) predict a survival fraction of $\sim$50 per cent. At fixed galaxy mass, survival is slightly more common in more massive hosts.} 
    \label{fig:survival_total}
\end{figure}

We begin by investigating the survival of all galaxies from their point of first accretion ($t_\text{branch}$). Our fiducial definition of survival requires that the galaxy is identified by \subfind{} at $z = 0$ and has a mass of at least $\mgalzz = 5\times10^8\,\msun$ (corresponding to $\approx\,$50 DM or $\approx\,$270 baryon particles); the effect of varying this threshold is explored below. The survival fraction of all galaxies ever accreted is plotted in {\color{blue}Fig.~\ref{fig:survival_total}} as a function of peak total galaxy mass $\mpeak$, in three halo mass bins. Solid lines represent the hydrodynamical simulations (with shaded bands representing binomial 1$\sigma$ uncertainties, as in Fig.~\ref{fig:accMode}), while the corresponding fractions from the DM-only simulations are shown by dotted lines. We have not matched individual galaxy pairs in the two simulation sets, because these may follow significantly different orbits due to amplifications of small differences in the cluster environment \citep{Prins_2018}.

The survival fraction is $\sim$50 per cent, with only a moderate dependence on galaxy or host mass. Perhaps surprisingly, it is slightly higher in massive clusters than groups (51 vs.~44 per cent when averaged over all $\mpeak > 10^{10}\,\msun$). It is also mildly higher around $\mpeak = 10^{12}\, \msun$ than at the highest and lowest galaxy masses, at least in clusters (up to 67 per cent). Averaged over our entire sample, 47 per cent of satellites with $\mpeak > 10^{10}\,\msun$ and $\mvir > 10^{12.5}\,\msun$ survive at $z = 0$. In Appendix \ref{app:resComp}, we demonstrate that these numbers are insensitive to an increase in mass resolution by a factor of eight, at least in low-mass groups and for $\mpeak \lesssim 3\times 10^{11}\,\msun$ (more massive objects are not sampled well by our high-resolution runs due to their smaller volumes).

A second key feature of Fig.~\ref{fig:survival_total} is that the survival fractions in the hydrodynamical simulations closely follow those in their DM-only counterparts. There are some minor differences, for example at $\mpeak \sim 10^{12}\, \msun$ in massive clusters -- where the inclusion of baryons increases the survival fraction by a few per cent -- and at the low-mass end ($\mpeak \lesssim 10^{11}\, \msun$), where the baryonic galaxies are slightly more susceptible to disruption at fixed $\mpeak$, possibly as a consequence of poor resolution (see above). Overall, however, \emph{the effect of baryons on galaxy survival is small}: if star formation and gas stripping separately have non-negligible impact, they happen to cancel each other almost exactly. 

The close agreement between the survival fractions in the hydrodynamical and DM-only simulations implies that the former should not contain many remnants that are (almost) completely devoid of dark matter and only survive because of their baryon content. To verify this, we have also computed the survival fractions above a \emph{dark matter} mass threshold of $5\times 10^8\,\msun$ in the hydrodynamical simulations\footnote{This threshold is not fully equivalent to $\mgalzz > 5\times10^8\, \msun$ in the DM-only (DMO) version, because the DM particles in the DMO simulations also account for the mass contributed by baryons and are therefore more massive, by a factor of $(1 - \Omega_\text{b}/\Omega_\text{m})^{-1} = 1.19$.}, which agree almost exactly with those from the equivalent threshold in total mass (not shown). 

This absence of (almost) purely baryonic remnants appears to be in tension with semi-analytic models, which typically require a large fraction of (baryonic) galaxies to survive the disruption of their dark matter subhalo in the form of `orphan' or `type-2' satellites (e.g., \citealt{Somerville_et_al_2008, Guo_et_al_2011, Henriques_et_al_2015}). In the \citet{Guo_et_al_2011} model applied to the Millennium-II simulation, for example -- which has almost exactly the same resolution as the Hydrangea DM-only runs -- 25 per cent of all satellite galaxies with $\mgalstarzz = 10^{9.5}\,\msun$ are orphans, and still almost 20 per cent at $\mgalstarzz = 10^{10.5}\,\msun$.    

The small net influence of baryons is also is at odds with the recent study of \citet{Chua_et_al_2017}, who found that, in the Illustris simulation, the inclusion of baryons \emph{reduces} the survival fraction by $\approx\,$5--20 per cent, at all masses. It is plausible that these differences reflect different sub-grid physics implementations, so that a destabilizing effect of gas stripping dominates in Illustris, while it is approximately cancelled by the cohesive effect of star formation in Hydrangea\footnote{Note that the absolute survival fractions in the DM-only simulation of \citet{Chua_et_al_2017} are significantly higher than in our Fig.~\ref{fig:survival_total}, because they do not explicitly include the pre-processing phase. We have verified that this does not account for the different impact of baryon physics in the two simulations.}.   

\subsection{Influence of the detection threshold}
\label{sec:survival_detThresh}

\subsubsection{Thresholds in total galaxy mass}
\begin{figure}
  \centering
    \includegraphics[width=\columnwidth]{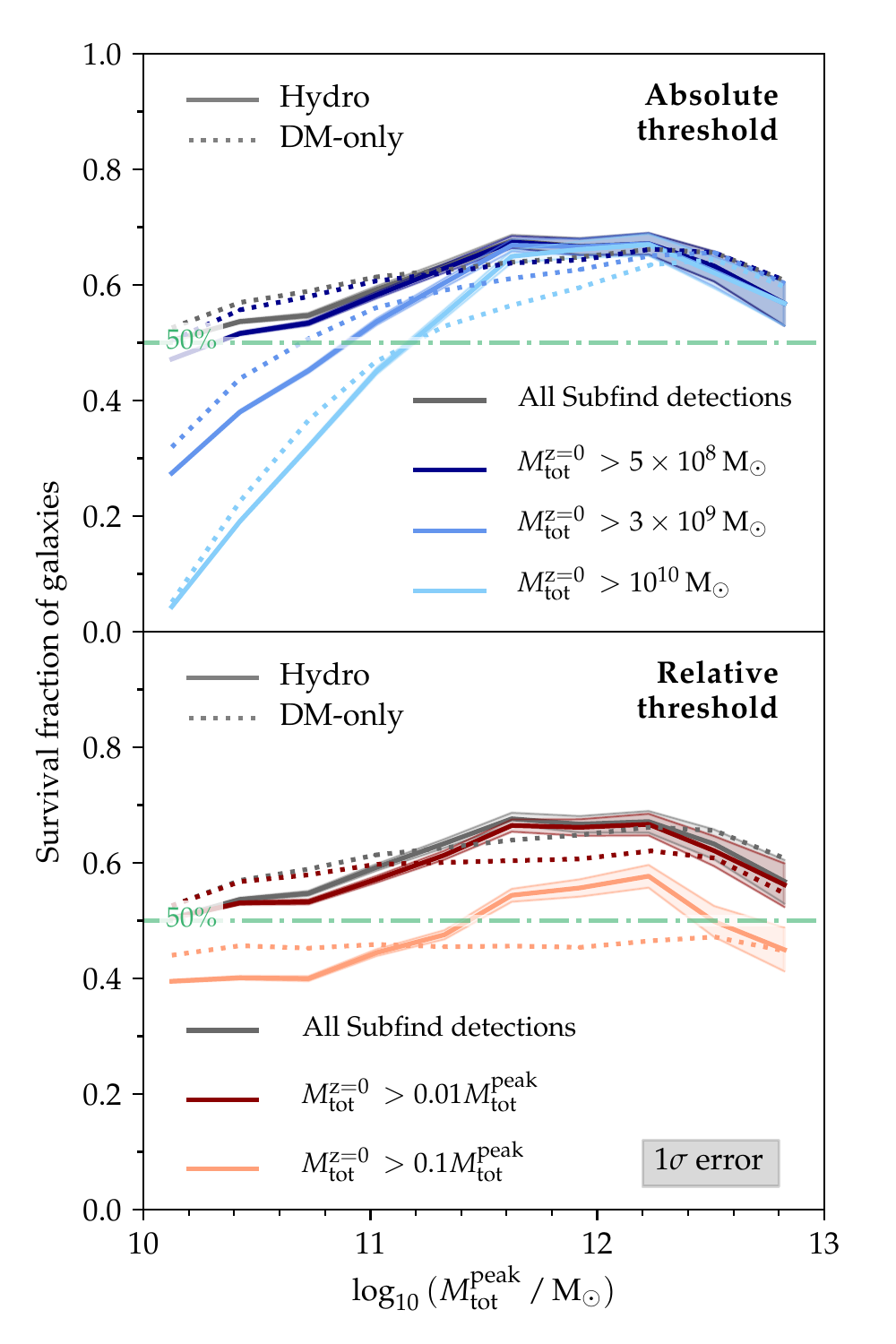}
       \caption{Dependence of satellite survival fractions on the imposed detection threshold in the hydrodynamical (solid lines) and DM-only simulations (dotted lines) of massive clusters. Grey lines show the total survival fraction, i.e.,~all galaxies detected by \textsc{subfind} at $z=0$. In the \textbf{top panel}, the dark, medium, and light blue lines show, respectively, the fraction of galaxies retaining a total mass of at least $5\times 10^8$, $3\times10^9$, and $10^{10}\, \msun$, respectively, at $z = 0$. The \textbf{bottom panel} gives the fraction of galaxies that retain at least 1 (dark red) and 10 per cent (light red) of their total peak mass at $z = 0$. All thresholds apart from this last one converge in the hydrodynamic simulations at $\mgalpeak > 3 \times 10^{11} \msun$. In contrast, many lower-mass galaxies -- and in the DM-only simulations even some Milky Way analogues -- only survive as low-mass remnants below $10^{10}\,\msun$.} 
    \label{fig:survival_detThresh}
\end{figure}

In Fig.~\ref{fig:survival_total}, we counted any galaxy as `surviving'  that was identified by \subfind{} at $z = 0$ and had a total mass of at least $5\times 10^8\,\msun$. To elucidate the sensitivity of our predictions to this threshold, we plot in Fig.~\ref{fig:survival_detThresh} the survival fractions with a number of other definitions; for clarity, only the massive cluster bin is shown, but we have verified that the qualitative conclusions also apply to lower-mass hosts. 

The top panel compares the survival fractions at our fiducial mass threshold of $\mgalzz = 5\times10^8\,\msun$ (dark blue, identical to the orange lines in Fig.~\ref{fig:survival_total}) to both those obtained from considering all \subfind{} detections at $z = 0$ as surviving (grey), and two stricter mass thresholds of $\mgalzz = 3\times 10^9$ and $10^{10}\,\msun$ (medium and light blue, respectively). As in Fig.~\ref{fig:survival_total}, we show results from the hydrodynamical simulations as solid, and from the DM-only runs as dotted lines. The lower panel shows the survival fractions above two \emph{relative} mass thresholds, requiring the galaxy to retain at least 1 per cent (dark red) or at least 10 per cent (light red) of their peak total mass. Recall that \subfind-derived satellite masses may be biased low, so that these lines should more accurately be interpreted as representing lower limits on the true surviving fractions.  

Compared to our fiducial threshold of $\mgalz = 5\times10^8\, \msun$ (dark blue lines in the top panel), the survival fractions hardly increase when including all \subfind{} detections (grey), in both the hydrodynamical and DM-only simulations; only at $\mgalpeak \lesssim 10^{11}\, \msun$ is there a difference of a few per cent. This indicates that $\mgalzz < 5\times 10^8 \msun$ remnants can, in principle, be resolved by our simulations, but also that they are very uncommon in the (peak) mass range considered here. This is confirmed in Appendix \ref{app:resComp}, where we show that the survival fractions of satellites with $\mgalpeak \gtrsim 3\times 10^{10}\,\msun$ in low-mass groups are unchanged when the mass resolution is increased, and the mass threshold for survival lowered, by a factor of eight. The more restrictive thresholds, on the other hand (medium and light blue), remove a successively larger fraction of galaxies with $\mgalpeak < 3\times10^{11}\, \msun$ that have a remnant in the $z = 0$ \subfind{} catalogue (69 per cent with $\mgalzz  < 10^{10}\,\msun$), indicating a continuous distribution of remnant masses between a lower limit ($\gtrsim 5 \times 10^8\,\msun$) and $\mgalpeak$. Our fiducial limit of $\mgalzz = 5\times10^8\,\msun$ is therefore a physically and numerically meaningful definition of galaxy survival in our simulations\footnote{A much lower threshold (e.g., $10^6\,\msun$) would be numerically meaningless because our simulations could not possibly resolve such a small remnant. A higher threshold would not do justice to the resolution of our simulations.}. At lower resolution, it may not be possible to identify remnants with $\mgalzz \lesssim 10^{10}\,\msun$, which could plausibly account for the higher disruption rates reported by, e.g., \citet{Jiang_vanDenBosch_2017}.     

An alternative criterion to distinguish between surviving and disrupted galaxies is the fraction of their peak mass retained at $z = 0$. As the bottom panel shows, a relative threshold of 1 per cent of the peak mass (dark red line) agrees to per cent level with the survival fraction from the entire \subfind{} catalogue in the hydrodynamical simulations. This implies a near-total absence of galaxies that lose more than 99 per cent of their mass but still survive as self-bound objects that can be detected at the resolution of our simulations. This is true even amongst the most massive galaxies ($\mgalpeak > 10^{12}\, \msun$) for which a remnant with one per cent of its peak mass would be well above the resolution limit of the simulations. In Paper II, we show that this is because massive galaxies predominantly merge with the core of the central group/cluster galaxy, rather than gradually dispersing into its halo. 

In contrast, a significant (but nevertheless minor) fraction of galaxies -- around 10 per cent in the hydrodynamical simulations, almost independent of mass -- are identified by \subfind{} at $z = 0$ but only retain less than one tenth of their peak mass (the difference between the light red and grey lines). These galaxies experienced strong mass loss (plausibly due to tidal stripping), but are nevertheless not disrupted completely. 

Although the DM-only versions (dotted lines in Fig.~\ref{fig:survival_detThresh}) yield broadly the same result as the hydrodynamical simulations discussed so far, there is an interesting second-order difference, especially at $\mgalpeak \sim 10^{12}\, \msun$. In this regime, the DM-only runs do produce a (small) population of galaxies that survive only as a very small remnant with mass below $10^{10}\,\msun$ or 1 per cent of their peak mass. This offset is particularly evident in the bottom panel, where the DM-only trends for both thresholds are almost flat, while they show a $\approx\,$50 per cent variation with $\mgalpeak$ in the hydrodynamical simulations. This suggests that baryons do have a non-negligible impact on mass stripping from satellites, but not on whether they ultimately survive as a (potentially very small) remnant. 

\subsubsection{Thresholds in stellar mass}

\begin{figure}
  \centering
    \includegraphics[width=1.05\columnwidth]{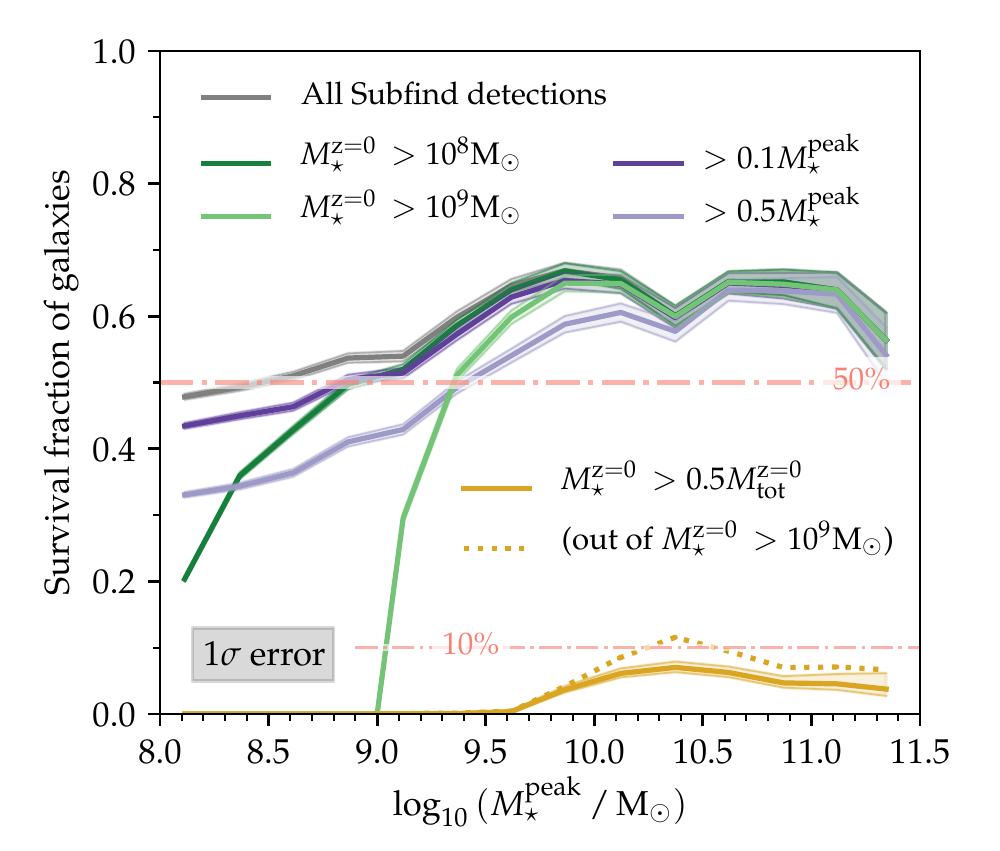}
       \caption{Dependence of the satellite survival fraction on \emph{stellar} peak mass ($\mgalstarpeak$) and detection threshold in the hydrodynamical simulations. The grey line shows the fraction of galaxies at a given $\mgalstarpeak$ that are detected by \subfind{} at $z = 0$, with binomial $1\sigma$ uncertainties marked by the shaded band. Dark and light green (purple) lines show the fraction whose stellar mass at $z = 0$ exceeds $10^8$ and $10^9 \msun$ (10 and 50 per cent of $\mgalstarpeak$), respectively. The yellow lines, near the bottom of the plot, give the fraction of star-dominated survivors ($\mgalstar > 0.5\, M_\text{tot}$ at $z = 0$) out of all galaxies (solid) and only those surviving with $\mgalstarzz > 10^9\,\msun$ (dotted). In the range plotted, virtually all surviving galaxies retain a resolved stellar remnant within 1 dex of their peak mass, but only a small subset are dominated by stars.} 
    \label{fig:survival_detThreshMstar}
\end{figure}

In Fig.~\ref{fig:survival_detThreshMstar}, we test similar thresholds in \emph{stellar} mass in the hydrodynamical simulations, and also classify galaxies by their stellar peak mass $\mgalstarpeakbr$. In terms of absolute thresholds (green lines), the result is qualitatively consistent with our findings for total mass: surviving galaxies with $\mgalstarpeak \gtrsim 3 \times 10^{9}\, \msun$ almost always retain a significant stellar remnant ($\mgalstarzz > 10^9\, \msun$ or $>$ 0.5 $\mgalstarpeak$ at $z = 0$), but many lower-mass galaxies drop\footnote{We note that this mass loss includes a contribution from stellar winds, in addition to stripping of stars through, e.g., tidal forces.} below a threshold of $10^9$ (and to a lesser extent also $10^8$) $\msun$. 

When considering relative thresholds, however (purple lines), it becomes clear that stellar mass loss from surviving satellites is considerably less severe than loss of total mass: even at $\mgalstarpeak = 10^8\, \msun$, only a few per cent are reduced to less than one tenth of their peak stellar mass (compare the grey and dark purple lines), and such strong loss hardly occurs at all above $10^9\, \msun$. Even only 50 per cent stellar mass loss is almost non-existent at the high-mass end ($\mgalstar > 2 \times 10^{10}\, \msun$) and only affects less than half the surviving lowest-mass galaxies (compare the grey and light purple lines). This is consistent with the findings of \citet{Bahe_et_al_2017a}, who found a median stripped stellar mass fraction from surviving galaxies in groups and low-mass clusters of $<$ 10 per cent, and with the works of \citet{Barber_et_al_2016} and van Son et al. (in prep.), who demonstrate that (massive) galaxies that lost around 90 per cent of their initial stellar mass are extreme outliers from the relations between stellar mass and black hole mass or stellar size. \emph{In terms of their stellar mass, satellite galaxy survival is therefore almost binary}: they either retain a large part of it, or they are lost completely.

Also shown in Fig.~\ref{fig:survival_detThreshMstar} is the fraction of galaxies that survive in stellar mass dominated form (i.e.,~with $\mgalstarzz  > 0.5\, \mgalzz$; yellow solid line) and the analogous fraction out of only those that survive with $\mgalstarzz > 10^9 \msun$ (yellow dotted line). Both are small, with only the latter reaching $\approx$10 per cent at $\mgalstar \sim 3\times10^{10}\, \msun$. Despite the much weaker loss of stellar than total mass, our simulations therefore predict that the vast majority of surviving galaxies, at any mass, remain dominated by their non-stellar component. Qualitatively, this agrees with the conclusions of \citet{Dolag_et_al_2009} based on lower-resolution simulations.

To summarise: the Hydrangea simulations predict that baryons have some impact on the mass loss of satellite galaxies, but are negligible with respect to their survival. The survival fraction is higher in more massive haloes -- up to 67 per cent for Milky Way analogue galaxies in massive clusters -- but still 44 per cent in low-mass groups. While many low-mass galaxies only survive as a small remnant with $\mgalzz < 10^{10}\, \msun$ -- but often still within a factor of $>$ 0.1 of their peak value in stellar mass -- at $z = 0$, more massive galaxies with $\mgalpeak > 3\times10^{11} \msun$ either disrupt completely, or retain a substantial core with $\mgalzz > 10^{10}\, \msun$ and $\mgalstarzz > 0.5 \mgalstarpeak$ at $z = 0$. Galaxies rarely survive in purely (or even mostly) stellar form.


\section{Influence of pre-processing, other satellites, and accretion time}
\label{sec:influence}

We now investigate different factors contributing to satellite disruption in more detail. The role of pre-processing (i.e., accretion onto a sub-group that is later accreted by their final host) is tested in Section \ref{sec:influence_preProc}, and that of satellite--satellite mergers in Section \ref{sec:influence_satSatMergers}. We then show how the survival fraction depends on accretion redshift (Section \ref{sec:influence_accTime}) and time elapsed since accretion (Section \ref{sec:influence_tDisrupt}), and conclude by investigating the distribution of galaxy disruption events over cosmic history (Section \ref{sec:tDisrupt}).

\subsection{Role of pre-processing} 
\label{sec:influence_preProc}

\subsubsection{Survival fractions of directly accreted and pre-processed galaxies}
\begin{figure}
  \centering
    \includegraphics[width=1.05\columnwidth]{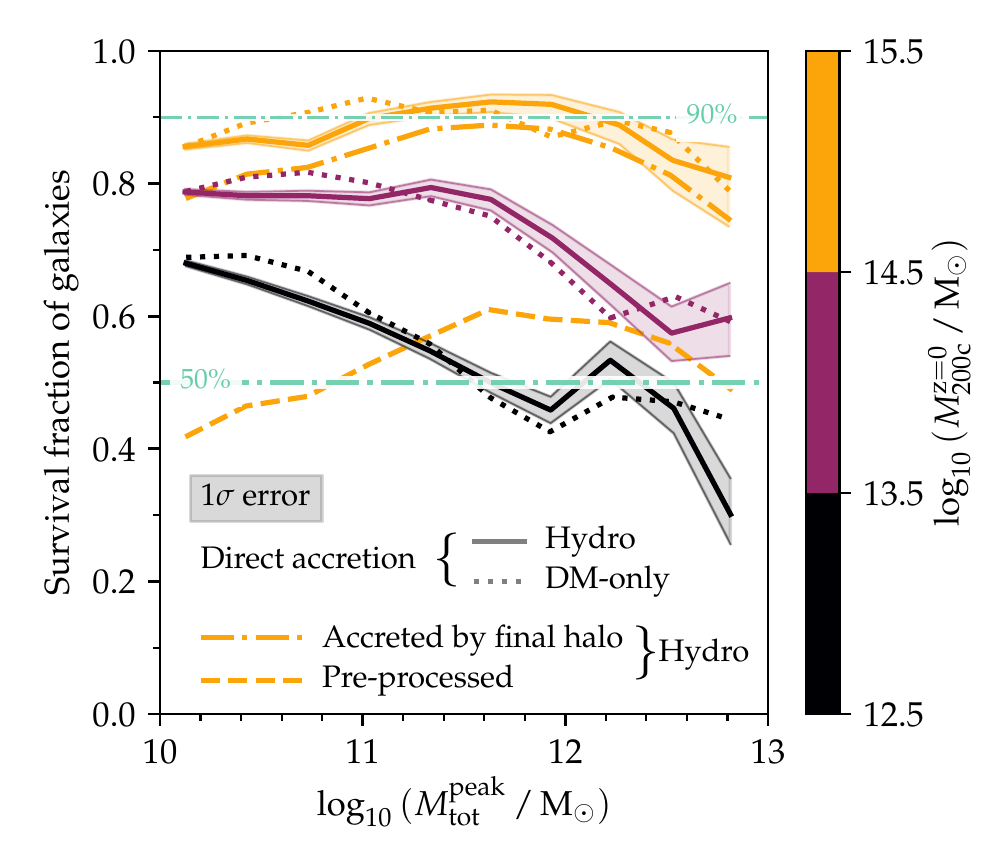}
       \caption{Survival fraction of galaxies ($\mgalzz > 5\times 10^8\,\msun$) that were directly accreted onto their final host, in the hydrodynamical (solid lines, shaded bands indicate binomial 1$\sigma$ errors) and DM-only simulations (dotted). For comparison, the survival fractions of all galaxies that reach their main host, and of pre-processed galaxies, are shown as dash-dotted and dashed lines, respectively; for clarity, we only show these for massive clusters ($M_\text{200c}^\text{z = 0} > 10^{14.5}\,\msun$) in the hydrodynamical simulations. Survival is more common for galaxies that are not pre-processed.}
    \label{fig:influence_nppSurvFrac_total}
\end{figure}

In Fig.~\ref{fig:influence_nppSurvFrac_total}, we repeat the survival analysis from Section \ref{sec:survival} (Fig.~\ref{fig:survival_total}), but this time we only consider galaxies that were not pre-processed, i.e., with $t_\text{main} = t_\text{branch}$. Different colours represent different host mass bins, and results from the hydrodynamical (DM-only) simulations are shown as solid (dashed) lines. 

It is evident that the survival fraction amongst these `directly accreted' galaxies is considerably higher than in the total population (c.f.~Fig.~\ref{fig:survival_total}): in massive clusters (orange), it reaches $\approx\,$85 per cent even at the low-mass galaxy end ($\mpeak \sim 10^{10}\, \msun$), and peaks above 90 per cent at $\mpeak \sim 3\times10^{11}\,\msun$. Even for groups (black), the survival fraction of directly accreted galaxies exceeds 60 per cent, albeit only at $\mpeak < 10^{11}\,\msun$. This contrasts starkly with the survival fractions for pre-processed galaxies, which are shown -- for clarity only for massive clusters in the hydrodynamical simulations -- as dashed lines in Fig.~\ref{fig:influence_nppSurvFrac_total} and lie in the range of $\approx\,$40--60 per cent. Pre-processing is evidently much more disruptive than the final host environment, consistent with the trend towards lower survival fractions in lower-mass (final) haloes.

Similar to the total satellite population, the survival fractions of directly accreted galaxies agree closely between DM-only and hydrodynamical simulations. The survival fractions of \emph{all} galaxies accreted by their final host is only $\loa\,$10 per cent lower than for their directly accreted subset, as shown for massive clusters by the orange dash-dotted line in Fig.~\ref{fig:survival_total}. Even higher is the survival fraction of only those galaxies that were a central immediately prior to $t_\text{main}$ (irrespective of whether they were previously pre-processed, not shown). As we demonstrate below, this is because most disruption of pre-processed galaxies occurs outside of their final host halo.

Massive clusters in particular therefore preserve a near complete `fossil record' of all galaxies with $\mpeak > 10^{10}\,\msun$ that have ever orbited within them. Keeping in mind that simulations may also disrupt satellite galaxies for numerical, rather than physical, reasons \citep{vanDenBosch_Ogiya_2018}, the true survival fractions may, in principle, be even higher than what is shown in Fig.~\ref{fig:influence_nppSurvFrac_total}. To test this, we compute in Appendix \ref{sec:vdbo} the fraction of surviving remnants that are numerically unreliable according to the criteria of \citet{vanDenBosch_Ogiya_2018}. Amongst massive galaxies ($\mpeak > 3\times 10^{11}\,\msun$), numerically unreliable remnants are rare ($\lesssim\,$1 per cent) in our simulations, but at $\mpeak \sim 10^{10}\, \msun$, up to one third of remnants may be unreliable. The survival fractions shown in Fig.~\ref{fig:influence_nppSurvFrac_total}, however, are not consistent with significant numerical disruption of low-mass satellites: e.g., they depend only weakly on galaxy mass. This suggests that \emph{numerical disruption of satellites is not common in our simulations}, at least at $\mpeak > 10^{10}\,\msun$.

\subsubsection{Where are pre-processed galaxies disrupted?}
\begin{figure}
  \centering
   \includegraphics[width=1.05\columnwidth]{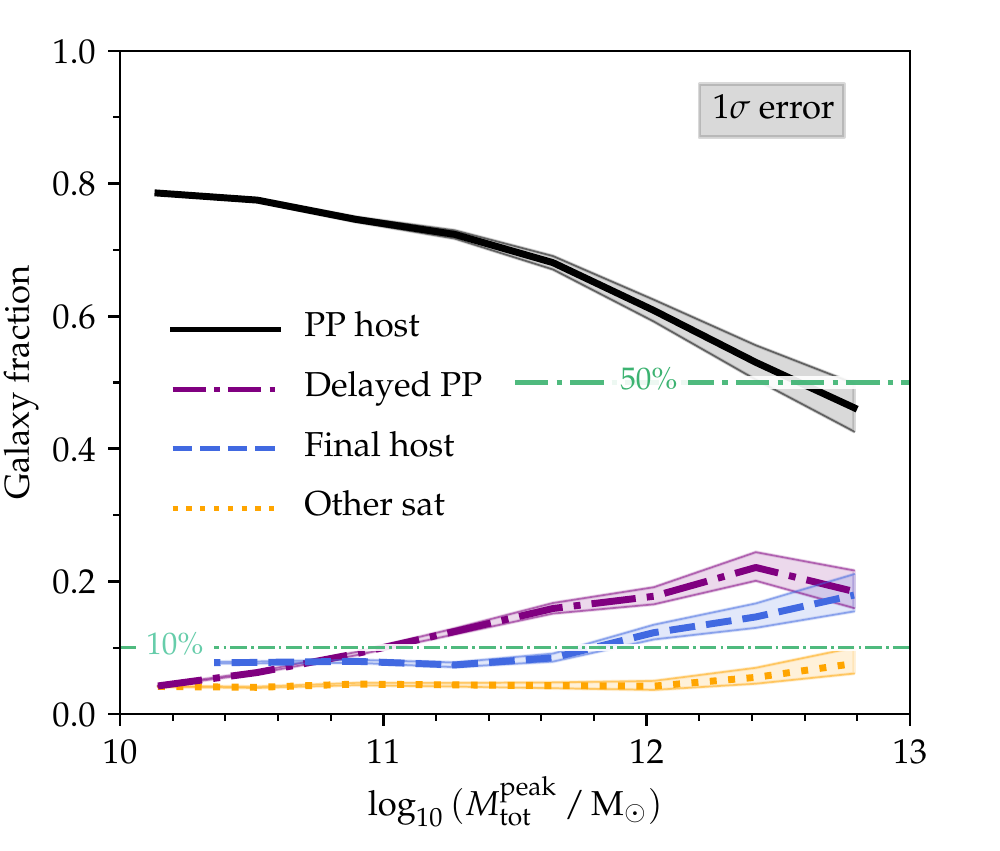}
       \caption{Merger routes of disrupted pre-processed satellite galaxies as a function of $\mgalpeak$ (all three host mass bins combined). The black solid line shows the fraction that merges with their pre-processing host before infall into their final halo ($t_\text{main}$). The purple dash-dotted and blue dashed lines show the fractions that undergo, after $t_\text{main}$, a `delayed' merger with their pre-processing host, and a merger with their final host, respectively. The orange dotted line represents mergers with another satellite, almost all of which occur during pre-processing. Shaded bands give binomial 1$\sigma$ uncertainties. The vast majority merge with their pre-processing host, either before or after reaching the final halo.}
    \label{fig:influence_mergerRoutes}
\end{figure}

Pre-processed galaxies can be disrupted either in their sub-group (prior to $t_\text{main}$), or later in their final host. In Fig.~\ref{fig:influence_mergerRoutes}, we disentangle these two scenarios, for simplicity combining all hosts with $\mvir > 10^{12.5} \msun$ into a single bin (we have verified that differences between different host masses are small). Different lines show the fractional contribution of different merger types to the disruption of pre-processed galaxies. Clearly dominant ($\approx\,$50--80 per cent, highest at lowest $\mpeak$) are mergers with the pre-processing host (black solid line), i.e.,~those that merged prior to $t_\text{main}$ with a galaxy that was previously the disrupted galaxy's central.

In addition, the next most common disruption route is also due to the pre-processing host, but only after it became itself a satellite of the (final) host halo (purple dash-dotted line). Although these are technically mergers between two satellites in the final halo, it is more appropriate to consider them as a case of `delayed pre-processing', since the infalling subgroup may retain its physical identity for some time after having been subsumed into its host. Including these, pre-processing hosts account for $\goa\,$70 per cent of all disruption of pre-processed galaxies, at all masses we probe. The remaining galaxies merge with their final host ($<$ 10 per cent at $\mpeak < 10^{12}\,\msun$, dashed blue line) or, even less commonly, with another unrelated satellite (orange dotted line), mostly during pre-processing. This is consistent with the recent study of \citet{Han_et_al_2018}, who inferred from a different set of simulations that pre-processing has a decisive impact on mass stripping from infalling galaxies, in particular when the mass ratio between galaxy and pre-processing host is low. 

\subsubsection{The contribution of pre-processing to galaxy disruption}

\begin{figure}
  \centering
    \includegraphics[width=1.05\columnwidth]{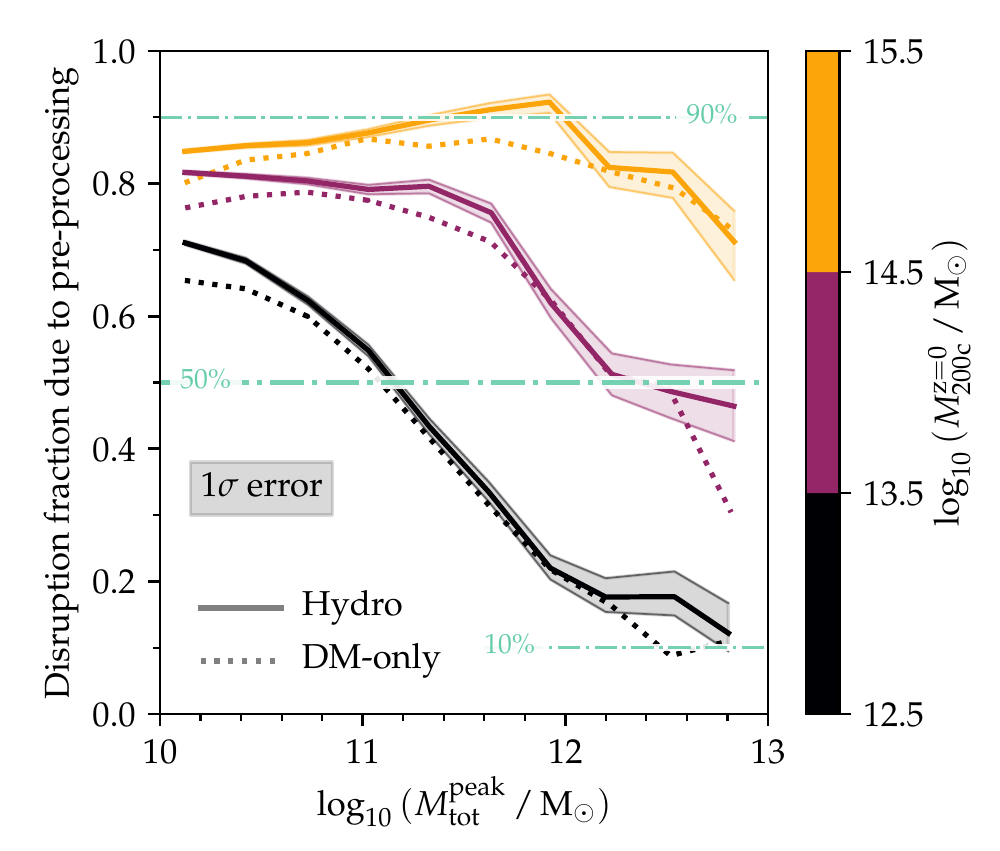}
       \caption{The fraction of all satellite disruption that is due to pre-processing (including delayed mergers), in the hydrodynamical simulations (solid lines) and their DM-only counterparts (dotted). Different host masses are represented by different colours (see the colour bar on the right). Pre-processing is by far the dominant cause of disruption in cluster galaxies with $\mgalpeak < 10^{12}\,\msun$, but it becomes much less relevant for more massive galaxies and those in lower-mass hosts.}
    \label{fig:influence_disruptionFractionByPP}
\end{figure}

To conclude our investigation of pre-processing, we show in Fig.~\ref{fig:influence_disruptionFractionByPP} the fraction of all satellite disruption that is due to pre-processing (including delayed mergers and mergers with other satellites prior to $t_\text{main}$), as a function of $\mpeak$ and $\mhost$. The combination of a higher pre-processed fraction at lower $\mpeak$ and higher $\mhost$ (Fig.~\ref{fig:accMode}), and their much lower survival fraction compared to directly accreted galaxies (Fig.~\ref{fig:influence_nppSurvFrac_total}) implies that the vast majority, $\approx\,$80--90 per cent, of all disruption at $\mpeak < 10^{12}\,\msun$ in massive clusters is the result of pre-processing. The fraction decreases somewhat towards higher masses, but pre-processing still accounts for $\approx\,$70 per cent of all disruption even at $\mpeak = 10^{13}\,\msun$. In lower-mass haloes, pre-processing is overall much less important, and only accounts for $\approx\,$20 per cent of the disruption of Milky Way analogues in groups. 

The DM-only simulations broadly agree with the hydrodynamical runs, but generally predict a slightly lower fraction of disruption that is due to pre-processing (by $\loa\,$5 per cent) and a slightly smoother transition from the flat part at low $\mpeak$ to the decline at high mass (especially in clusters). This suggests that baryons have a (small) disruptive effect in situations where the mass contrast between the satellite and host is not too large; we investigate this further in Paper II.  
 
To summarize, we have found that \emph{pre-processing plays a crucial role in disrupting galaxies}, particularly in clusters where it accounts for the vast majority of all disruption ($\approx\,$90 per cent at $\mpeak \sim 10^{12}\,\msun$ and $\mhost > 10^{14.5}\,\msun$). Galaxies accreted directly onto their final host survive to $\goa\,$85 per cent in massive clusters, and still $\approx\,$80 per cent in lower-mass clusters at $\mpeak \leq 3\times 10^{11}\,\msun$. Pre-processing disruption mostly involves mergers with the central galaxy of the subgroup. The lowest-mass haloes are therefore the most efficient in disrupting satellites at fixed $\mpeak$, plausibly as a consequence of dynamical friction, while massive galaxy clusters should preserve a near-complete record of all galaxies (at least with $\mpeak > 10^{10}\,\msun$) that they have ever accreted.  

\subsection{Role of satellite--satellite mergers} 
\label{sec:influence_satSatMergers}

\begin{figure}
  \centering
    \includegraphics[width=1.05\columnwidth]{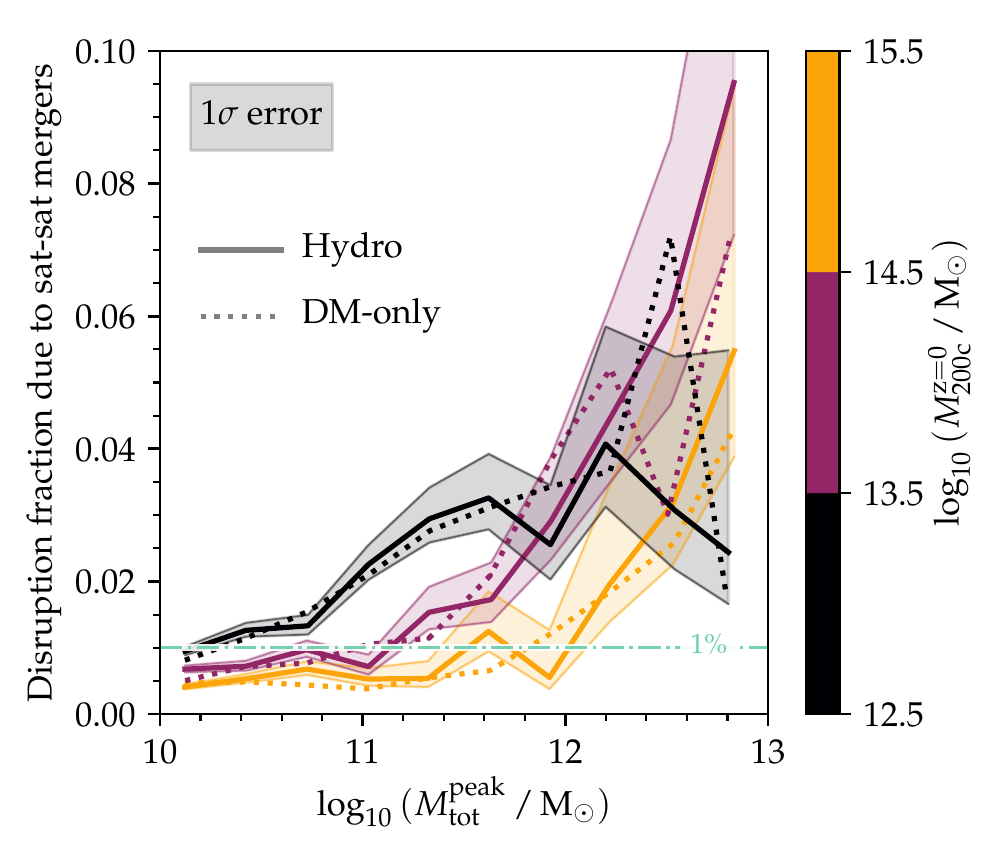}
       \caption{The fraction of non-surviving directly accreted satellites that are disrupted by mergers with other satellites. Hydrodynamical simulations are represented by solid lines (with shaded bands indicating binomial $1\sigma$ uncertainties), DM-only runs by dotted lines. For clarity, the $y$-axis range is reduced compared to the other plots. Satellite--satellite mergers are very uncommon (particularly in massive clusters): only at the highest masses ($\mgalpeak \gtrsim 10^{12}\,\msun$) do they account for $\approx\,$10 per cent of disruption events.}
    \label{fig:disruptionFractionBySat}
\end{figure}

We had noted above that satellite--satellite mergers are rather uncommon for pre-processed galaxies. Their role in the (final) host haloes themselves is explored in Fig.~\ref{fig:disruptionFractionBySat}, where we show the fraction of all disruption events amongst directly accreted galaxies ($t_\text{branch} = t_\text{main}$) that are due to mergers with other satellite galaxies. We exclude cases where this other satellite was previously the galaxy's central (due to central--satellite swaps, which is only relevant for massive galaxies in groups).

The key feature is that satellite--satellite mergers in massive haloes are extremely rare; note that the $y$-axis range is reduced to [0, 0.1] in order to highlight any deviations from zero at all. At $\mgalpeak < 10^{12}\,\msun$, they account for less than one per cent of disruption events in massive clusters, and still $\loa\,$3 per cent in groups. Only amongst the most massive galaxies are they slightly more relevant, with fractions of up to 10 per cent in low-mass clusters at $\mgalpeak \approx 10^{13}\,\msun$. What disruption occurs in massive haloes (see above) is therefore almost exclusively due to mergers with the central galaxy (including dispersal into its halo, as we test in Paper II). We note that interactions between satellites may nevertheless contribute significantly to their mass loss and ultimate dispersal (see, e.g., \citealt{Moore_et_al_1996, Marasco_et_al_2016}). At all galaxy and host masses that we consider, the predictions from hydrodynamical and DM-only simulations agree to within the statistical uncertainties, which rules out a significant impact of baryon physics on this merger channel.

\subsection{Evolution of surviving fraction with accretion time} 
\label{sec:influence_accTime}

\begin{figure*}
	\centering
	\includegraphics[width=2.1\columnwidth]{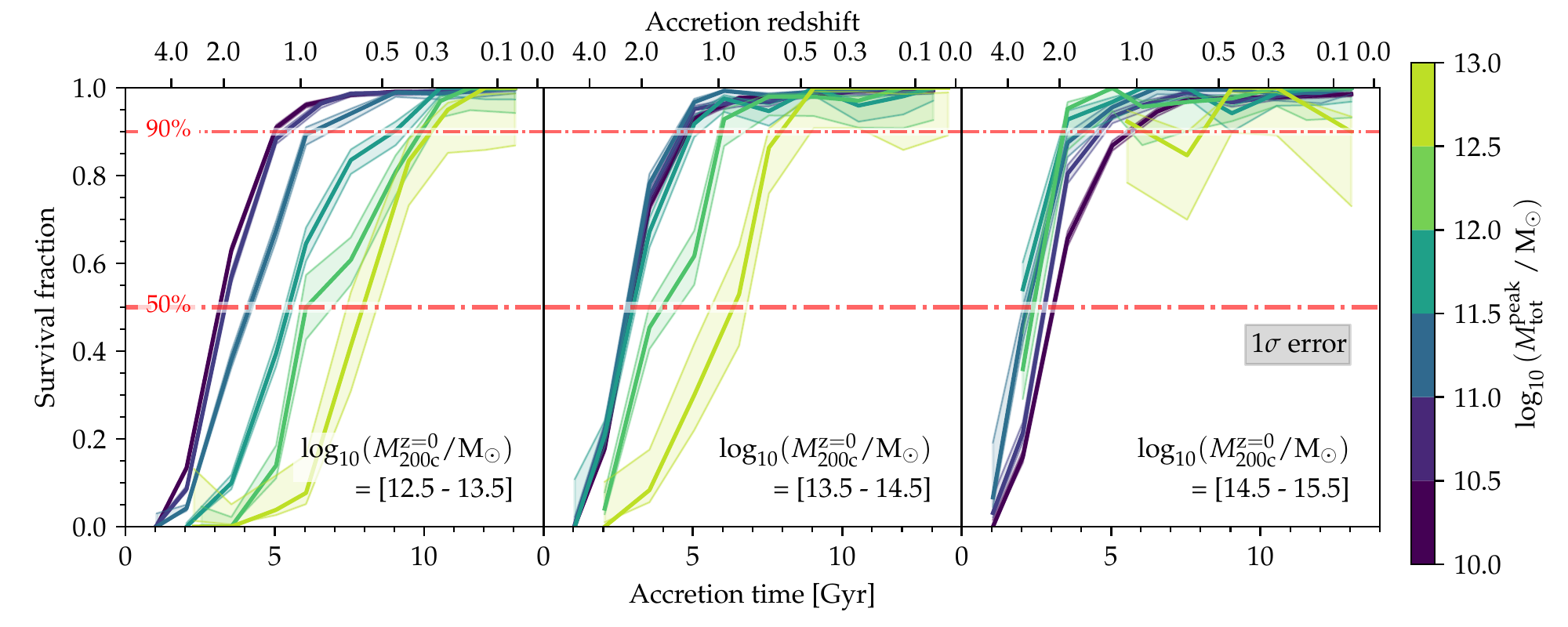}
	\caption{The fraction of non-pre-processed galaxies accreted at a given redshift ($\zacc$) that survive to $z = 0$ (with $\mgalzz > 5\times 10^8\,\msun$), in the hydrodynamical simulations. Only bins with at least ten galaxies are shown, which is why not all lines extend to the earliest accretion times. Different panels show different host mass ranges, as indicated in the bottom-right corners. In all three panels, the survival fraction of low-mass galaxies approaches unity for galaxies accreted at $z_\text{acc} \lesssim 1$, and then drops rapidly towards earlier accretion times. More massive galaxies, and those in less massive groups, are still disrupted at lower $z_\text{acc}$.} 
	\label{fig:survival_accTime}
\end{figure*}

We now examine the influence of accretion time on galaxy survival. For ease of interpretation, we focus here on directly accreted galaxies. In Fig.~\ref{fig:survival_accTime}, galaxies are split into three host mass bins (three different panels, $\mvir$ increasing from left to right) and six bins in galaxy peak mass $\mpeak$ (different coloured lines, increasing from purple to green). Each line traces the fraction of galaxies that survive (with $\mgalzz > 5\times 10^8\,\msun$) as a function of accretion time ($t_\text{acc}$), or equivalently redshift ($z_\text{acc}$). For clarity, only the hydrodynamical simulations are shown, but we have verified that the DM-only runs give very similar results.

The dominant trend of all lines in Fig.~\ref{fig:survival_accTime} is that galaxies that were accreted later are more likely to survive to $z = 0$, in agreement with previous work (e.g., \citealt{DeLucia_et_al_2004, Gao_et_al_2004}). At $\zacc \approx 0$ the survival fraction approaches unity, as should be expected. The few per cent of galaxies that were accreted very early, on the other hand ($\zacc \goa 4$, see Fig.~\ref{fig:accTimes}), almost never survive to $z = 0$.

Within each bin of host and galaxy mass (individual lines in Fig.~\ref{fig:survival_accTime}), the survival fraction always transitions quite rapidly from $\sim$0 to $\sim$1, over a period of typically only a few Gyr. The accretion time (measured from the Big Bang) at which the survival fraction reaches 50 per cent ($t_\text{trans}$) depends in general on both $\mpeak$ and $\mhost$. In low-mass groups (left-hand panel), $t_\text{trans} \approx 2.5$ Gyr ($z \approx 2$) for the lowest-mass galaxies (purple) and then increases fairly gradually to $t_\text{trans} \approx 7.5$ Gyr ($z \approx 0.6$) at $\mpeak > 10^{12.5}\, \msun$ (yellow-green). While the lowest-mass galaxies therefore already survive to 90 per cent at $\zacc = 1.3$, those with the highest masses only reach this point at $\zacc = 0.3$.

The dependence of $t_\text{trans}$ on galaxy mass is noticeably less strong in more massive hosts. In low-mass clusters ($\mhost \sim 10^{14}\, \msun$; middle panel of Fig.~\ref{fig:survival_accTime}) the lowest-mass galaxies follow almost exactly the same trend as in groups, but not until $\mpeak = 10^{12}\, \msun$ is there a noticeable shift towards later $t_\text{trans}$. Consequently, even the most massive galaxies reach 50 (90) per cent survival already at $\zacc = 0.9$ ($\zacc = 0.6$). 

In massive clusters (right-hand panel), any differences with $\mpeak$ are very small, but there is a slight shift towards \emph{even earlier} $t_\text{trans}$ with increasing galaxy mass, at least for those bins where our simulations contain enough galaxies to identify $t_\text{trans}$. This shift may reflect the enhanced ability of more massive galaxies to withstand tidal stripping, while their mass is still so far below that of the host cluster that, e.g., dynamical friction does not cause accelerated disruption in the same way as in lower-mass hosts. Milky Way analogues ($\mpeak \sim 10^{12}\, \msun$) therefore reach 90 per cent survival already at $\zacc = 2.0$ and 96 per cent of all galaxies with $\mpeak > 10^{10}\,\msun$ and $\zacc < 2$ survive at $z = 0$. The small fraction of galaxies that are disrupted in massive clusters are therefore predominantly those that were accreted the earliest.

\subsection{From accretion to disruption: rapid, delayed, or continuous?}
\label{sec:influence_tDisrupt}

A natural question to ask is whether the relatively rapid transition from disruption- to survival-dominated accretion redshifts is indicative of a long, mass-dependent delay between accretion and disruption. In other words, galaxies accreted just after $t_\text{trans}$ may survive at $z = 0$ because they have (just) not been a satellite for long enough, while those accreted just before could have been disrupted very recently. We now demonstrate that such a delay time argument cannot be invoked as the reason for the lower survival fraction of early-accreted galaxies. 

\begin{figure}
   \centering
   \includegraphics[width=1.03\columnwidth]{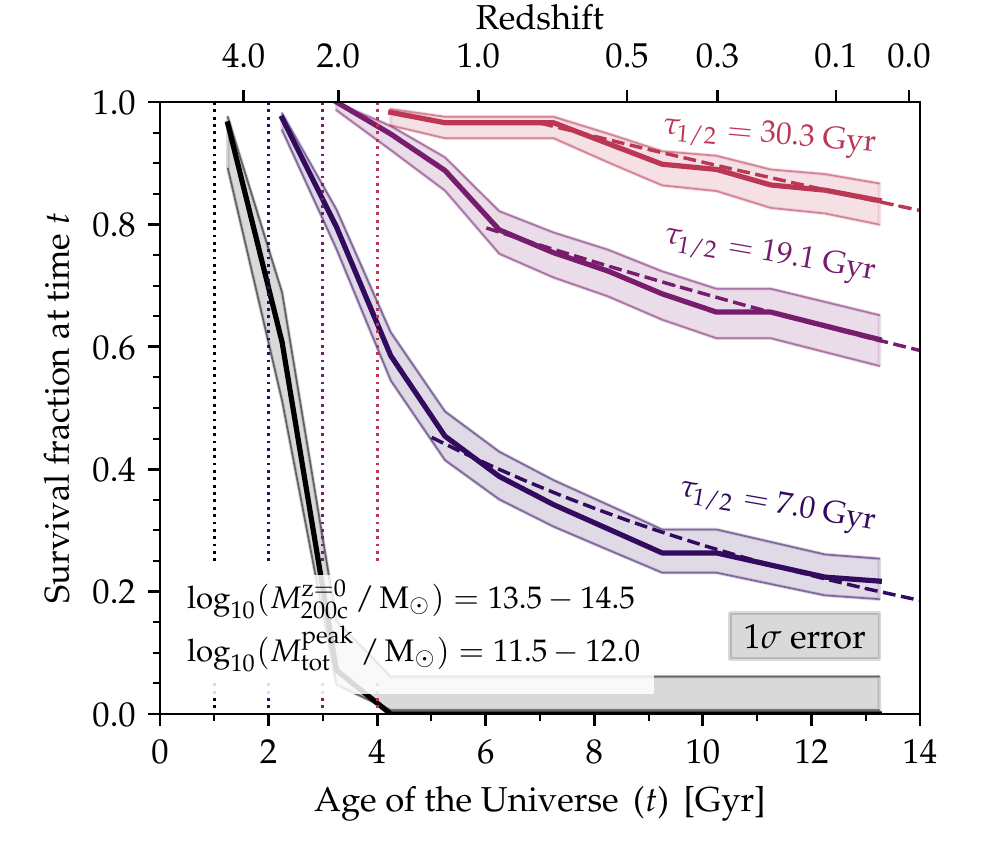}
   \caption{Connection between accretion and disruption time for one bin in galaxy and host mass (see bottom-left corner) in the hydrodynamical simulations. We select galaxies that were accreted within 4 intervals of $\Delta t = 500$ Myr, centered on the times indicated by vertical dotted lines, and show their survival fraction (above a total mass threshold of $5\times10^8\,\msun$) as a function of time. Thin dashed lines represent the best-fit late-time exponential decay model, with corresponding half-life time $\tau_{1/2}$ as indicated on the right. The earliest-accreted galaxies have all disrupted rapidly after accretion. Later generations show a progressively shallower decline in survival fraction, with $\tau_{1/2} \gg t_\text{Hubble}$ at $\zacc < 2$.} 
    \label{fig:survivalTime}
\end{figure}

For this purpose, Fig.~\ref{fig:survivalTime} shows the survival fraction of galaxies as a function of cosmic time $t$, i.e., the fraction with $\mtot (t) > 5\times 10^8\,\msun$, . We select galaxies that were not pre-processed in four bins of $\Delta t_\text{acc} = 500$ Myr, with centres indicated by the vertical dotted lines. For clarity, we focus on only one bin in galaxy mass ($\mpeak = 10^{11.5}$--$10^{12.0}\, \msun$) and host mass (low-mass clusters) in the hydrodynamical simulations. For each bin in accretion time, the correspondingly coloured solid line shows the fraction of galaxies still alive at time $t$, and the bands the corresponding $1\sigma$ binomial uncertainties. 

It is immediately evident that there is no universally long delay between accretion and disruption, particularly at high $z_\text{acc}$ (black/indigo). The disruption rate (i.e.,~the line slope) is greatest within the first few Gyr after accretion and then flattens off. In the earliest accretion bin ($z_\text{acc} > 4$; black), all galaxies are disrupted within 3 Gyr of accretion, while a successively higher fraction of later-accreted galaxies survive at least this long. At $t > t_\text{acc} + 3\, \text{Gyr}$, the survival fraction decays approximately exponentially with $t$. The best fits are given by the dashed lines, with a systematically increasing half-life time $\tau_{1/2}$ for lower $z_\text{acc}$. At $z_\text{acc} < 2$, $\tau_{1/2}$ exceeds (significantly) the available time until $z = 0$, which naturally explains why most of these galaxies survive until today. 

The strong dependence of the survival fraction on accretion redshift (Fig.~\ref{fig:survival_accTime}) is therefore the result of the disruption efficiency decreasing (strongly) with time. It is conceivable that this reflects the lower host halo masses at higher $\zacc$, but we have verified that our results are not markedly changed when galaxies are instead binned by their host mass \emph{at accretion}, as long as it remains\footnote{In other words, excluding situations better described as minor or major galaxy mergers, rather than accretion of satellites.}  $\gtrsim\,$1 dex above $\mpeak$. Instead, the fact that the half-life times shown in Fig.~\ref{fig:survivalTime} scale with accretion redshift approximately as $(1+z_\text{acc})^{-3/2}$ -- the expected scaling of the dynamical time with redshift \citep{McGee_et_al_2014} -- suggests that the low survival fraction of early-accreted galaxies is due to different orbital conditions imprinted at accretion. In Paper II, we show that early-accreted galaxies lose mass more rapidly because they have (much) shorter orbital periods, while massive galaxies are more strongly dragged towards the host centre at high $z_\text{acc}$ and can therefore merge more efficiently with the (growing) central cluster galaxy.    

\subsection{Galaxy disruption times}
\label{sec:tDisrupt}
\begin{figure}
   \centering
   \includegraphics[width=1.03\columnwidth]{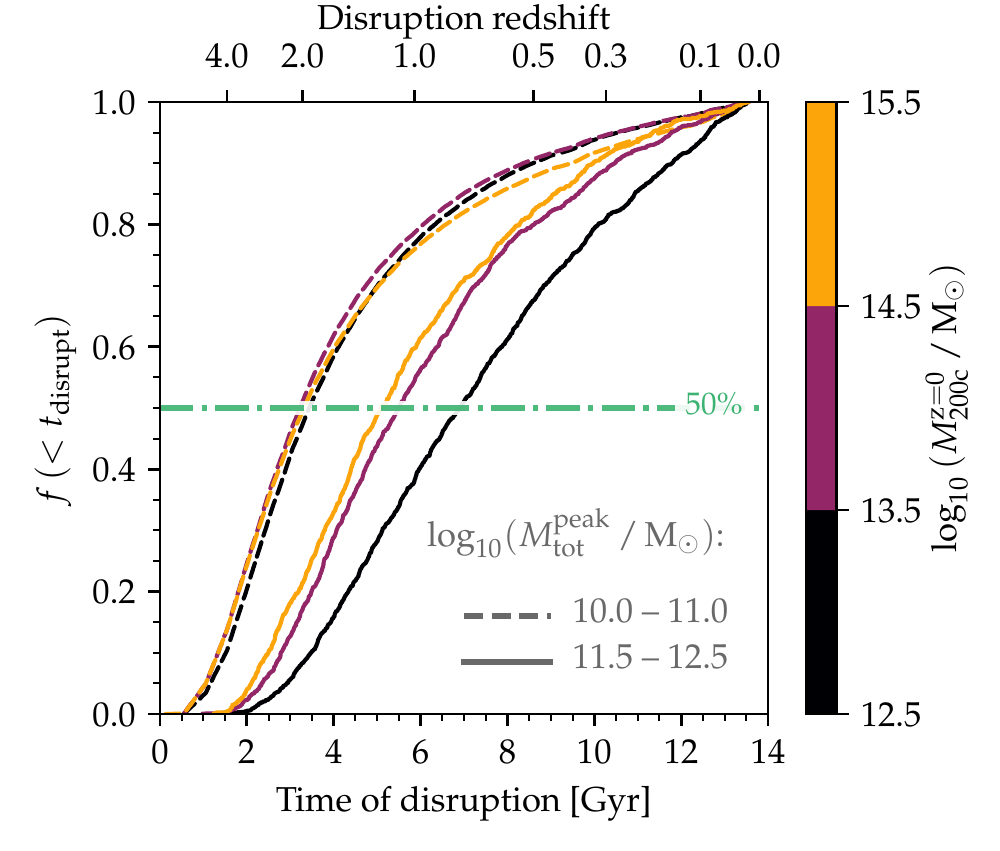}
   \caption{Distribution of galaxy disruption times. Shown is the fraction of (ultimately disrupted) galaxies that fall below the survival threshold of $5\times10^8\,\msun$ prior to time $t_\text{disrupt}$, in two bins of peak galaxy mass (different line styles) and three host mass bins (different colours). No selection is made with regards to pre-processing. Disruption was most prevalent at redshift $z \approx 1$--3, but with a broad tail extending to $z = 0$. Massive galaxies were disrupted later, but the influence of host mass is small.} 
    \label{fig:disruptionTimes}
\end{figure}

We have so far only distinguished galaxies by their accretion times, but a related question of interest -- particularly for connection with observational work -- is when galaxies actually disrupt. This is shown in {\color{Blue}Fig.~\ref{fig:disruptionTimes}}, which gives the cumulative fraction of (non-surviving) galaxies that were disrupted (i.e., fell below our mass threshold of $5\times10^8\,\msun$) prior to a given time $t_\text{disrupt}$. The three bins in host mass are represented by differently coloured lines; two bins in galaxy mass are distinguished by different line styles. As in Fig.~\ref{fig:accTimes}, all times are offset by a random value of up to $\pm$250 Myr to suppress artificial discreteness due to the limited number of snapshots. 

The distribution is qualitatively similar to that of the branch accretion times shown in the top panel of Fig.~\ref{fig:accTimes}, consistent with the picture that most galaxies are disrupted during pre-processing, soon after first accretion. In line with their lower accretion redshifts, more massive galaxies (solid lines) are disrupted slightly later, with median disruption redshifts of $z = 2$ and $\approx\,$1 for $\mpeak = 10^{10}$--$10^{11}$ and $10^{11.5}$--$10^{12.5} \msun$, respectively. The lower-mass bin shows no dependence of $t_\text{disrupt}$ on $\mhostzz$ at all, but there is a slight tendency towards later disruption in groups than clusters for more massive galaxies (by $\loa\,$2 Gyr), consistent with the equivalent trends in $t_\text{branch}$. Due to the typically long delay between accretion and disruption for most galaxies accreted at intermediate and low redshifts (Fig.~\ref{fig:survivalTime}), disruption is still prevalent in the low-redshift Universe, in particular amongst massive galaxies ($\mpeak \sim 10^{12}\,\msun$), for which $\goa\,$10 per cent of all disruption events occur at $z < 0.3$.


\section{Biases between surviving and disrupted galaxies} 
\label{sec:bias}

For the final part of our analysis we test whether there are any differences in pre-infall properties between galaxies that are disrupted and those that survive, at fixed (total) $\mpeak$. Such differences could cause subtle biases between (surviving) cluster and field galaxies without any actual galaxy transformation process. To pre-empt the answer, we did not find any strong differences of this kind in terms of either the baryonic or dark matter properties of galaxies -- at least those accreted around $z \approx 2$ -- and can therefore rule out such `differential disruption' as a significant contributor to the observed differences between field and cluster galaxies in the local Universe.

A complication in comparing the pre-infall properties of disrupted and surviving galaxies is that, as we have found above (Fig.~\ref{fig:survival_accTime}), disrupted galaxies were preferentially accreted earlier than survivors. Because the relations of, e.g.,~stellar mass and star formation rate with halo mass evolve with redshift (see, e.g.,~\citealt{Furlong_et_al_2015} and references therein), a comparison between \emph{all} disrupted and surviving galaxies would show strong differences that are purely the result of this redshift bias. A meaningful comparison is therefore only possible between galaxies with similar accretion redshift $z_\text{acc}$ and furthermore -- due to the finite number of galaxies in our simulation -- only around the $z_\text{acc}$ where the survival and disruption fractions are comparable (i.e., $z_\text{acc} \approx 2$).  

\begin{figure}
  \centering
    \includegraphics[width=\columnwidth]{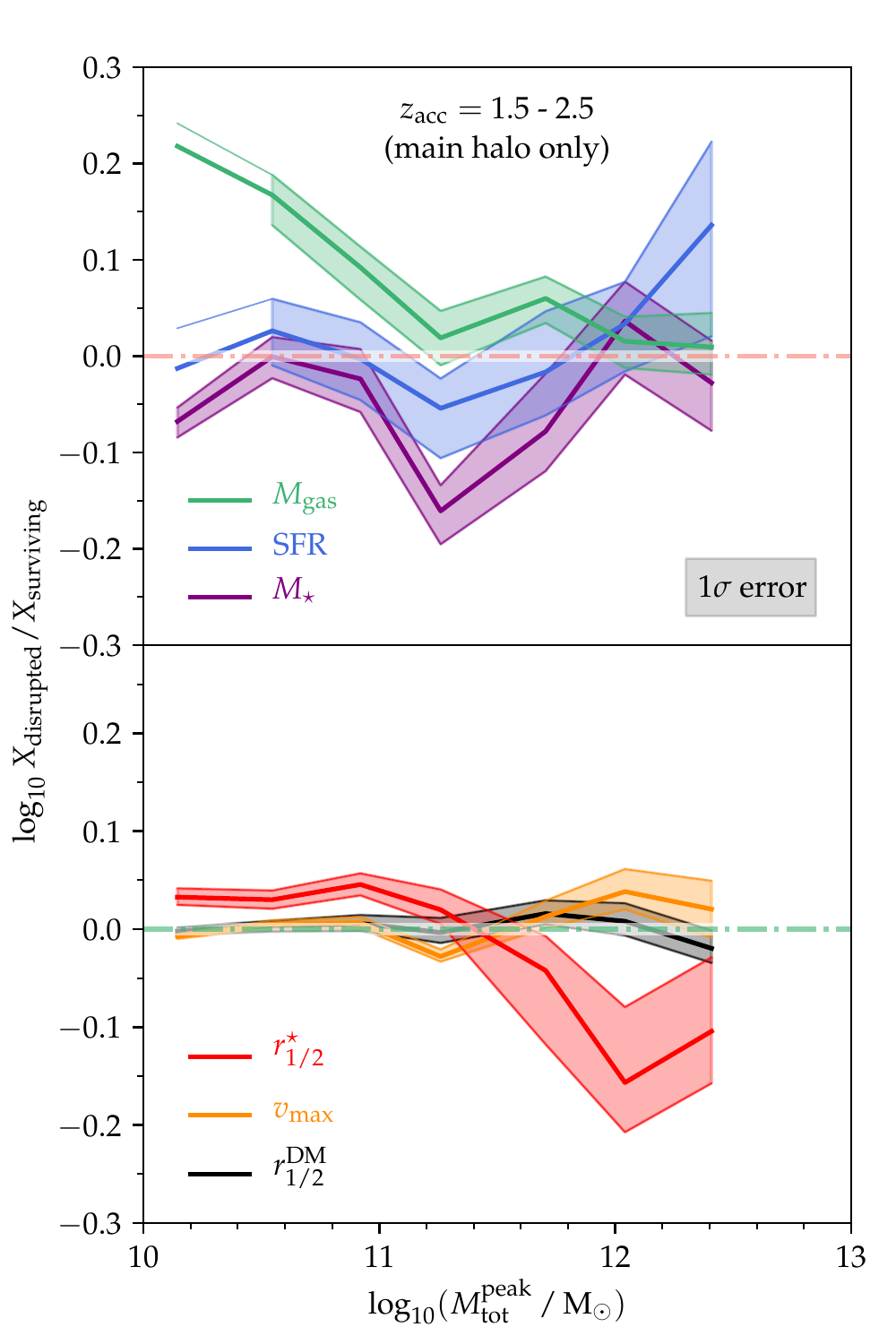}
       \caption{\textbf{Top panel:} bias of disrupted galaxies in baryonic properties prior to accretion, compared to surviving galaxies (at a mass threshold of $5\times10^8\,\msun$ for survival). Stellar mass is shown in purple, star formation rate in blue, and gas mass in green. Only galaxies directly falling into their final host between $z = 2.5$ and 1.5 are shown to avoid indirect biases due to different accretion times. Only bins with at least ten surviving and disrupted galaxies are shown. At fixed $\mpeak$, disrupted galaxies had slightly higher gas mass, and slightly lower stellar mass, than surviving galaxies. \textbf{Bottom panel:} the same for DM half-mass radius (black), stellar half-mass radius (red), and maximum circular velocity (orange). Out of these, only the stellar half-mass radius shows a significant (but relatively small) difference between disrupted and surviving galaxies.}
    \label{fig:survivalBias}
\end{figure}

In the top panel of {\color{blue}Fig.~\ref{fig:survivalBias}}, we show the gas mass (green), star formation rate (blue), and stellar mass (purple) in the snapshot before the main accretion time for disrupted galaxies that were directly accreted between $z_\text{acc} = 1.5$ and 2.5. The DM half-mass radius (black), stellar half-mass radius (red), and maximum circular velocity (orange) are compared in the bottom panel. All values are normalised to their analogues for surviving galaxies: we compute the median and $1\sigma$ uncertainty for disrupted and surviving galaxies as a function of $\mpeak$ and then plot their logarithmic ratio. The $1\sigma$ errors shown as shaded bands are here computed as the difference between the median and the 16$^\text{th}$/$84^\text{th}$ percentiles, divided by $\sqrt{N}$ where $N$ is the number of galaxies per bin. Note that for SFR and $M_\text{gas}$, the $16^\text{th}$ percentile is equal to zero in the lowest-mass bin, so that we cannot compute a meaningful (logarithmic) lower error boundary. 

The key feature of Fig.~\ref{fig:survivalBias} is the absence of any clear, strong differences between disrupted and surviving galaxies. There is a mildly significant negative bias in stellar mass, i.e.,~in the sense that disrupted galaxies contained less stellar mass prior to accretion than equally-massive surviving galaxies, but only by $\loa$\,0.1 dex. Similarly, there is a mild positive bias in gas mass, at least for low-mass galaxies ($\mpeak < 10^{11} \msun$). This is consistent with a picture in which there are (small) individual effects of gas stripping (e.g., \citealt{Saro_et_al_2008}) and (past) star formation (e.g., \citealt{Weinberg_et_al_2008}), which largely cancel each other on average. 

The pre-accretion stellar half-mass radius of disrupted low-mass galaxies ($\mpeak \lesssim 3 \times 10^{11} \msun$) is marginally (but significantly) larger than for surviving galaxies, consistent with the expectation that less compact galaxies are more susceptible to tidal stripping. Interestingly, our simulations predict the opposite trend for more massive galaxies, where disrupted galaxies were, on average, slightly more compact prior to accretion. This is further evidence for two different disruption channels for low- and high mass galaxies, as we discuss in Paper II. No consistent and significant difference is seen for SFR, DM half-mass radius ($r_{1/2}^\mathrm{DM}$), or maximum circular velocity ($v_\mathrm{max}$).

The key implication is that \emph{whether a galaxy survives or not depends at best weakly on its internal properties}. This fits in with our earlier conclusion that the survival fraction is similar between DM-only and hydrodynamical simulations, and does not depend strongly on galaxy mass. The almost an order of magnitude higher stellar mass fractions of (surviving) satellites in clusters \citep{Bahe_et_al_2017b}, in particular those accreted early \citep{Armitage_et_al_2018}, are therefore predominantly a consequence of dark matter being stripped more efficiently than stars from satellite galaxies (see Section \ref{sec:survival_detThresh}), rather than star formation enhancing the likelihood of survival. We caution, however, that we could only do this test for galaxies in a relatively narrow and high range of accretion redshifts. A significantly larger cluster sample would be required to check whether the same conclusion holds for galaxies that were accreted later. 

Finally, we note that we have also considered the equivalent biases for pre-processed galaxies (with quantities calculated at $t_\text{branch}$, not shown). Most features are qualitatively consistent with Fig.~\ref{fig:survivalBias}, but there appears to be a stronger negative bias in stellar mass (approximately -0.15 dex), and a small but significant negative bias in $r_{1/2}^\mathrm{DM}$ (approximately -0.06 dex) for disrupted pre-processed galaxies. This may, however, simply be a manifestation of indirect bias due to large-scale environmental influence of the host\footnote{Galaxies whose pre-processing begins closer to their final host have a higher chance of survival (due to the shorter time before their pre-processing host is itself accreted) and are more strongly affected by large-scale environmental influence of their final host. Directly accreted galaxies are not subject to this bias, because their hosts affected all of them approximately equally at the point of accretion.}. Again, we would require a larger simulation volume to control for this indirect effect and test whether internal galaxy properties are causally connected to survival in the pre-processing phase.


\section{Summary and Discussion}
\label{sec:summary}

We have investigated the disruption of galaxies in groups and clusters with the aid of the Hydrangea simulations, a suite of cosmological, hydrodynamical/$N$-body zoom-in simulations of 24 galaxy clusters and their large-scale environments. From the evolutionary histories of individual simulated galaxies with a peak (i.e., maximum past) total (baryons plus dark matter) mass of $\mpeak > 10^{10}\,\msun$ -- corresponding to a peak stellar mass $\mgalstarpeak \gtrsim 5\times 10^7\,\msun$ -- that we have computed with an updated tracing procedure, we have searched for galaxies that were accreted by a group/cluster in the past and identified those as `surviving' that still correspond to distinct subhaloes with total mass above $5\times 10^8\,\msun$ at $z = 0$. Our main conclusions may be summarised as follows:

\begin{enumerate}
\item Averaged over the entire history of the Universe, our simulations predict that 47 per cent of all satellite galaxies with peak total mass $\mpeak \geq 10^{10}\, \msun$ that were accreted onto groups or clusters ($\mvir > 10^{12.5}\,\msun$) survive to the present day. The survival fraction increases somewhat with host halo mass and is rather insensitive to galaxy mass. The fraction is highest (67 per cent) for galaxies with $\mpeak \sim 10^{12}\, \msun$ in massive clusters, and differs only marginally ($\loa\,$5 per cent) between simulations with and without baryons (Fig.~\ref{fig:survival_total}).

\item Many surviving galaxies have lost a large fraction of their $\mgalpeak$ by $z = 0$, and may therefore not be counted as surviving with higher mass thresholds and/or in lower-resolution simulations. However, hardly any galaxies in the hydrodynamical simulations survive with less than 1 per cent of their $\mgalpeak$, even where such a remnant would be well resolved (Fig.~\ref{fig:survival_detThresh}). Hence, once a galaxy hast lost $\gg$ 90 per cent of its peak total mass, its chance of survival is very small.

\item Stellar mass loss from surviving galaxies is less severe than total mass loss. Even at very low peak stellar masses ($\mgalstarpeak \sim 10^8\, \msun$) and including mass loss from stellar evolution, only a few per cent of galaxies survive with less than one tenth of their peak stellar mass. At $\mgalstarpeak > 10^{10}\, \msun$, even 50 per cent stellar mass loss is rare. In terms of stellar mass, survival is therefore almost binary: either a significant fraction is retained, or the galaxy is lost completely. Nevertheless, only $\loa\,$10 per cent of surviving galaxies are stellar-mass dominated at $z = 0$, even at the most favourable $\mgalstarpeak \approx 2\times 10^{10}\, \msun$ (Fig.~\ref{fig:survival_detThreshMstar}).

\item Most galaxy disruption in clusters, and at $10^{10}\,\msun \leq \mpeak < 10^{11}\,\msun$ also in groups, occurs during pre-processing. At $\mpeak \sim 10^{12}\,\msun$, 90 per cent of all disrupted galaxies in massive clusters ($M_\text{200c}^\text{z = 0} > 10^{14.5}\,\msun$) were pre-processed (Figs.~\ref{fig:influence_mergerRoutes} and \ref{fig:influence_disruptionFractionByPP}). The survival fraction of galaxies that were directly accreted by their final host is as high as $\approx\,$90 per cent (at $\mpeak = 10^{11.5} \msun$ in a massive cluster), with only per cent level variations between hydrodynamical and DM-only simulations (Fig.~\ref{fig:influence_nppSurvFrac_total}). The most massive host haloes are therefore the least efficient in disrupting satellites of a given mass, and vice versa.

\item The survival fraction of satellite galaxies depends strongly and non-linearly on their accretion redshift ($\zacc$). In massive clusters, and at $\mpeak < 10^{11}\, \msun$ even in low-mass groups, $\goa\,$95 per cent of non-pre-processed galaxies with $\zacc \leq 1$ survive to $z = 0$. Towards higher $\zacc$, the survival fraction drops steeply and universally becomes negligible for $\zacc \goa 4$. Below the scale of massive clusters ($\mhost < 10^{14.5} \msun$), this transition from low ($<$ 10 per cent) to high ($>$ 90 per cent) survival fractions occurs at lower $\zacc$ for galaxies with higher $\mpeak$ and those in lower-mass hosts: at fixed $\zacc$ and $\mhost$, the lowest-mass galaxies are therefore the most likely to survive (Fig.~\ref{fig:survival_accTime}). This redshift dependence is the result of a strong evolution in the disruption efficiency with $\zacc$, rather than reflecting a uniformly long delay time between accretion and disruption (Fig.~\ref{fig:survivalTime}). 

\item The disruption of galaxies continues until $z = 0$. Half of all non-surviving galaxies with $\mpeak \sim 10^{12}\, \msun$ are disrupted after $z = 1$ (including during pre-processing); for clusters ($M_\text{200c}^\text{z = 0} > 10^{13.5}\,\msun$), ten per cent of these disruption events occur after $z = 0.3$ (Fig.~\ref{fig:disruptionTimes}). 

\item The survival of galaxies is not strongly correlated with their internal properties before accretion, at least for those accreted in the interval $1.5 \leq \zacc \leq 2.5$. Compared to survivors with the same $\mpeak$, disrupted galaxies contained only slightly more gas ($\loa\,$0.2 dex) and slightly less stellar mass ($\loa\,$0.1 dex). Stellar half-mass radii show a slight, mass-dependent bias between disrupted and surviving galaxies; star formation rate, maximum circular velocity, and dark matter half-mass radius display no significant offsets. The observed differences between cluster and field galaxies at $z \approx 0$ are therefore unlikely the result of biased survival amongst the former (Fig.~\ref{fig:survivalBias}).

\end{enumerate}

According to these findings, the disruption of satellite galaxies is not a ubiquitous feature of cosmological galaxy cluster simulations, at least not at $\mpeak > 10^{10}\,\msun$ and at the relatively high mass resolution of Hydrangea ($\sim$10$^6$ and $10^7\, \msun$ for baryons and DM, respectively). In contrast to recent predictions from idealised $N$-body experiments \citep{vanDenBosch_Ogiya_2018}, galaxies with the lowest (peak) mass are in fact the most likely ones to survive to $z = 0$ at any $\zacc$ and $\mhost$ (with the possible exception of the most massive clusters, where galaxies of all masses we have considered display a similarly high survival fraction). The mass range that we have probed extends well below the scale at which baryonic properties of galaxies become affected by poor resolution ($\mpeak \sim 10^{11} \msun$; \citealt{Schaye_et_al_2015}). This suggests that artificial disruption of satellites is not a major roadblock for cosmological hydrodynamical simulations.

Although massive galaxy clusters give rise to strong tidal and ram pressure forces, our simulations predict that this is in fact the environment in which the \emph{smallest} fraction of satellite galaxies are destroyed. Instead, they should contain a near-complete `fossil record' of all galaxies that have ever orbited within them, whereas $\approx\,$1/3 of satellites in low-mass groups are disrupted before $z = 0$. Despite their rarity, massive clusters therefore constitute a valuable laboratory to study the effect of environmentally-induced galaxy transformations over time. These findings are consistent with the observational detection of an upturn in the satellite luminosity function at the faint end in clusters (e.g.~\citealt{Lan_et_al_2016}), which suggests that low-mass satellites are indeed able to survive and accumulate in massive haloes.  

There are two regimes where our simulations do predict a significant fraction of satellites to be disrupted: pre-processing in lower-mass groups, which then later assemble into a more massive group or cluster, and satellites accreted at high redshift, where disruption was evidently much more widespread, and more swift, than in the present-day Universe. This agrees with the observational evidence for widespread (dwarf) galaxy disruption during the early stages of cluster formation \citep{Lopez-Cruz_et_al_1997}. We defer further exploration of these trends to a follow-up paper, where we show that they are the consequence of enhanced mergers between satellite and central galaxies, and a strong evolution of the orbital timescale of galaxies, with increasing $\zacc$ (see also \citealt{Han_et_al_2018}). Both effects highlight the impact of the cosmological environment of groups and clusters on the predicted evolution of their member galaxies.

Our simulations suggest that the role of baryons in determining the survival of satellite galaxies -- but not the degree of stripping they experience -- is small, which is important in two ways. Firstly, it rules out `biased survival' as a significant contributor to the environmental differences that are observed in the local Universe: in principle, e.g.,~the relative overabundance of red, quenched galaxies could also have stemmed from a preferential disruption of their blue, star-forming cousins. Our findings therefore corroborate the hypothesis that these differences are the result of individual galaxies being transformed by their environment, through processes such as ram-pressure stripping, strangulation, or tidal stripping. Secondly, the small impact of baryons implies that pure $N$-body simulations can, at least in principle, predict the survival of galaxies with reasonable accuracy. 

We finally emphasize that negligible total disruption of satellites in massive clusters, as predicted by our study, is not incompatible with (significant) mass loss from surviving satellites. Indeed, we have shown that many low-mass galaxies only survive as small remnants with total (but typically not stellar) mass well below their peak values. In future work, we will investigate in more detail how this mass loss is connected to the build-up and growth of central group and cluster galaxies, and of their extended dark matter and stellar haloes.

\section*{Acknowledgments}
We thank the reviewer for their report, which improved the presentation of the results in this paper. We thank Lydia Heck for expert computational support with the Cosma machine in Durham, which was used for part of the work presented here. YMB acknowledges funding from the EU Horizon 2020 research and innovation programme under Marie Sk{\l}odowska-Curie grant agreement 747645 (ClusterGal) and the Netherlands Organisation for Scientific Research (NWO) through VENI grant 016.183.011. CDV acknowledges financial support from the Spanish Ministry of Economy and Competitiveness (MINECO) through grants AYA2014-58308 and RYC-2015-1807. The Hydrangea simulations were in part performed on the German federal maximum performance computer~``HazelHen'' at the maximum performance computing centre Stuttgart (HLRS), under project GCS-HYDA / ID 44067 financed through the large-scale project~``Hydrangea'' of the Gauss Center for Supercomputing. Further simulations were performed at the Max Planck Computing and Data Facility in Garching, Germany. This work also used the DiRAC Data Centric system at Durham University, operated by the Institute for Computational Cosmology on behalf of the STFC DiRAC HPC Facility (www.dirac.ac.uk). This equipment was funded by BIS National E-infrastructure capital grant ST/K00042X/1, STFC capital grant ST/H008519/1, and STFC DiRAC Operations grant ST/K003267/1 and Durham University. DiRAC is part of the National E-Infrastructure. \bibliographystyle{mnras}
\bibliography{bibliography_disruption}


\begin{appendix}

\section{The \textsc{Spiderweb} tracing algorithm}
\label{app:spiderweb}

\begin{figure}
  \centering
    \includegraphics[width=\columnwidth]{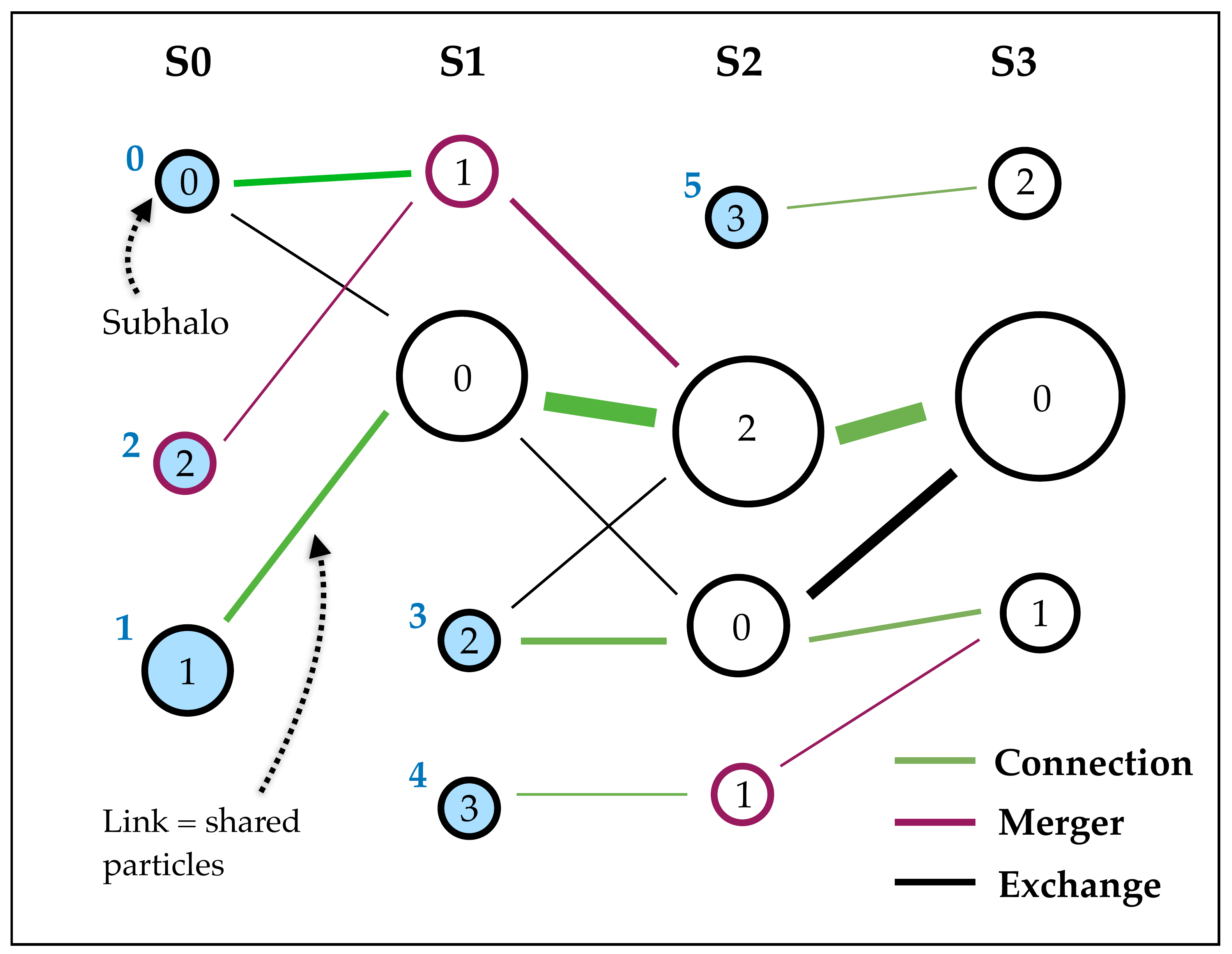}
       \caption{Example tracing situation with 14 subhaloes (circles with inscribed number representing their arbitrary ID and size indicating their mass) in 4 snapshots S0--S3 (different columns). Lines represent links, i.e.,~particle overlaps between subhaloes, whose width scales with link particle number (for simplicity assuming that all particles have equal mass). Green lines represent links that connect successive subhaloes of the same galaxy. The first subhalo of each galaxy is shaded blue, with a number to the left indicating the (arbitrary) galaxy ID. Purple circles indicate the last subhalo of a galaxy about to disrupt/merge, with the corresponding purple link pointing to the galaxy's carrier in the next snapshot. Black lines represent exchange links between galaxies.}
    \label{fig:appA_galSH}
\end{figure}

In this Appendix, we provide a detailed description of the \spiderweb{} algorithm that we have used to trace simulated galaxies through time, a significantly updated version of the procedure described in \citet{Bahe_McCarthy_2015} and \citet{Bahe_et_al_2017b}. The fundamental assumption is that simulated galaxies are physical structures that persist through time and are therefore, in general, present in multiple snapshots. In each snapshot, an (identified) galaxy corresponds to exactly one subhalo in the \subfind{} catalogue. Tracing galaxies through time therefore equates to identifying those subhaloes in successive snapshots that represent the same galaxy (see Fig.~\ref{fig:appA_galSH} for a schematic illustration).  

\subsection{Extraction of links from subhalo catalogues}
We make use of the Lagrangian nature of the Hydrangea simulations, which allows us to identify the same particle in successive snapshots $i$ and $j$. Any subhaloes in $i$ and $j$ that have particles in common may, in principle, represent the same galaxy. We therefore begin by extracting all such `links' (i.e., particle overlaps) between subhaloes in $i$ and $j$, including particles of all types (not just dark matter). 

Intuitively, it may be more natural to only consider one link per subhalo in $i$ (as is done in the schemes of, e.g., \citealt{Rodriguez-Gomez_et_al_2015} and \citealt{Qu_et_al_2017}). However, our more general choice is justified in the regime of groups and clusters, where galaxies may not only grow, but also lose mass through tidal and/or hydrodynamic stripping. As illustrated in Fig.~\ref{fig:appA_galSH} (galaxy 3 between S2 and S3), this can lead to the majority of particles from one galaxy being transferred to another (e.g.,~the central cluster galaxy), and so would require predicting which particles are least likely to be transferred. Our approach is, instead, to test whether the main link (with the largest particle overlap, see below) leads to a viable descendant, and consider alternative links if this is not the case. In this way, we aim to trace individual galaxies for as long as possible.

\begin{figure}
  \centering
    \includegraphics[width=\columnwidth]{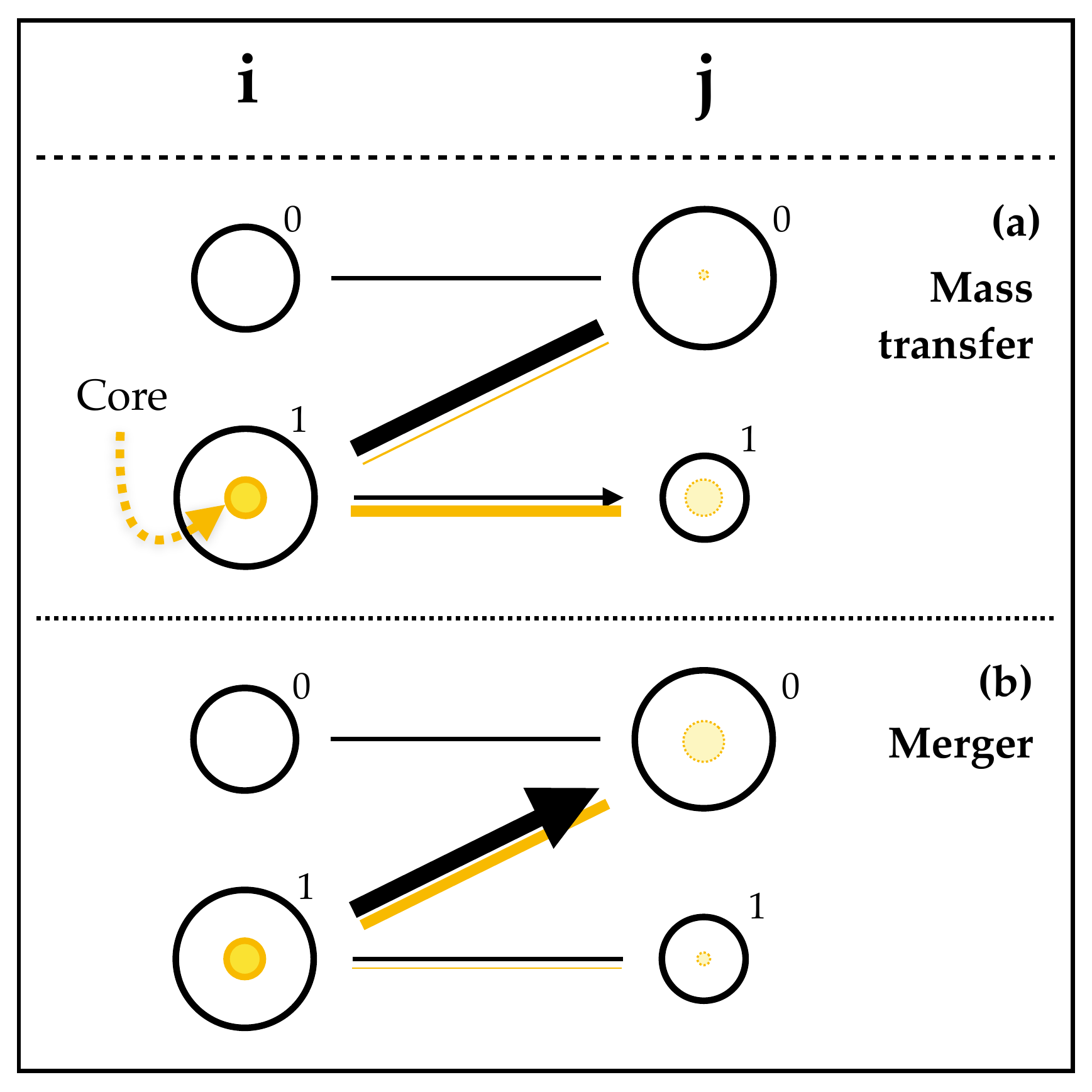}
       \caption{Identification of mass transfer by considering subhalo cores. Black circles represent two subhaloes each in two successive snapshots $i$ and $j$. The total particle overlap between them, as indicated by the width of the connecting black lines, is identical in the top and bottom configurations and could be interpreted as either mass transfer or a merger. The small yellow circle in subhalo $i_1$ represents its most bound `core' particles. As indicated by the yellow lines, these transfer differently in the two scenarios: under mass transfer (top), they remain mostly within their own galaxy and end up in $j_1$ (as indicated by the faint yellow circles in the $j$ subhaloes), but in a merger they end up (mostly) in subhalo $j_0$.}
    \label{fig:appA_coreTransfer}
\end{figure}

Nevertheless, we also give special consideration to a small set of `core' particles in each subhalo, defined as the 5 per cent most bound collisionless particles (i.e.,~excluding gas), limited to a maximum number of $10^5$. We have found that this is necessary to correctly trace galaxies in situations where a large fraction of particles are transferred from one galaxy to another as a result of a swap in the central/satellite classification between the two\footnote{Due to the way in which \subfind{} associates particles to satellite galaxies, there can be a large population of `ambiguous' particles that are preferentially assigned to the central, and therefore change subhalo membership if the central/satellite classification is swapped.}. As illustrated in the top panel of Fig.~\ref{fig:appA_coreTransfer}, this could lead to a transfer of galaxy ID from one object to the other ($i_1 \rightarrow j_0$), leaving one subhalo without descendant ($i_0$) and one without progenitor ($j_1$). Under the plausible assumption that the core particles are least likely to be affected by such an artificial particle transfer, they represent a robust tracer for their galaxy (yellow lines in Fig.~\ref{fig:appA_coreTransfer}) and can therefore distinguish this case from a similar situation in which two galaxies merge (bottom panel of Fig.~\ref{fig:appA_coreTransfer}).

Although this reasoning would suggest that the core should be as small as possible -- ideally containing only the few most-bound particles -- there are two arguments against a very small core. Firstly, small regions of a galaxy, such as a spiral arm or a central clump near a massive black hole, can occasionally become self-bound and dense enough that they appear as a separate entry in the subhalo catalogue. With a very small core, there is a risk that the majority of core particles become members of such a spurious subhalo, which would then be (wrongly) identified as the descendant. Secondly, we have found that within the most bound few per cent of particles the ordering in terms of binding energy fluctuates noticeably between snapshots. In other words, only a small fraction of, e.g., the 0.01 per cent most bound particles in $i$ are also the 0.01 per cent most bound in $j$, whereas at a threshold of $\approx\,$1--5 per cent, this fraction approaches unity. We have found that our choice of core fraction, a compromise between these competing constraints, produces stable tracing results across the full range of subhalo masses encountered in our simulations.

For each link, we record the subhalo to which it is connected in $i$ (which we call its `sender') and in $j$ (its `receiver'), as well as its total number of particles ($N$), their total mass ($M$), and number of particles that form the core of its sender ($N_\text{core}$, which may be zero). This information is then used in the following steps to deduce which links connect the same galaxy between snapshots, and which represent interactions between two different galaxies. We note that there are typically only slightly more links than subhaloes (within $\sim$50 per cent). 
      
\subsection{Compensation of prior mass exchanges}
\begin{figure}
  \centering
    \includegraphics[width=\columnwidth]{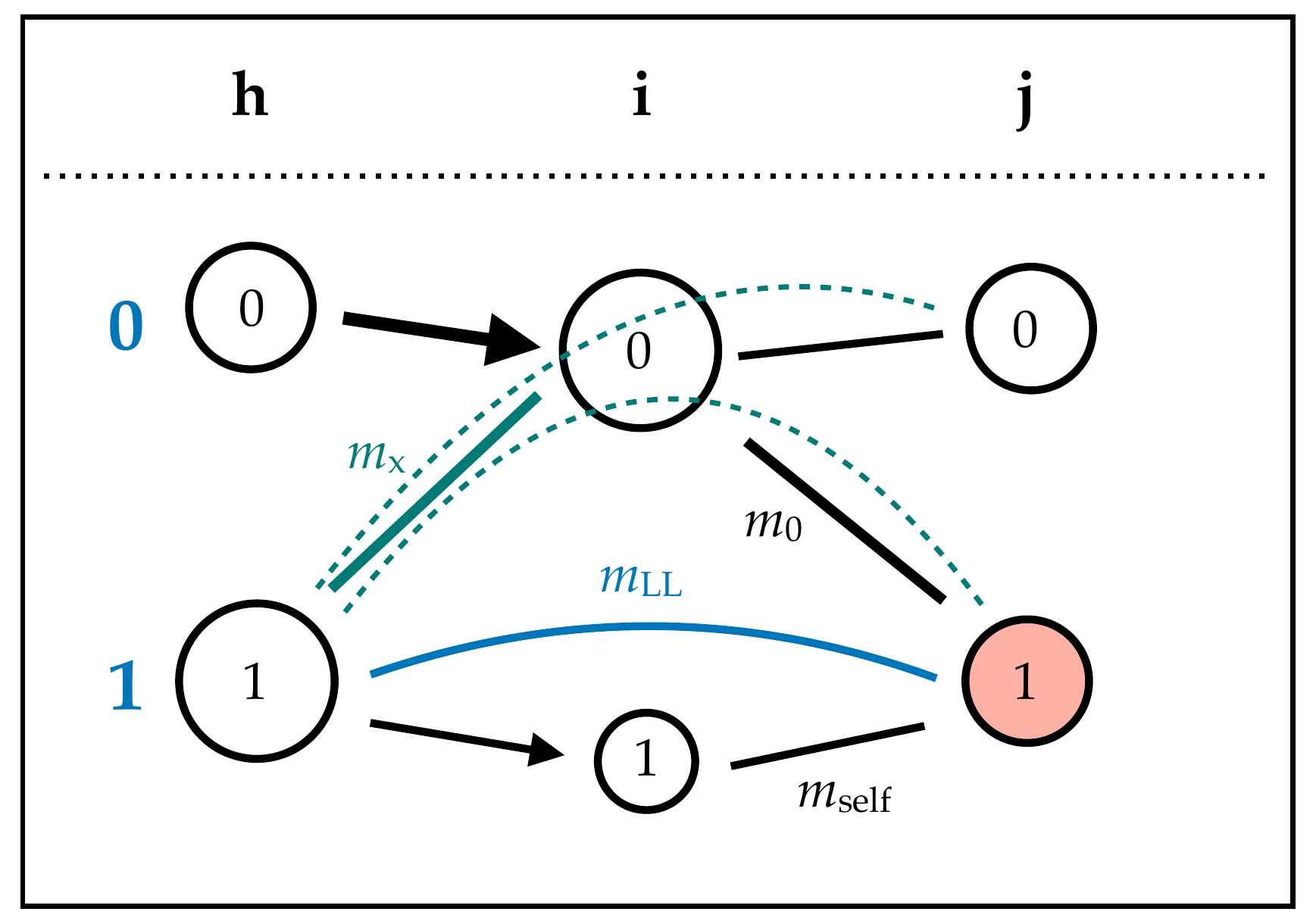}
       \caption{Accounting for prior mass exchanges between interacting galaxies. Subhalo $j_1$ (red) may be the descendant of either $i_0$ (galaxy 0) or $i_1$ (galaxy 1). Prior to the snapshot interval $i$--$j$, these galaxies have already exchanged mass (turquoise link $h_1$--$i_0$). These particles may continue along either of the two curved dashed turquoise lines, i.e., to $j_0$ or $j_1$, but only the latter case is relevant in determining the progenitor of $j_1$. As described in the text, the fraction following this path is estimated from the three link masses indicated as $m_\text{x}$, $m_\text{self}$, and $m_\text{LL}$. Lines with arrowheads represent connections following the same galaxy.}
    \label{fig:appA_histComp}
\end{figure}

In the simplest scenario, each subhalo in $j$ would receive only one link, in which case it could be unambiguously identified as representing the same galaxy as that link's sender in $i$. In reality, however, there will frequently be situations where a subhalo in $j$ receives several links, because it amalgamates matter from several subhaloes in $i$ (as a result of mergers or mass exchange). In this situation, illustrated in Fig.~\ref{fig:appA_histComp}, it is less obvious to decide which link to select as the one leading back to the progenitor of the target galaxy ($j_1$, highlighted in red). 

Physically, it is desirable to rank candidate progenitors in order of the mass that they contribute to the target galaxy in $j$. A complication with this approach is that galaxies may exchange mass -- both physically and numerically -- over an extended period of time. Neither the total subhalo masses in $i$, nor the mass of their links to the target in $j$, may therefore be a fair proxy of what fraction of the target galaxy is actually contributed by each progenitor candidate. A common way to correct for this, first suggested by \citet{DeLucia_Blaizot_2007}, is to rank the candidate progenitors by their `branch mass', i.e., the total mass of their progenitors in all previous snapshots.

Here, we follow a different strategy, exploiting the fact that the link network provides us with a complete record of all galaxy interactions prior to snapshot $i$. We can therefore reconstruct the net prior mass exchange between each pair of candidate progenitors, and then adjust the masses of their links to the target appropriately. 

However, as illustrated in Fig.~\ref{fig:appA_histComp}, the link network does not provide full information about the correlation between links in different snapshot intervals. For instance, the particles in the exchange link between snapshots $h$ and $i$ (solid turquoise line) may, in the interval $i$--$j$, either be carried on towards $j_0$ (the subhalo at the top), or to the target currently under consideration ($j_1$), as indicated by the two dashed lines. In principle, it is possible to distinguish between these cases by explicitly comparing particle IDs in different links, but this would add unjustified complexity to the code.

Instead, we estimate the connection between links by comparing the particle IDs between subhaloes in \emph{non-adjacent} snapshots, i.e., $h$ and $j$ in the current example. These `long links' directly measure how many particles have been transferred from subhalo $h_1$ to the target ($j_1$), but contain no information about which subhaloes (if any) these particles were associated with in $i$. We therefore estimate the total mass that has been `bypassed' around a subhalo ($i_1$ in Fig.~\ref{fig:appA_histComp}) via another (here, $i_0$) as $m_\text{by} = m_\text{LL} - m_\text{self}$ (limited to the interval [0, $m_x$]), where $m_\text{LL}$, $m_x$, and $m_\text{self}$ are, respectively, the masses of the long link ($h_1$--$j_1$), the exchange link ($h_1$--$i_0$), and the direct link $i_1$--$j_1$ (see Fig.~\ref{fig:appA_histComp}). If there is more than one subhalo in $i$ along which mass could be routed from $h_1$ to $j_1$, we compute the ratio $f_\text{by} = m_\text{by} / \Sigma_m$ (where $\Sigma_m$ is the sum of link masses from $h_1$ to all such subhaloes in $i$), limited to the interval [0, 1], and assume that a fraction $f_\text{by}$ of each individual exchange links is routed towards the target $j_1$.

This direct accounting scheme is only performed for up to four snapshot intervals prior to $i$--$j$ (i.e., typically 2.5 Gyr prior to $j$). In situations where galaxies have already interacted before this point, we count the full mass of these `old' exchange links. This is justified because such long-lasting interactions typically affect satellites orbiting a much more massive host, where there is almost guaranteed to be no confusion about the correct progenitor--descendant identification. Furthermore, the weighting factors $f_\text{by}$ computed for recent interactions are typically close to unity, with a (mass-weighted) average of $\approx\,$0.8.

\begin{figure}
  \centering
    \includegraphics[width=\columnwidth]{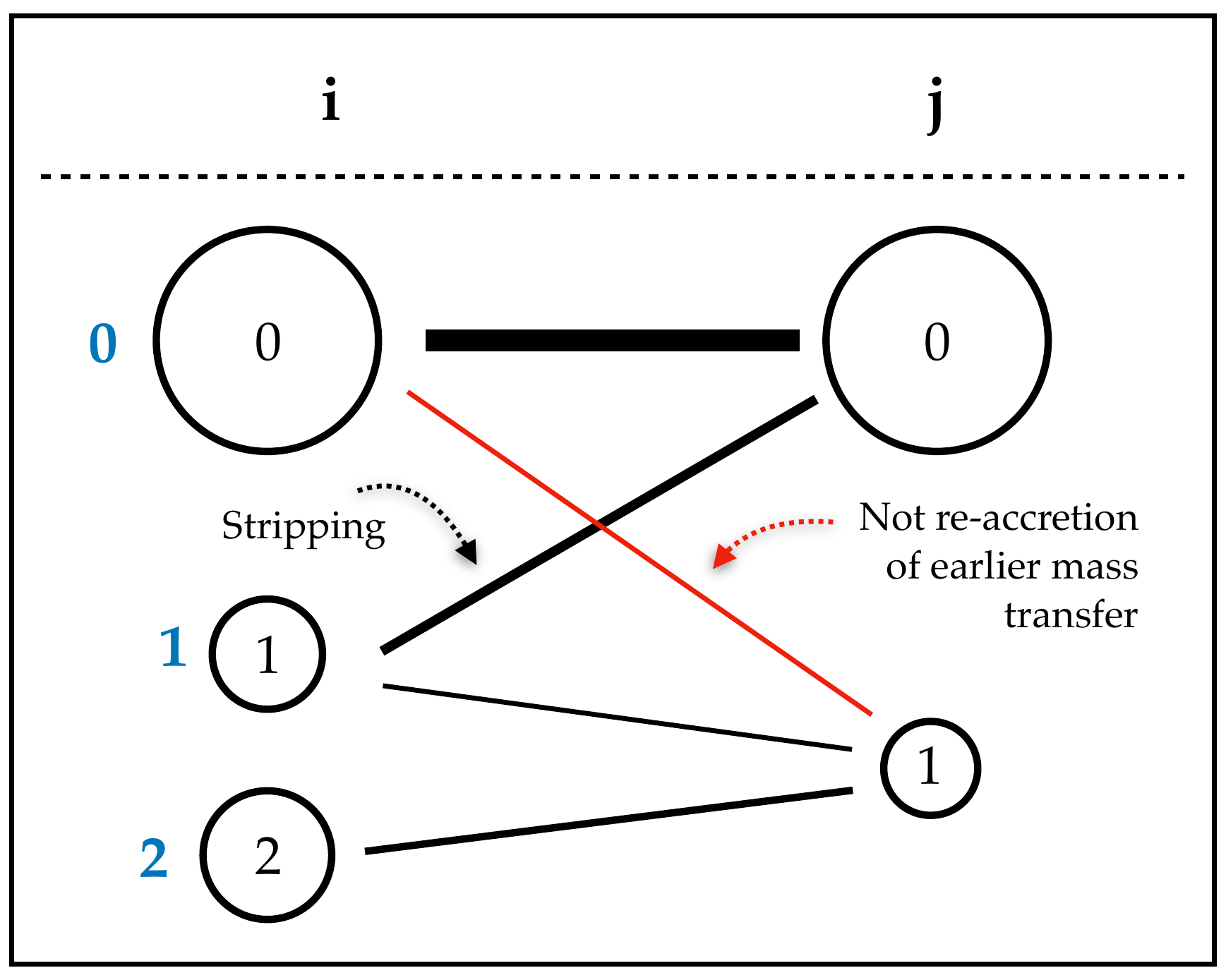}
       \caption{A situation in which transfer compensation is not allowed. Galaxy 1 transfers the majority of its mass to galaxy 0 ($i_1$--$j_0$), so it is unphysical to assume that it regains previously transferred mass at the same time ($i_0$--$j_1$). This can affect the choice between subhaloes $i_1$ and $i_2$ as progenitor of $j_1$.}
    \label{fig:appA_histCompBlock}
\end{figure}

As a result, we obtain a $N \times N$ matrix ($\mathsf{X}$) that contains for any pair of the $N$ candidate progenitors in $i$ ($N = 2$ in the example of Fig.~\ref{fig:appA_histComp}) an estimate of the total mass that was previously transferred from one to the other, and is now transferred to the target $j_1$. In other words, $X_{0, 1}$ contains (an estimate of) the mass in the link $i_1$--$j_1$ that should actually be counted towards the mass of link $i_0$--$j_1$. We then check whether this reassignment of mass is physically justified: galaxies that are currently undergoing significant stripping are unlikely to simultaneously re-accrete mass from other galaxies. As illustrated in Fig.~\ref{fig:appA_histCompBlock}, we therefore test for each candidate link to the target ($j_1$) whether it carries at least 2/3 the number of core particles ($n_C$) in the link with the highest $n_C$ from the same sender (which may be the link itself). If this is not the case, we judge that any galaxy connected along this link would be unlikely to re-gain particles and therefore set the corresponding entry in the exchange matrix $\mathsf{X}$ to zero.

As a final consistency check, we compute for each candidate progenitor in $i$ the total mass that it needs to return to the other progenitors. If this sum exceeds the total mass of its link to the target ($m_0$ in Fig.~\ref{fig:appA_histComp}), all exchanges are scaled down such that their sum is equal to this link mass. At last, we then define a `compensated link mass' $m_\text{comp}$ equal to the original link mass, reduced by the total mass returned to other links and increased by returns from other links. These compensated masses represent an estimate of the true contribution of each candidate galaxy to the target.

\subsection{Ranking and filtering of links}
To determine the order in which each link should be considered when finding the descendant of its sender, and the progenitor of its receiver, it is necessary to rank all links sent, and likewise all those received, by one subhalo according to some priority criterion. 

For sender ranking, we order links by their \emph{number of core particles} ($n_C$). As explained above, this ensures that the most bound particles, which are least affected by numerical mass transfers, are given the highest weight in determining the descendant of a galaxy. Note that we do not weight particles by mass here. This is because some core particles (notably BHs) can be orders of magnitude more massive than others, but we are here treating the particles as tracers to determine the most plausible descendant subhalo. To limit the effect of small-number statistics, we group all links with $n_C < 3$ and rank them at the bottom according to their \emph{total} number of particles. Because link selection proceeds from the highest ranked downwards (see next section), those links are typically irrelevant for determining the evolution of galaxies, except in very low-mass systems close to the resolution limit.

Analogously, all links to the same receiver subhalo are ranked according to their \emph{compensated mass} (see previous section). The reason for employing mass weighting here, and including gas particles, is that we are now interested in determining the galaxy that has contributed the most to the subhalo under consideration, so that all particles should be included and more massive particles \emph{should} carry more weight. Note that, for particle species that can change their mass over the course of the simulation, all masses are consistently determined at the later of the two snapshots ($j$). We first select those links with $n_C \geq 3$ and rank them in inverse order of their compensated mass, giving highest priority, to those with the highest $m_\text{comp}$. All links with $n_C < 3$ are then ranked in a second group at the bottom, again in inverse order of $m_\text{comp}$. We also assign an analogous receiver rank based on the original, uncompensated link masses.  
 
As a final step before selecting the links that connect subhaloes belonging to the same galaxy, we need to filter the links to exclude those that would lead to physically questionable connections. As discussed above, a key feature of our approach is to include the possibility of connecting subhaloes along links that carry only a minority of the (core) particles from the sender subhalo. This can be physically motivated in the case of strong stripping. It can also, however, lead to physically undesirable situations in which a small part of a galaxy that is temporarily identified as an independent subhalo -- which we refer to as a `spectre' in the following -- is selected as a descendant. 

\begin{figure}
  \centering
    \includegraphics[width=\columnwidth]{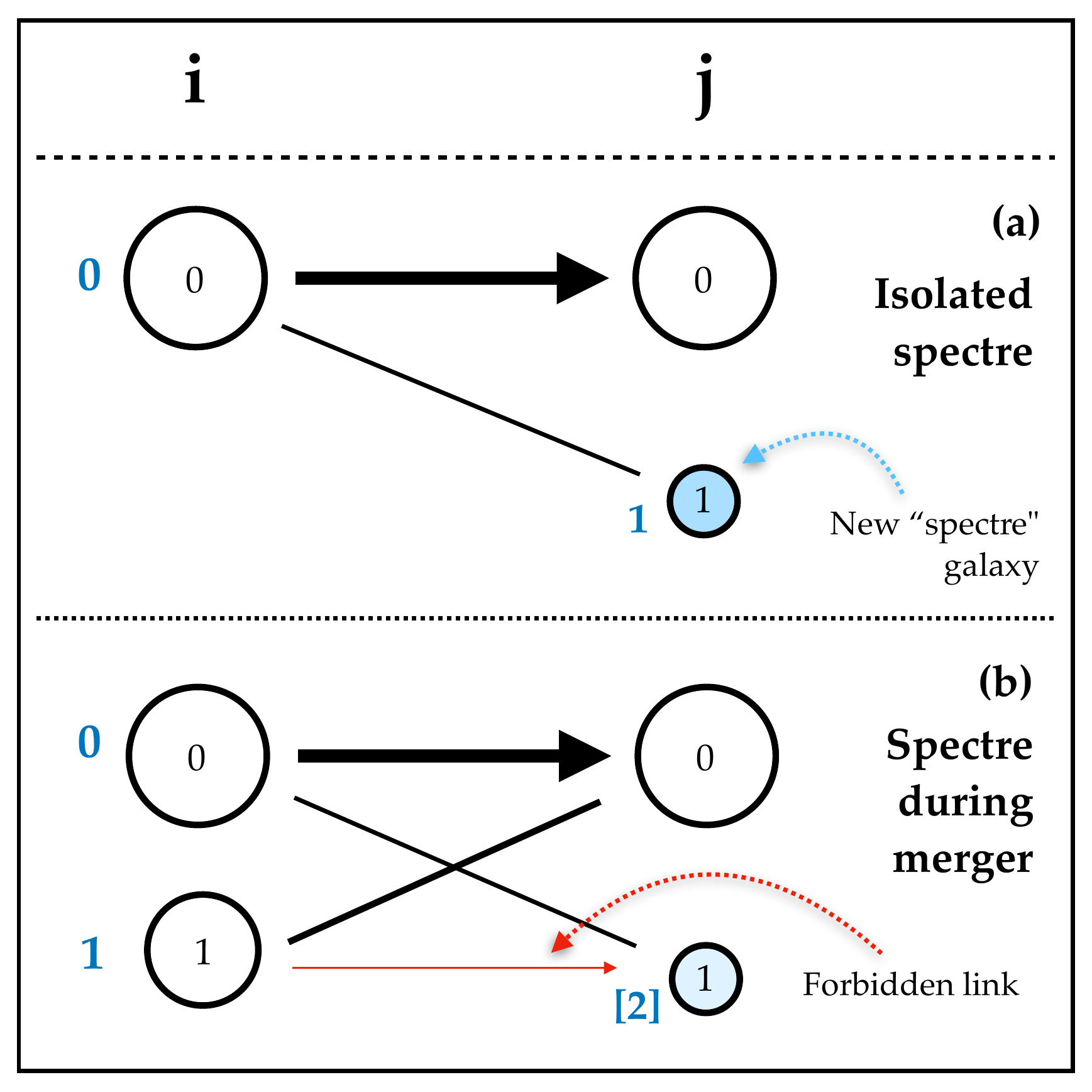}
       \caption{Exclusion of links to prevent connections to spurious `spectre' galaxies. \textbf{Top:} formation of a spectre in isolation, which is easily identified as a new galaxy. \textbf{Bottom:} when a spectre forms at the same time as a merger, it may be mis-identified as the descendant of the merging galaxy (1). The link $i_1$--$j_1$ is therefore marked as forbidden.}
    \label{fig:appA_spectres}
\end{figure}

To illustrate this possibility, consider the situation depicted in Fig.~\ref{fig:appA_spectres}. The top panel shows a simple scenario in which a spectre is formed from part of an existing galaxy. Because the link to the spectre (blue) carries only a small fraction of the core particles, it is ranked below the link to the main subhalo (indicated by an arrow). The latter is therefore identified as the galaxy's progenitor, while the spectre becomes a new galaxy (see below). 

The situation becomes more complex in the bottom panel. Here, the spectre is generated during a galaxy merger, and contains matter from both merging galaxies. Both subhaloes in $i$ send their highest-ranked link towards subhalo $j_0$, but because link $i_0$--$j_0$ contains more mass than $i_1$--$j_0$, $i_0$ is selected as the progenitor of $j_0$. However, $i_1$ has a second link to $j_1$ (the spectre), which could therefore undesirably be selected as its descendant.

To exclude such mis-identifications, we mark a link as `forbidden' if it carries neither an appreciable fraction of its sender subhalo's core nor of its total particles (i.e., $n_C < 2/3 n_C^\text{max}$ and $n < 2/3 n^\text{max}$, where $m$ and $n$ are the link mass and number of particles, and $n_C^\text{max}$ and $n^\text{max}$ the maximum number of core and total particles sent from subhalo $i_1$ along any link, respectively), and additionally satisfies either $m < 2/3 m_\text{recv}^\text{max}$ or $n < 2/3 n_\text{recv}^\text{max}$ (with $m_\text{recv}^\text{max}$ and $n_\text{recv}^\text{max}$ the maximum mass and particle number received along any link at subhalo $j_1$, respectively). These criteria capture both the situation depicted in the bottom panel of Fig.~\ref{fig:appA_spectres} -- in which all four conditions are satisfied -- and more subtle situations in which subhalo $i_1$ contributes the majority of mass, but not particles to the spectre (typically, this occurs if a massive BH particle originally belonging to $i_1$ becomes part of the spectre). At the same time, it does not exclude physically plausible scenarios, for instance if galaxy 1 were severely stripped ($m$ and $n$ close to $m_\text{recv}^\text{max}$ and $n_\text{recv}^\text{max}$, respectively) or re-accreted mass that was (physically or numerically) temporarily ascribed to $i_0$ ($n_C$ or $n$ close to $n_C^\text{max}$ or $n^\text{max}$).  
 
\subsection{Select connecting links}
Once all links are ranked, and those that are not physically plausible excluded, those links that connect the progenitor and descendant subhaloes of the same galaxy are selected. As discussed above, our approach is to consider several possible links for each subhalo and attempt connections first along the highest-ranked (most plausible) one, and then successively lower-ranked alternatives if the former were unsuccessful. The highest-priority class of links are clearly those with the highest sender- \emph{and} receiver-rank (which we denote as 0). For subsequent levels, there is an ambiguity between sender-rank 0, but receiver-rank 1, and sender-rank 1, but receiver-rank 0 (and analogous for lower levels). We here prioritise the former, which effectively prefers connecting galaxies in such a way that they retain the largest possible fraction of their core particles, rather than accrete the smallest possible fraction of mass from other objects. In practice, we have found that there is hardly any difference between these two ordering options.

We therefore iterate through successively lower sender ranks and consider all those links whose sender- and receiver-subhaloes have both not yet been connected. All links that are the only ones leading to their respective receiver subhalo can be selected to connect its two associated subhaloes as part of the same galaxy. 

For subhaloes in $j$ that receive multiple links in the current iteration, the most straightforward solution would be to select the one with the highest receiver rank. However, to increase the robustness of the tracing results, we first test whether the link with the highest receiver rank based on compensated mass is the same as that obtained with the uncompensated, original masses. If so, the situation is unambiguous and the highest receiver-rank link is connected. 

\begin{figure}
  \centering
    \includegraphics[width=\columnwidth]{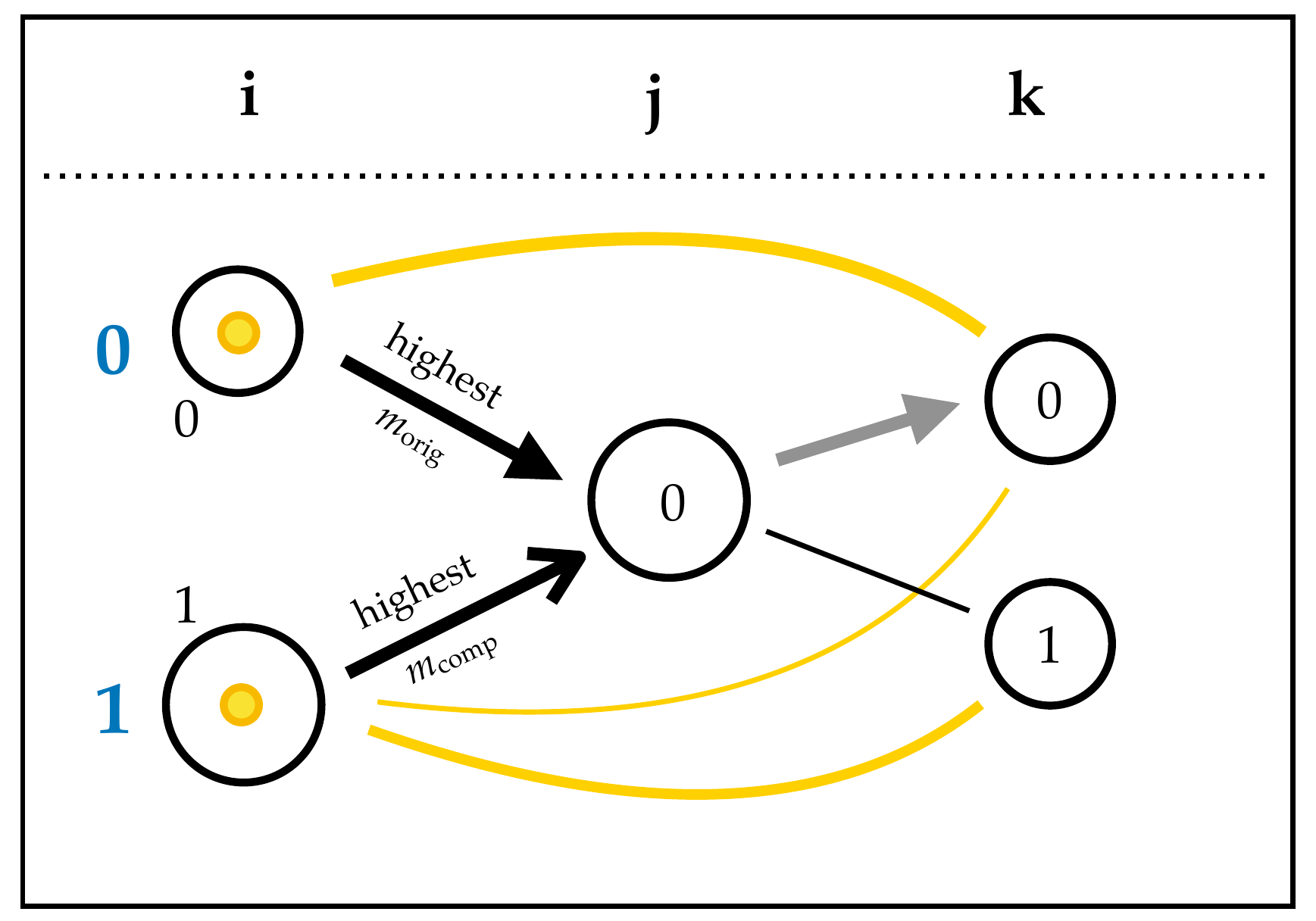}
       \caption{Decision between ambiguous progenitors. Subhalo $j_0$ could be the descendant of $i_0$ (which contributes the largest amount of mass, $m_\text{orig}$) or $i_1$ (whose mass contribution adjusted for prior exchanges, $m_\text{comp}$, is greatest). In the following snapshot ($k$), $k_0$ is the likely descendant of $j_0$, but only $i_0$ sends most of its core particles to this subhalo. The progenitor of $j_0$ is therefore chosen to be $i_0$ (continuing galaxy 0), while galaxy 1 is merged (and will likely be re-connected to $k_1$ in the next snapshot, as described in Section \ref{app:tracing_longlinks}). Under all other circumstances, $i_1$ would be selected as the progenitor of $j_0$.}
    \label{fig:appA_selection}
\end{figure}

If the two estimates differ -- as depicted in Fig.~\ref{fig:appA_selection} -- we proceed to the \emph{next} snapshot interval ($j$--$k$) and identify the `likely descendant' of the subhalo currently under consideration. The motivation behind this is to select the progenitor that maximises the long-term particle overlap between different subhaloes of the galaxy. The links between $j$ and $k$ are analysed in the same way as between $i$ and $j$, but without mass compensation. We next test whether there are long links between the two respective sender subhaloes in $i$ and the likely descendant in $k$ (i.e.,~whether they share any particles), and if so, whether the long-link corresponding to the $i$--$j$ link with the higher original receiver rank ($i_0$ in the example of Fig.~\ref{fig:appA_selection}) has a higher sender rank than its alternative ($i_1$). If this is the case, disregarding the mass compensation leads to a better long-term particle consistency, so the corresponding link is selected ($i_0$--$j_0$). In all other situations (including if $j$ is the last snapshot of the simulation), we select the link with the higher compensated receiver rank. Reassuringly, this covers the vast majority of cases: only in $\loa\,$10 per cent of ambiguous situations (i.e.~with different links labelled as highest-ranked by compensated and original mass) is the selected one that which is highest-ranked by original mass. Overall, the receiver ranks computed from compensated and original mass differ in only $\approx\,$0.5 per cent of selected links in each snapshot interval.

\subsection{Connect temporary non-identifications}
\label{app:tracing_longlinks}
At this point, all subhaloes in $i$ and $j$ that can plausibly be identified as representing the same galaxy are connected. However, it is still possible that a subhalo in $j$ could not be connected although it represents an existing galaxy: subhaloes are occasionally missed by \subfind{}, especially against the dense background of a massive galaxy cluster. An orbiting satellite galaxy may therefore be (temporarily) left without a counterpart in the subhalo catalogue in snapshot $i$. Uncorrected for, such galaxies would appear to spuriously disrupt and then form as new galaxies a short time later. This is clearly not an appropriate description of their actual evolution.

\begin{figure}
  \centering
    \includegraphics[width=\columnwidth]{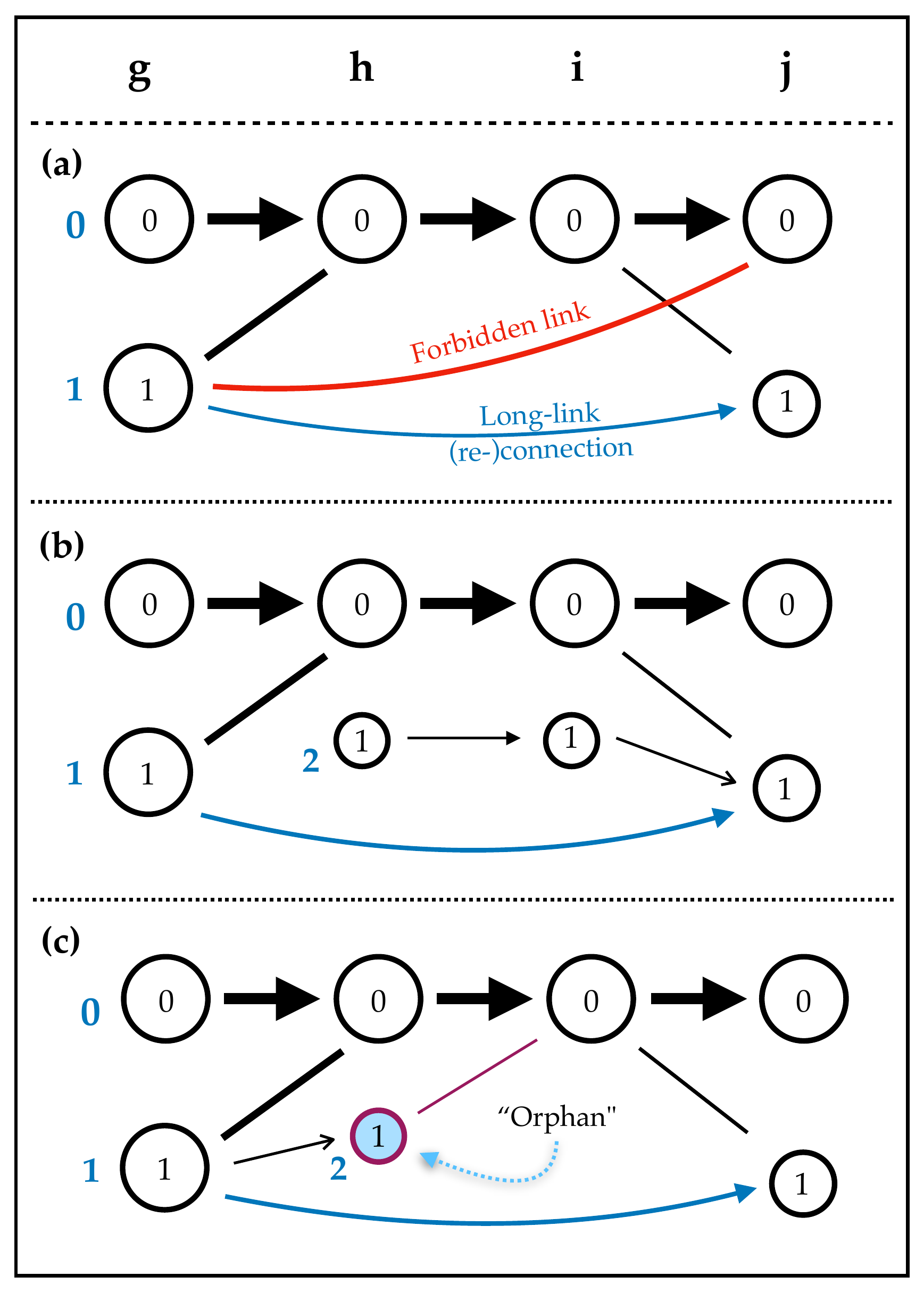}
       \caption{Three example situations involving the re-connection of subhaloes via long-links. Panel a \textbf{(top)} depicts the simplest case in which a subhalo without descendant ($g_1$) is re-connected to a subhalo without progenitor ($j_1$). As described in the text, link $g_1$--$j_0$ is forbidden because it would invert the survival order established in snapshot $h$. Panel b (\textbf{middle}) illustrates a long-link connection that `overrides' a (weaker) direct link from galaxy 2 (subhalo $i_1$). Panel c (\textbf{bottom}) shows a case in which an originally identified descendant ($h_1$) is disconnected and turned into an `orphan' that is only alive in one snapshot, because the long-link $g_1$--$j_1$ allows the continued tracing of galaxy 1 to snapshot $j$.}
    \label{fig:appA_longLinks}
\end{figure}

To prevent such mis-classifications, we make use of the already mentioned long links and retrospectively connect galaxies that were left without a descendant in an earlier snapshot ($z > z_i$) to a subhalo in $j$ (see Fig.~\ref{fig:appA_longLinks}). Such long-link connections are enabled over up to 5 snapshot intervals -- if a galaxy can still not be connected after this period, it is marked as disrupted. The selection of long-links is performed in analogy to the steps for direct links described above\footnote{The only difference is that, in computing their compensated masses, all links are allowed to re-gain particles. This is because galaxies may transition from stripping to re-accretion over the longer time intervals probed by long links.}. Because connections along long links should be an exception, rather than the rule, a number of additional constraints on their eligibility are imposed. All of them aim to limit the selection of long links to cases where they are clearly required:

\begin{enumerate}
\item The sender subhalo must not currently have a descendant, or if it does, this descendant subhalo could in turn not be connected. The first case covers the standard situation of a galaxy temporarily disappearing from the catalogue. The second, less common, case arises in situations where a small part of a disappearing galaxy (e.g.,~a spectre) is still identified as a separate subhalo, but is not strongly enough linked to the galaxy when it re-appears to establish a connection (bottom panel of Fig~\ref{fig:appA_longLinks}). In this case, we have the option of re-establishing a link from the last snapshot in which the galaxy was properly identified. The original descendant ($h_1$) is then disconnected and turned into an `orphan' galaxy that only exists in one snapshot.

\item The link must contain at least three core particles. This is to exclude connections that represent only marginal (core) particle overlap, which is not justified in the exceptional situation of linking across multiple snapshots.

\item The link must have a higher compensated mass than a (potential) currently connected shorter link to the same receiver subhalo. This is because we want to permit re-connections also in cases where a galaxy has, during its absence from the subhalo catalogue, accreted a smaller galaxy. Naively, the latter would be identified as its progenitor, but if the long link contributes more mass, it should be connected instead (middle panel of Fig.~\ref{fig:appA_longLinks}).

\item The link must not be received by any subhalo that is (backwards) connected to a subhalo that the sender subhalo already sends a link to (see the top panel of Fig.~\ref{fig:appA_longLinks} for a schematic illustration, in which the upper long-link, coloured in red, satisfies this criterion). This condition imposes that once a galaxy has been mis-classified as merged with another (as would happen if it has been missed by the subhalo finder against a dense group/cluster background), it cannot at a later point be identified as the progenitor of that galaxy. We have found that this is necessary to prevent unintended situations in which two galaxies of similar mass that have physically merged both `survive' by alternately skipping snapshots, often for many Gyr. 

\item If more than one long links satisfy the above constraints per sender subhalo -- i.e.,~if there is more than one option to re-connect a galaxy that has temporarily disappeared -- only that with the highest sender rank is allowed, or others that contain at least 2/3 of the number of core particles of that link (and which therefore offer a comparably strong connection).

\end{enumerate}

\subsection{New galaxies}
Once eligible long links are connected, each subhalo in $j$ that can be identified as continuing a pre-existing galaxy is connected with its progenitor, from which they inherit a unique galaxy ID (an identifier that is the same for all, and only those, subhaloes that represent the same galaxy; see Fig.~\ref{fig:appA_galSH}). Typically, some subhaloes are still not connected at this stage, because they represent newly formed galaxies. They are therefore assigned new galaxy IDs, which may be passed on to their descendants in subsequent snapshots\footnote{In the first snapshot, no subhalo can have a descendant and so each is assigned a new galaxy ID.}. 

While the majority of these new galaxies are relatively small, isolated objects that have just emerged above the detection threshold of \subfind, a subset of them are typically spectres, anti-hierarchically formed transient substructures within larger galaxies. Since they typically form in baryon-dominated regions, they can reach up to $\sim$$10^9\,\msun$ in stellar mass and could therefore be confused with genuine galaxies in stellar-mass selected samples. To avoid such contamination, we flag new galaxies as (likely) spectres if they receive at least one link from another galaxy and less than half their particles (by number or mass) were unbound in the previous snapshot. At high $z$, almost all newly emerging galaxies are genuine, but the fraction of spectres increases steadily with time and reaches $\approx\,$25 per cent at $z = 0$. In this paper, we have consistently excluded galaxies that were flagged as spectres.\footnote{Because most spectres are dominated by stars, they constitute an appreciable fraction in stellar mass limited galaxy samples, approximately 15 per cent at $\mgalstar > 10^9 \msun$.} 

\subsection{Carrier list}
The steps described above are repeated for all snapshots in the simulation. At the end of this process, every subhalo in any of the 30 snapshots corresponds to exactly one galaxy, and each galaxy to at most one subhalo in each snapshot. Typically, there are a factor of a few more galaxies than there are subhaloes at the final snapshot, because the majority does not survive\footnote{We note that this is not in contradiction to our findings in the main part of this paper: the majority of galaxies have lower peak masses than those that we have analysed, and/or disrupt in lower-mass haloes.} to $z = 0$.

As a final step, \textsc{Spiderweb} identifies the `carrier' of each disrupted galaxy, i.e., the galaxy that inherits the largest fraction of its (core) particles in the first snapshot after the galaxy has been lost. Note that such a carrier may not exist for all galaxies: if, e.g., gradual mass loss brings them below the \subfind{} detection threshold, its particles may all be unbound (not assigned to any subhalo) in the next snapshot. In many other cases, however, including those of interest in this paper, a galaxy is lost because it dissolves (merges) into a more massive galaxy. In this situation, most of its particles are still members of a galaxy, which can therefore be identified as the dissolved galaxy's `carrier'. 

To keep track of these mergers, we define a `carrier ID' for each galaxy, which is initially equal to its galaxy ID. Once a galaxy is disrupted, its carrier ID is updated to that of the galaxy which receives its highest (sender-)rank link, i.e., which carries the largest share of its (core) particles and is therefore the most plausible merger target. Any other galaxies that it had itself accreted in the past, and whose carrier IDs were therefore equal to its own, are likewise updated. On the one hand, this enables an easy identification of `where a galaxy ends up' at a given snapshot after its disruption. On the other hand, it also provides a simple method of determining all progenitors of a galaxy: those are simply all galaxies whose carrier ID is equal to the galaxy's own ID. 

As an illustration, Table \ref{tab:carrier} shows the carrier list for the scenario depicted in Fig.~\ref{fig:appA_galSH} in each of the four snapshots. Note that galaxy 2 undergoes two mergers, so its carrier ID is first changed to 0 (in S1) and then to 1 (in S2). The three galaxies that end up in subhalo 0 in S3 (which represents galaxy 1) all have the same carrier ID (1). Galaxies 1, 3, and 5 retain its own galaxy ID as carrier ID because they are still alive in S3.

\begin{table}
\caption{Example carrier list for the situation depicted in Fig.~\ref{fig:appA_galSH} in each of the four snapshots shown.}
\label{tab:carrier}

\begin{tabular}{l|cccccc}
Snapshot & Gal.~0 & Gal.~1 & Gal.~2 & Gal.~3 & Gal.~4 & Gal.~5 \\ \hline
S0 & 0 & 1 & 2 & --- & --- & --- \\  
S1 & 0 & 1 & 0 & 3  & 4   & --- \\  
S2 & 1 & 1 & 1 & 3  & 4   & 5 \\  
S3 & 1 & 1 & 1 & 3  & 3   & 5 \\  
\end{tabular}
\end{table}


\section{Robustness of subhalo identification}
\label{app:robustness}

In Fig.~\ref{fig:survival_detThresh}, we had shown that only a small fraction of galaxies with $\mgalpeak > 10^{10}\, \msun$ that are detected in the \subfind{} catalogue at $z = 0$ have a total mass below $5\times 10^8\, \msun$ at $z = 0$. It is conceivable that there are additional galaxies that do (physically) survive -- possibly with lower mass -- but which are missed in the \subfind{} catalogue for numerical reasons (either the limited resolution of our simulations, or shortcomings of the \subfind{} algorithm).

\subsection{Possibility of insufficient resolution}
\label{app:resComp}
To test the possibility that surviving subhaloes may be missed due to the finite resolution of the Hydrangea simulations, we have repeated our analysis on three simulations from the \eagle{} project that model the evolution of a (25 cMpc)$^3$ cube at two different levels of resolution. One, L0025N0376/Ref, uses the same mass and spatial resolution as the Hydrangea simulations analysed in the main part of this paper\footnote{The Ref model uses slightly different parameters for subgrid AGN feedback than the AGNdT9 model used for Hydrangea, but this is of no significance to the resolution test presented here.}. Two others (L0025N0752/Ref and L0025N0752/Recal) have a mass (spatial) resolution that is higher by a factor of eight (two), where the latter also uses recalibrated simulation parameters to achieve a similarly good match to the galaxy stellar mass function as the lower-resolution counterpart (see \citealt{Schaye_et_al_2015}). Due to their limited volume, these simulations only contain around a dozen low-mass groups ($\mhost = 10^{12.5}$--$10^{13.5}\,\msun$), with correspondingly larger statistical uncertainties than in the Hydrangea analysis.

\begin{figure}
  \centering
    \includegraphics[width=1.05\columnwidth]{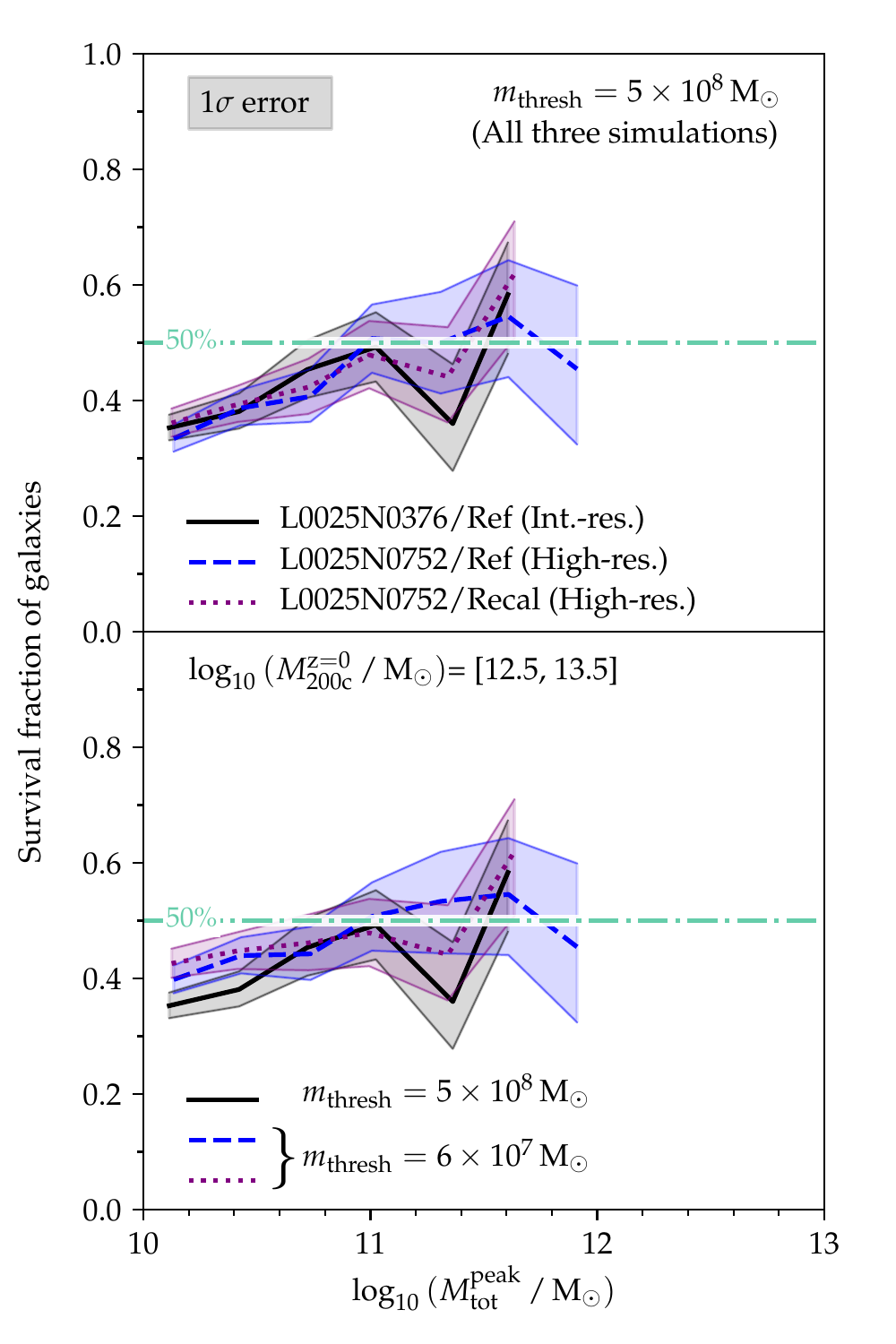}
       \caption{Survival fraction of satellites in low-mass groups in three (25 cMpc)$^3$ simulations from the \eagle{} project. L0025N0376/Ref (black solid line) uses the same resolution as the Hydrangea simulations analysed in the main part of this paper, while the other two (blue dashed and purple dotted lines, corresponding to different choices of subgrid parameters) have eight times better mass resolution. In the \textbf{top} panel, the same survival threshold ($5\times10^8\,\msun$) is applied to all three, while the \textbf{bottom} panel uses an eight times lower threshold for the higher-resolution simulations. In the first case, the higher resolution has no impact, but with a lower mass threshold the fraction of surviving galaxies increases slightly at the low-mass end.}
    \label{fig:appB_resComp}
\end{figure}

In Fig.~\ref{fig:appB_resComp}, we show the predicted survival fractions of low-mass group satellites as a function of their peak mass in these three simulations, in analogy to Fig.~\ref{fig:survival_total} for Hydrangea. The default-resolution simulation L0025N0376/Ref is shown as a solid black line, while the two higher-resolution versions L0025N0752/Ref and L0025N0752/Recal are represented by blue dashed and purple dotted lines, respectively. The top panel applies the same survival threshold of $5\times 10^8\,\msun$ to all three and reveals near-perfect agreement between the two resolution levels (irrespective of whether the subgrid parameters in the high-resolution version are re-calibrated or not). At least within the relatively low galaxy and halo masses accessible with these (25 cMpc)$^3$ simulations, the survival fractions above this threshold are therefore insensitive to the finite resolution of our simulations.

In the bottom panel, we explore the effect of lowering the survival mass threshold by a factor of eight for the two high-resolution simulations. Only at the lowest galaxy masses that we probe ($\mgalpeak \sim 10^{10}\,\msun$) does this increase the survival fraction, by up to 20 per cent (in a relative sense) from $35 \pm 2$ to $40 \pm 2$ or $43 \pm 3$ per cent (in the Ref and Recal models, respectively)\footnote{Fig.~\ref{fig:survival_detThresh} already indicated that a few per cent of these low-mass galaxies survive with a mass below our fiducial threshold at the intermediate resolution level of Hydrangea. We have verified, however, that the low-mass survival fraction is still raised in the higher-resolution simulations even when a uniform threshold of $6\times 10^7\,\msun$ is applied to all three simulations.}. At higher masses -- i.e., in the regime where baryonic processes are approximately converged \citep{Schaye_et_al_2015, Crain_et_al_2017} -- there is no indication that a significant population of galaxy remnants is missed because of the finite resolution of the Hydrangea simulations.

In summary, we conclude that our fiducial survival fractions (above a threshold of $5\times 10^8\,\msun$) are insensitive to an increase in resolution, and that they represent the total survival fractions for satellite galaxies with $\mgalpeak \gtrsim 3\times 10^{10}\,\msun$.

\subsection{Possibility of missed subhaloes} 
\label{app:SubfindRobustness}
To test the possibility that \subfind{} may be have missed (resolved) subhaloes, we have iteratively recomputed the bound mass of all those galaxies that are \emph{not} present in the \subfind{} catalogue. Starting from the dark matter, star, and black hole particles that constitute each galaxy in the last snapshot in which it is identified\footnote{We do not include gas because this component is expected to be efficiently removed through ram pressure.}, we compute the gravitational potential $\phi_i$ and kinetic energy $K_i$ of each particle $i$ (with respect to the mass-weighted average velocity of the particles in the most negative decile in gravitational potential). Any particle whose binding energy $\epsilon_i = \phi_i + K_i$ is positive is removed, and the iteration continued until less than 0.05 per cent of particles are removed in any one iteration.

\begin{figure}
  \centering
    \includegraphics[width=\columnwidth]{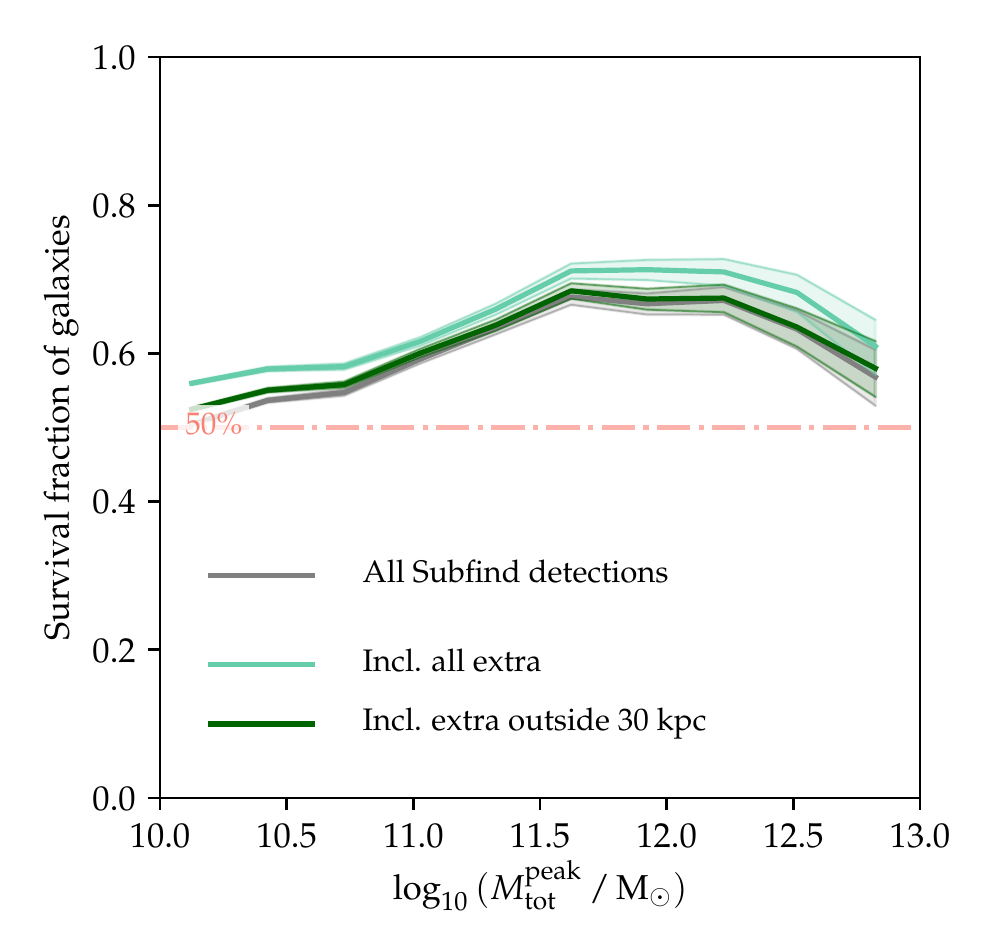}
       \caption{Fraction of galaxies in massive clusters ($\mhost > 10^{14.5}\,\msun$) that survive to $z = 0$. The grey line includes all galaxies present in the \subfind{} catalogue. The light green line adds galaxies which retain a self-bound remnant with at least ten particles when starting from all particles that are part of the galaxy in its last snapshot. The dark green line limits those recovered detections to only those that lie away from subhaloes in the \subfind{} catalogue (see text for details). There is no significant population of surviving galaxies that is missed by \subfind, particularly with the more restrictive definition (dark green).}
    \label{fig:appB_robustness}
\end{figure}

The result is shown as the light green line in Fig.~\ref{fig:appB_robustness}, which shows the fraction of galaxies in massive clusters ($\mhost > 10^{14.5}\, \msun$) that are either present in the original \subfind{} catalogue at $z = 0$ or for which the recomputation yielded a remnant with at least ten bound particles. For comparison, the grey line shows the fraction of only those galaxies identified by \subfind{} (as in Fig.~\ref{fig:survival_detThresh}). It is evident that the re-computation only increases the survival fraction by at most $\approx\,$5 per cent, implying that most physically surviving subhaloes are indeed detected by \subfind. 

In addition, the recomputation described above considered each galaxy individually and therefore represents an upper limit to the fraction of galaxies surviving as \emph{independent} self-bound structures. Some of them, while self-bound, may in fact be an indistinguishable part of a more massive galaxy, in the same way as a random selection of stars from the Milky Way's bulge may be self-bound to each other without constituting a separate galaxy. To estimate the impact of this effect, the dark green line in Fig.~\ref{fig:appB_robustness} includes only those galaxies with a remnant from the re-computation that lies outside of min(30 kpc, $R_{1/2}^\text{star}$), where $R_{1/2}^\text{star}$ is the stellar half-mass radius from any subhalo in the \subfind{} catalogue. In effect, this limits new detections to the outer halo of more massive galaxies. With this stricter definition, the difference between the recomputed and \subfind{} catalogue disappears almost entirely. Although this may in turn be overly restrictive -- some of the galaxies recovered in the recomputed catalogue within min(30 kpc, $R_{1/2}^\text{star}$) may in fact be genuine, independent, survivors -- we conclude from Fig.~\ref{fig:appB_robustness} that \subfind{} robustly identifies the vast majority of surviving galaxies in a group/cluster environment, and that our results in this paper are therefore not an artefact of our particular subhalo finder.


\section{Comparison to idealised experiments}
\label{sec:vdbo}

Our simulations predict substantial survival of even low-mass galaxies ($\mpeak \sim 10^{10}\,\msun$) in massive haloes, especially if they were accreted at $z < 2$. This appears to be in tension with the idealised $N$-body experiments of \citet{vanDenBosch_Ogiya_2018}, in which numerical disruption occurs even for satellites that are initially resolved by $\geq 10^5$ particles (corresponding to $\mpeak \gtrsim 10^{12}\,\msun$ at our resolution). For a quantitative comparison to their work, we use the criteria in their equations (21) and (22): these specify the minimum mass fraction that a satellite must retain to avoid numerical artefacts from inadequate force softening and particle discreteness noise, respectively, as
	\begin{equation}
	f_\text{bound, min}^\text{softening} = 1.12\, \frac{c^{1.26}}{f^2(c)} \left(\frac{\epsilon}{r_\text{s,\,0}} \right)^2\,\,\, \text{and}
	\label{eq:e1}
	\end{equation}   
	\begin{equation}
	f_\text{bound, min}^\text{discreteness} = 0.32 \left(\frac{N_\text{acc}}{1000} \right)^{-0.8}, 
	\label{eq:e2}
	\end{equation}
	where $c = r_\text{s,\,0}/r_\text{200c}$ is the concentration parameter of the galaxy's DM halo, $r_\text{s,\,0}$ its NFW scale radius \citep{Navarro_et_al_1996}, $f(c) = \ln(1+c) - c/(1+c)$, $\epsilon$ is the force softening length of the simulation (here equal to 0.7 proper kpc), and $N_\text{acc}$ the number of particles bound to the galaxy at the time of accretion. 
	
In our simulations, we take $N_\text{acc}$ from the \subfind{} catalogue at the last snapshot before accretion ($t_\text{branch}$). Instead of fitting NFW profiles, we estimate $c$ (and hence $r_\text{s,\,0}$) from the redshift-dependent $M_\text{200c}$--$c$ relation of \citet[their appendix B1]{Correa_et_al_2015}, including log-normal scatter with $\sigma = 0.11$ dex. For galaxies that were a central prior to accretion, we use the $M_\text{200c}$ of its FOF group, for others, we estimate $c$ from $\mpeak$. 

\begin{figure}
  \centering
    \includegraphics[width=\columnwidth]{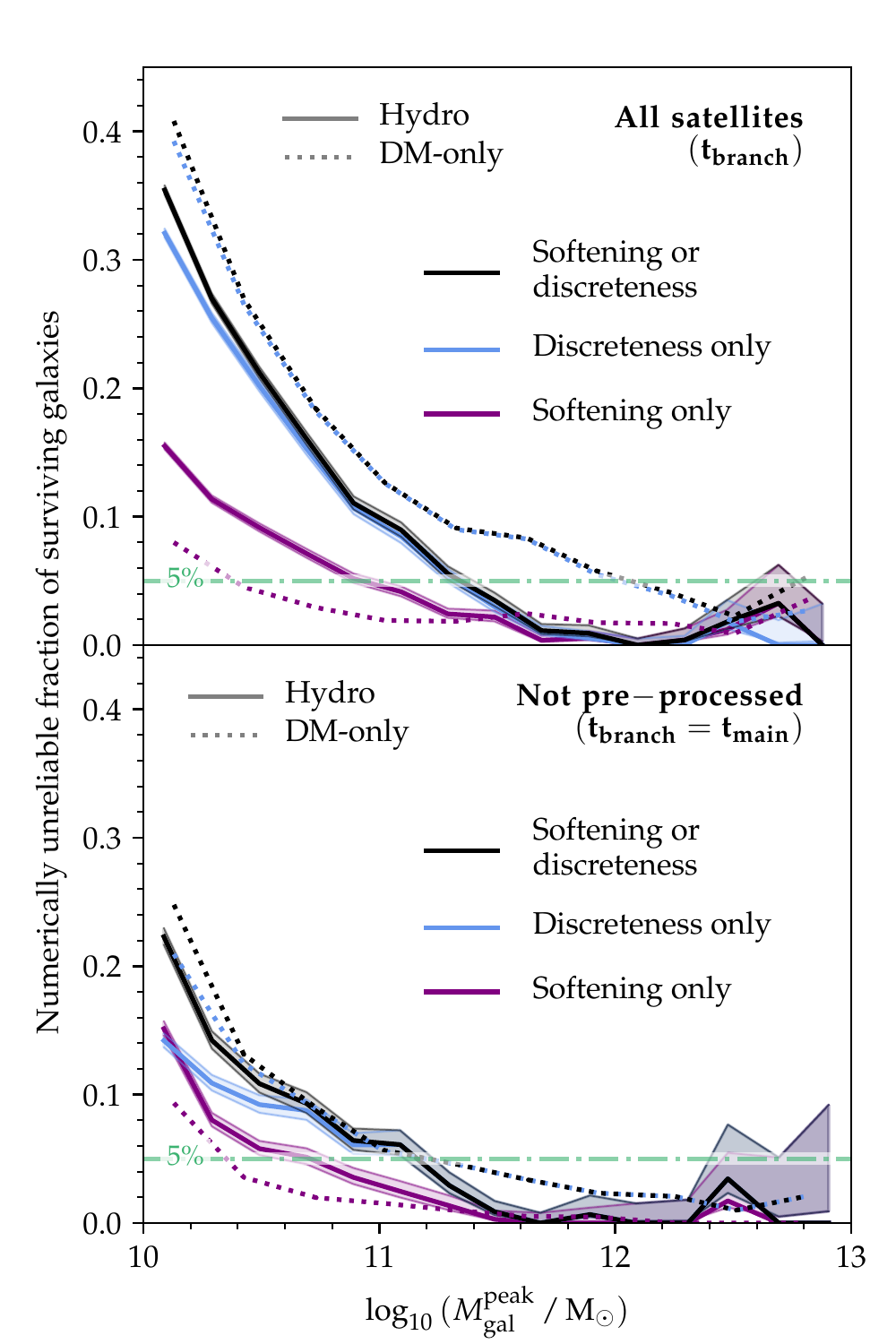}
       \caption{The fraction of surviving galaxies ($M_\text{tot}^\text{z = 0} > 5\times10^8 \msun$) whose $z = 0$ remnants are affected by numerical artefacts, according to the criteria of \citet{vanDenBosch_Ogiya_2018}. In the \textbf{top panel} we show all surviving satellites (including those that were pre-processed) and calculate pre-accretion properties at $t_\text{branch}$. In the \textbf{bottom panel} only galaxies that were not pre-processed are shown, with pre-accretion properties calculated at $t_\text{main}$. In both cases, the blue, purple, and black lines show, respectively, the fraction of survivors that violate the discreteness noise criterion, the softening criterion, or either of them. Results from the hydrodynamical simulations are shown as solid lines (with binomial $1\sigma$ errors as shaded bands), while DM-only simulations are represented by dotted lines. At $\mgalpeak < 3\times 10^{11}\, \msun$, numerical artefacts should play a non-negligible role, but this is not reflected in the survival fractions.}
    \label{fig:vdBO}
\end{figure}

{\color{blue}Fig.~\ref{fig:vdBO}} shows the fraction of surviving ($M_\text{tot}^\text{z = 0} > 5\times10^8 \msun$) galaxies in massive clusters ($\mhostzz > 10^{14.5}\,\msun$) whose remaining bound fraction ($f_\text{bound} = M_\text{tot}^\text{z = 0}/M_\text{tot}^\text{peak}$) is below the requirements in equations \eqref{eq:e1} and \eqref{eq:e2}, i.e., those that are expected to be susceptible to numerical artefacts. Galaxies affected by inadequate force softening are shown by purple lines, those subject to particle discreteness noise in blue, and black lines give the fraction of galaxies violating either constraint. In the top panel, all satellites are included, while the bottom panel only shows galaxies that were not pre-processed.
	
At $\mpeak \gtrsim 3\times 10^{11}\,\msun$, the fraction of remnants violating either numerical reliability constraint is close to zero in the hydrodynamical simulations (black solid lines). For these, numerical artefacts would only occur for bound fractions below 1 per cent, which are extremely rare (Fig.~\ref{fig:survival_detThresh}). We therefore conclude that massive galaxies are unaffected by numerical disruption, because they are \emph{physically} disrupted before losing sufficient mass to become numerically unreliable. We note that DM-only simulations do predict a small fraction ($\approx\,$10 per cent) of survivors even at $\mpeak = 10^{12}\, \msun$ that were so severely stripped that they should have become numerically unreliable, at least when pre-processed galaxies are included (dotted black/blue lines in the top panel; see also Fig.~\ref{fig:survival_detThresh}).

In lower-mass galaxies, the fraction of numerically unreliable remnants increases rapidly, and reaches 35 (23) per cent of all (non-pre-processed) satellites at $\mpeak = 10^{10}\,\msun$. The dominant driver is susceptibility to discreteness noise, with softening by itself only affecting a few per cent of galaxies except at $\mpeak < 3\times10^{10}\,\msun$. Because the discreteness noise threshold is independent of concentration, this means that the extent of (overall) numerical unreliability (black lines) is not significantly affected by our simplified approach of estimating $c$ from the \citet{Correa_et_al_2015} relation\footnote{We have verified that using the relation of \citet{Dutton_Maccio_2014}, or even shifting the $M_\text{200c}$--$c$ relation upwards by $1\sigma = 0.11$ dex does not cause an appreciable difference in the overall fraction of numerically unreliable galaxies.}. In the DM-only runs, the fraction of unreliable remnants of low-mass galaxies is even slightly higher.

These numerically unreliable survivors should be accompanied by similarly massive galaxies that were numerically disrupted. In our simulations, we instead find near-complete survival of low-mass galaxies in massive clusters (Figs.~\ref{fig:influence_nppSurvFrac_total} and \ref{fig:survival_accTime}), and only small differences with satellite mass in all host mass bins (Fig.~\ref{fig:survival_total}) that also tend to be shallower at the lowest masses, rather than steepening as seen in the fraction of unreliable survivors. Finally, the survival fractions of low-mass galaxies are (slightly) lower in the hydrodynamical simulations compared to the DM-only runs (Figs.~\ref{fig:survival_detThresh} and \ref{fig:influence_nppSurvFrac_total}), although they have a slightly lower fraction of unreliable remnants. All this suggests that numerical disruption of satellites is rare even when the bound fraction falls below the \citet{vanDenBosch_Ogiya_2018} thresholds.

There are (at least) two possible explanations for this. Firstly, it might be that many low-mass galaxies are actually affected by numerical artefacts in our simulations, but that these are mild and only cause some unphysical mass loss, rather than any appreciable number of extra disruption events (even at a threshold of $0.1\,\mgalpeak$ as explored in Fig.~\ref{fig:survival_detThresh}). Secondly, the criteria of \citet{vanDenBosch_Ogiya_2018} may be overly conservative in the more realistic situations produced by our simulations. It is conceivable, for instance, that most of the mass loss is due to transient events, such as encounters with other satellites (`galaxy harrassment'; \citealt{Moore_et_al_1996}) or pre-processing, while the tidal field of the host itself is too weak to induce numerical inaccuracies. More work would be required to test these scenarios in detail.

\end{appendix}

\bsp	
\label{lastpage}
\end{document}